\documentclass[twocolumn]{aastex61}
\pdfoutput=1 
\usepackage{amsmath,amstext}
\usepackage[T1]{fontenc}
\usepackage{apjfonts} 
\usepackage[figure,figure*]{hypcap}
\usepackage{gensymb}
\usepackage{subfloat}
\usepackage{multirow}

\shorttitle{Coherent Radio Bursts on M Dwarfs} 
\shortauthors{Villadsen et al.}

\begin{document}

\title{Ultra-Wideband Detection of 22 Coherent Radio Bursts on M Dwarfs}

\correspondingauthor{Jackie Villadsen}
\email{jvillads@nrao.edu, jackievilladsen@gmail.com}

\author[0000-0003-3924-243X]{Jackie Villadsen}
\altaffiliation{Jackie Villadsen is a Jansky Fellow of the National Radio \\ Astronomy Observatory.}
\affil{National Radio Astronomy Observatory \\
	   520 Edgemont Rd \\
	   Charlottesville, VA 22903}
\affil{California Institute of Technology \\
		     1200 E California Blvd, MC 249-17 \\
             Pasadena, CA 91125}

\author[0000-0002-7083-4049]{Gregg Hallinan}
\affil{California Institute of Technology \\
		     1200 E California Blvd, MC 249-17 \\
             Pasadena, CA 91125}

\begin{abstract}
Coherent radio bursts detected from M dwarfs have some analogy with solar radio bursts, but reach orders of magnitude higher luminosities. These events trace particle acceleration, powered by magnetic reconnection, shock fronts (such as formed by coronal mass ejections, CMEs), and magnetospheric currents, in some cases offering the only window into these processes in stellar atmospheres.  We conducted a 58-hour, ultra-wideband survey for coherent radio bursts on 5 active M dwarfs.  We used the Karl G. Jansky Very Large Array (VLA) to observe simultaneously in three frequency bands covering a subset of 224-482~MHz and 1-6~GHz, achieving the widest fractional bandwidth to date for any observations of stellar radio bursts. We detected 22 bursts across 13 epochs, providing the first large sample of wideband dynamic spectra of stellar coherent radio bursts.  The observed bursts have diverse morphology, with durations ranging from seconds to hours, but all share strong (40-100\%) circular polarization. No events resemble solar Type~II bursts (often associated with CMEs), but we cannot rule out the occurrence of radio-quiet stellar CMEs. The hours-long bursts are all polarized in the sense of the x-mode of the star's large-scale magnetic field, suggesting they are cyclotron maser emission from electrons accelerated in the large-scale field, analogous to auroral processes on ultracool dwarfs.  The duty cycle of luminous coherent bursts peaks at 25\% at 1-1.4 GHz, declining at lower and higher frequencies, indicating source regions in the low corona.  At these frequencies, active M dwarfs should be the most common galactic transient source.

\end{abstract}


\section{Introduction}\label{section:intro}
The Sun produces intense bursts of emission at low radio frequencies, which caused it to become the~second detected astrophysical radio source after the Galactic center \citep{hey1946,southworth1945,orchiston2005}.  These bursts are classified according to their morphology in the time-frequency plane (``dynamic spectrum''), reviewed in \citet{dulk1985} and \citet{bastian1998}. There are a large number of types of solar radio bursts, with durations from milliseconds to weeks, at frequencies from kHz to GHz. Many of the bursts have high brightness temperatures, strong circular polarization, and/or narrow bandwidth, indicating that they are powered by coherent emission mechanisms, plasma emission or electron cyclotron maser (ECM) emission.   As a picture has emerged for the origin of the different types of bursts, they have proved to have significant diagnostic power for processes in the solar corona and interplanetary medium, including Earth-impacting (``geoeffective'') events.

Two tracers of geoeffective space weather events are solar Type II and Type III bursts, which occur at frequencies from hundreds of MHz down to kHz.  These bursts occur at the local plasma frequency and its second harmonic, and their frequency declines over time as a coronal source moves outwards into lower densities.  Solar Type II bursts trace coronal shock fronts, moving at speeds of order 1000~km\,s$^{-1}$, passing through a decade in plasma frequency in a matter of minutes.  These shocks may be driven by coronal mass ejections (CMEs) or by flare blast waves \citep{vrsnak2008}. Solar Type III bursts are produced by electron beams moving at speeds of order 10\% of the speed of light, with radio frequency decreasing exponentially on a timescale of seconds; when these bursts occur at low frequencies, they indicate that energetic electrons (and likely protons as well) are escaping the solar corona on open magnetic field lines.  Both of these classes of bursts are correlated with solar space weather events that impact Earth: coronal mass ejections and solar energetic particle (SEP) events \citep{kouloumvakos2015}. Analogous events, tracing bulk motion of plasma, may be detectable from nearby stars; and based on Sun-as-a-star data, such stellar radio bursts can be used to recover approximate dynamical properties of CMEs \citep{crosley2017}. This suggests that low-frequency radio observations can be used to characterize stellar mass loss and the space weather environment experienced by extrasolar planets.

Based on planet occurrence statistics, many of the nearest habitable-zone Earth-size planets should be found around M~dwarfs \citep{dressing2015}, a prediction borne out by the planet detections around Proxima Cen \citep{anglada-escude2016}, TRAPPIST-1 \citep{gillon2017}, and Ross~128 \citep{bonfils2017}.  However, the space weather environment around M dwarfs may differ dramatically from our own solar system, complicating considerations of exoplanet habitability \citep{shields2016}. In contrast to solar-type stars, M dwarfs retain their youthful rapid rotation \citep{newton2016} and vigorous magnetic activity \citep{west2008} for hundreds of millions to billions of years. Even old, slowly rotating M dwarfs can have significant magnetic spot coverage \citep{newton2016} and powerful flares \citep{france2016}. Active M dwarfs have a high rate of energetic flares \citep{lacy1976}, with potentially dramatic consequences for planetary atmospheres: coronal mass ejections can compress the planetary magnetosphere and erode the atmosphere \citep{khodachenko2007,lammer2007,kay2016,patsourakos2017} and stellar energetic particles can penetrate deep into the atmosphere and interact chemically \citep{segura2010,airapetian2016,tilley2017,howard2018}.  There is limited observational evidence of CMEs and energetic protons around stars other than the Sun, so models of the impact of stellar ejecta on exoplanetary atmospheres must rely on extrapolating solar relationships, between flares and CMEs \citep{yashiro2009,aarnio2011,drake2013,osten2015,odert2017}, magnetic flux and CMEs \citep{cranmer2017}, and flares and energetic protons \citep{belov2007,cliver2012,atri2017,youngblood2017}, to apply to the high-energy flares and strongly magnetized coronae of active M dwarfs.

There are both observational and theoretical reasons to expect that solar scaling relationships for CMEs and energetic protons may not apply to active M dwarfs.  Observations of Lyman~$\alpha$ absorption in the astrosphere of active M dwarf EV~Lac indicate that it has a mass loss rate similar to the Sun \citep[assuming isotropic mass loss;][]{wood2005}, in contrast to the flare-based prediction of \cite{osten2015} that it would be up to 500 times greater due to frequent CMEs.
The strong global field of active M dwarfs \citep[kG strength -][]{morin2008} may confine or prevent eruptions \citep{vidotto2016,odert2017,alvarado-gomez2018}, as has been suggested for the few strong solar flares that are not accompanied by CMEs \citep{thalmann2015,sun2015}.  In addition, the most intense solar energetic particle events that hit Earth are caused by particle acceleration at CME shocks \citep{reames2013}, so solar flare-proton scaling relationships will not apply well to active stars if energetic flares are not accompanied by CMEs or if those CMEs do not form shocks to accelerate particles; shocks may be rare due to high Alfv\'en speeds in the strongly magnetized stellar coronae. With all these complicating factors, there is a need for direct stellar observations to test for the presence of CMEs, shocks, and energetic particles.

Beyond radio observations, various other observational techniques provide evidence suggestive of stellar eruptions. Chromospheric and coronal emission lines sometimes display blueshifted emission components during flares on M~dwarfs \citep[e.g.,][]{houdebine1990,fuhrmeister2004,leitzinger2011,vida2016}.
Surveys with many hours of stellar observations \citep{leitzinger2014,vida2016} have detected these blueshifted features at a rate much lower than the rate of energetic flares, but \cite{leitzinger2014} attribute this low rate of CME detections to sensitivity limits.

\cite{harra2016} searched for ways of differentiating between eruptive and non-eruptive energetic solar flares, using optical, EUV, and X-ray signatures potentially observable in stars.  They ruled out long flare decay timescales \citep[suggested as evidence of a stellar eruption on AU~Mic;][]{cully1994,katsova1999} and identified only one reliable indicator of eruptive flares: coronal dimming, when coronal EUV line emission decreases as a CME carries away coronal material \citep[detectable in Sun-as-a-star observations;][]{mason2014}. The closest stellar analog to this so far is a post-flare reduction in EUV continuum and transition region line emission on active M dwarf EV~Lac \citep{ambruster1986}. Stellar eruptions have also been invoked to explain diverse cases of absorption of stellar chromospheric line emission \citep{collier_cameron1989,bond2001} and coronal X-ray continuum \citep{haisch1983,favata1999}.


Another observational signature of CMEs may be their impact on the circumstellar environment.  CMEs have been proposed as drivers of observed variability in a hot Jupiter atmosphere \citep{lecavelier2012} and circumstellar disks \citep{osten2013,boccaletti2015}.

Finally, wideband radio spectroscopy of stellar coherent radio bursts can track outwards-moving coronal sources, offering the potential for direct observations of stellar eruptions. Coherent radio bursts are abundant on active M dwarfs, dominating the population of radio flares observed on these stars below 5~GHz \citep{bastian1990review}. However, given the strong global magnetic field and rapid rotation of active M~dwarfs, it is not clear whether these bursts can be interpreted within the solar paradigm.  Dynamic spectroscopy has helped to characterize these bursts, finding features such as striae with rapid frequency drift \citep{osten2008}, rapid time variability, and strong circular polarization that provide evidence that ECM may be responsible for many of these bursts.  A number of coherent radio bursts in stellar atmospheres are also attributed to plasma emission \citep{stepanov2001,osten2006}, a promising sign that events analogous to the solar Type II and Type III plasma-frequency bursts may occur in stellar atmospheres.  In particular, \cite{osten2006} detected a 1.4-GHz burst on AD~Leo with gradual frequency drift that may correspond to a disturbance in the low corona moving outwards at speeds of 1200-12000~km\,s$^{-1}$, as expected for a CME or a flare blast wave.

Searches for stellar coherent radio bursts at low frequencies (<1~GHz, the frequency range for observing solar Type II and III radio bursts) have been inconclusive: while early observations yielded bright events coincident with optical flares \citep[e.g.,][]{lovell1963}, these events may have been due to RFI. Later experiments imposed additional requirements to distinguish between stellar bursts and RFI, including: interferometric observations \citep[finding two hours-long 408-MHz bursts on YZ~CMi;][]{davis1978}, absence of the signal in off-source beams \citep[identifying a handful of candidate events; e.g.,][]{abdul-aziz1995,abranin1998}, and spectral shape of the burst \citep[yielding a short 430-MHz burst on YZ~CMi;][]{bastian1990}. Other low-frequency observations produced marginal or uncertain detections \citep[e.g.,][]{jackson1990}. Recent Low-Frequency Array (LOFAR) beam-formed observations of YZ~CMi at 10-190~MHz yielded no detections \citep{crosley2016}, while \cite{lynch2017} detected multiple 154-MHz bursts on UV~Cet using interferometric imaging with the Murchison Widefield Array (MWA). VLA observations of EQ~Peg at 230-470~MHz  detected two coherent bursts \citep{crosley2018vla}, but no Type~II-like events in spite of an optical flare rate implying that this star may produce CMEs at least once per hour \citep{crosley2018multiwavelength}.

While many observational methods provide hints of stellar eruptions, none so far have yielded unequivocal evidence of an eruption nor a systematic collection of candidate CME events.  This is due to limitations on sensitivity, wavelength range, and/or available observing time of existing facilities, as well as the possibility that M~dwarfs do not produce CMEs according to the solar paradigm.  Recent and upcoming improvements in radio facilities mean that radio observations may now be able to detect a population of stellar space weather events.

\renewcommand{\arraystretch}{1.00}
\begin{deluxetable*}{ccccccc}
\tablecaption{Summary of Observations.\label{table:obs_summ}}
\tablehead{\colhead{Star} & \colhead{Date} & \colhead{Duration (hrs)} & \colhead{Frequency (GHz)} & \colhead{Flux Calibrator} & \colhead{Gain Calibrator} & \colhead{VLA Configuration}}
\startdata
\\[-1.5ex]
AD~Leo 	& 2013 Apr 24	& 2	& 2-6\tablenotemark{a} & 3C147 & J1111+1955	& D					\\
		& 2013 May 26	& 2	& 1-6 				& 3C147	& J1111+1955	& DnC				\\
		& 2013 Jun 2	& 2	& 1-6 				& 3C147	& J1111+1955	& DnC$\rightarrow$C	\\
		& 2013 Jul 27	& 2 & 1-6 				& 3C147	& J1111+1955	& C					\\
		& 2015 May 12	& 2	& 0.22-0.48, 1-4	& 3C286	& J1111+1955	& B$\rightarrow$BnA	\\
		& 2015 May 13	& 2	& 0.22-0.48, 1-4	& 3C286	& J1111+1955	& B$\rightarrow$BnA	\\
		& 2015 Jul 5	& 4	& 0.22-0.48, 1-4\tablenotemark{b}	& 3C286	& J1111+1955	& A		\\
		& 2015 Jul 19	& 4	& 0.22-0.48, 1-4\tablenotemark{b}	& 3C286	& J1111+1955	& A		\\
		& 2015 Sep 5	& 4	& 0.22-0.48, 1-4\tablenotemark{b}	& 3C286	& J1111+1955	& A		\\ [1.5ex]
UV~Cet	& 2013 May 26	& 2	& 1-6			& 3C48	& J0204-1701	& DnC				\\
		& 2013 Jun 2	& 2	& 1-6			& 3C48	& J0204-1701	& DnC$\rightarrow$C	\\
		& 2013 Aug 24	& 2	& 1-6			& 3C48	& J0204-1701	& C					\\
		& 2015 May 15	& 2	& 1-2\tablenotemark{a}	& 3C147	& J0204-1701 & B$\rightarrow$BnA	\\
		& 2015 May 16	& 2	& 0.22-0.48, 1-4	& 3C147	& J0204-1701 	& B$\rightarrow$BnA		\\
		& 2015 Jul 4	& 4	& 0.22-0.48, 1-4\tablenotemark{b}	& 3C147	& J0204-1701 	& A			\\
		& 2015 Jul 18	& 4	& 0.22-0.48, 1-4\tablenotemark{b}	& 3C147	& J0204-1701 	& A			\\
		& 2015 Sep 6	& 4	& 0.22-0.48, 1-4\tablenotemark{b}	& 3C147	& J0204-1701 	& A			\\ [1.5ex]
EQ~Peg	& 2015 May 15	& 2	& 0.22-0.48, 1-4	& 3C147	& J2254+2445	& B$\rightarrow$BnA		\\
		& 2015 May 16	& 2	& 0.22-0.48, 1-4 & 3C147 & J2254+2445 & B$\rightarrow$BnA \\ [1.5ex]
EV~Lac	& 2015 May 12	& 2	& 0.22-0.48, 1-4	& 3C286	& J2202+4216	& B$\rightarrow$BnA		\\
		& 2015 May 13	& 2	& 0.22-0.48, 1-4 & 3C286 & J2202+4216 & B$\rightarrow$BnA	\\ [1.5ex]
YZ~CMi	& 2015 May 11	& 2	& 0.22-0.48, 1-4	& 3C147	& J0745+1011	& B$\rightarrow$BnA		\\
		& 2015 May 13	& 2	& 0.22-0.48, 1-4	& 3C147	& J0745+1011 & B$\rightarrow$BnA \\ [1.5ex]
\enddata
\tablenotetext{a}{Narrower bandwidth is due to bad data in some bands during these observations.}
\tablenotetext{b}{The 4-hour blocks have simultaneous VLBA observations at 8.2-8.5~GHz.}
\end{deluxetable*}

The recently upgraded Karl G. Jansky Very Large Array (VLA) offers wide instantaneous bandwidth and low observing frequencies, making it a powerful instrument for searching for stellar eruptions.
We have conducted a 58-hour survey of a sample of 5 active M dwarfs with the VLA, using the VLA's subarray capabilities to achieve ultra-wide bandwidth over a subset of frequencies from 224-482~MHz and 1-6~GHz.  We analyze the results of this survey in a series of three papers.  In this paper (Paper I), we present the ensemble of coherent radio bursts detected by the survey, consider the physical origins of two populations of events (short-duration bursts and long-duration bursts), and predict rates of stellar coherent radio bursts for upcoming transient surveys.  Papers II and III analyze in detail a series of coherent bursts on UV~Cet and AD~Leo, respectively, accompanied by simultaneous 8.2-8.5~GHz high-resolution imaging with the Very Long Baseline Array (VLBA).

\section{Observations}

The survey consists of 58 hours of VLA observations in 2013 and 2015, summarized in Table~\ref{table:obs_summ}.  The observations consist of 2- and 4-hour observing blocks on individual targets (these durations include 20-25\% overhead time on calibrators), for a total of 14 hours in 2013 observing two targets at 1-6~GHz, and 44 hours in 2015 observing five targets at 0.22-0.48 and 1-4~GHz.

\subsection{Target Stars}

The targets are well-known flare stars of spectral type M3.5V to M6V, summarized in Table~\ref{table:targets}: AD~Leonis, binary BL \& UV~Ceti \citep[semi-major axis 2.1'';][]{kervella2016}, binary EQ~Pegasi AB \citep[semi-major axis 6.9'';][]{heintz1984}, EV~Lacertae, and YZ~Canis~Minoris. Both binaries are unresolved by most of the VLA observations; in this paper the names ``UV~Cet'' and ``EQ~Peg'' refer to the binary systems unless otherwise specified.

The targets are main sequence stars whose rapid rotation rates (Table~\ref{table:targets}) suggest an age less than $\sim$2~Gyr \citep{newton2016}.  To the best of our knowledge, these stars do not show indicators of extreme youth such as low surface gravity or lithium \citep[ruled out for EV~Lac by][]{shkolnik2009}.  These stars represent an early evolutionary stage that many mid-M~dwarfs should inhabit for hundreds of millions of years, affecting the long-term evolution of planetary atmospheres.

The targets produce a high rate of energetic flares across the electromagnetic spectrum, including $>10^{34}$-erg superflares on AD~Leo \citep{hawley1991} and YZ~CMi \citep{kowalski2010} that likely occur about once a month. \cite{lacy1976} measured optical U-band flare frequency distributions for all of our targets, finding that the stars produce flares with U-band energy $E_U>5\times10^{30}$~erg every $\sim$1-15~hours (depending on which star).  This corresponds to a bolometric flare energy of $\sim5\times10^{31}$~erg \citep{osten2015}, above which 100\% of solar flares are associated with CMEs \citep{yashiro2006,osten2015}. \cite{crosley2016} use solar flare-CME scaling relationships to predict that $E_U\sim2\times10^{30}$~erg is the minimum flare energy for an associated CME to form a shock (a necessary condition to produce a Type~II burst) at a height of $2R_*$ in a simple coronal model of YZ~CMi. Thus if these stars follow solar flare-CME scaling relationships, they may produce super-Alfv\'enic (shock-forming) CMEs every few hours. However, as discussed in Section~\ref{section:intro}, there is a gap in understanding as to how the extreme coronal conditions of these stars impact the relationship between flares and CMEs. Notably, \cite{alvarado-gomez2018} predict that strong global magnetic fields may suppress CMEs for bolometric flare energies less than $6\times10^{32}$~erg, and reduce the velocity of even the more energetic CMEs, potentially preventing shock formation.

The high rate of magnetic energy release in the outer atmospheres of these stars powers coronae that are both hotter and denser than the Sun's (Table~\ref{table:targets}).  X-ray observations show the targets have quiescent coronal temperatures of $\sim$3-10~MK, in contrast to the Sun's 1-3~MK, and electron densities about 100 times greater than in the Sun's corona.  The stars also have large-scale magnetic fields very different from the Sun's, thanks to their rapid rotation and fully convective dynamo.  While photospheric magnetic field strengths in active regions saturate at kilo-Gauss levels in both the Sun and active M~dwarfs, the covering fraction of these fields is much larger for active M dwarfs, about 50-100\% \citep{shulyak2014}, compared to $\sim$1\% for the Sun.  A significant fraction of the target stars' magnetic flux is in the large-scale field \citep{reiners2009topology}. \cite{morin2008} and \cite{kochukhov2017} have mapped the large-scale magnetic field of all the target stars, finding that the large-scale components of the field (dipole, quadrupole, ...) have a combined strength of hundreds of Gauss to a few kilo-Gauss, in contrast to a few Gauss on the Sun \citep{vidotto2016}.

The observed stars are known to produce broadband slowly-variable quiescent emission, attributed to incoherent non-thermal gyrosynchrotron radiation, with flares on timescales of minutes to days \citep[reviews:][]{bastian1990review,gudel2002}.  While the observed data set could also be used to study incoherent radio emission from flares, this paper focuses instead on the search for coherent radio bursts, where the signature of coherent processes is seen in narrow bandwidths or complex spectral structure, rapid time variability and high flux densities implying high brightness temperatures, and strong circular polarization of up to 100\%.

\renewcommand{\arraystretch}{1.00}
\begin{deluxetable*}{cccccccccccc}
\tablecaption{Properties of Target Stars.\label{table:targets}}
\tablehead{
& \multicolumn{7}{c}{Observed properties} & & \multicolumn{3}{c}{Hydrostatic equilibrium model} \\
\cline{2-8}
\cline{10-12}
\colhead{Star} & \colhead{Spectral Type} & \colhead{Distance} & \colhead{$P_{rot}$} & \colhead{Mass} & \colhead{Radius} & \colhead{$T_{cor}$} & \colhead{log $n_e$} & & \multicolumn{2}{c}{Density scale height} & \colhead{$\nu_p$ at $r=2R_*$} \\
& & \colhead{pc} & \colhead{days} & \colhead{$M_\odot$} & \colhead{$R_\odot$} & \colhead{MK} & \colhead{cm$^{-3}$} & & \colhead{Mm} & \colhead{$R_*$} & \colhead{MHz}
}
\startdata
AD~Leo 	& M3V 		& 4.7	& 2.24	& 0.42	& 0.44	& 2-10 		& 10.4		& & 115	& 0.38	& 420	\\ [1.2ex]
EQ~Peg A & M3.5V 	& 6.2	& 1.06	& 0.39	& 0.41	& $\sim$3	& 10.6		& & 62	& 0.29	& 280	\\ [-0.3ex]
EQ~Peg B & M4.5V 	&		& 0.40	& 0.25	& 0.30	& $\sim$3	& <11.0		& & 90	& 0.43	& 500	\\ [1.2ex]
EV~Lac	& M3.5V 	& 5.1	& 4.38	& 0.32	& 0.30	& 2-10		& 10.3-10.7	& & 70	& 0.34	& 360	\\ [1.2ex]
YZ~CMi	& M4V 		& 6.0	& 2.78	& 0.31	& 0.29	& 4-7		& 10.5		& & 68	& 0.34	& 360	\\ [1.2ex]
BL~Cet	& M5.5V 	& 2.7	& 0.24	& 0.12	& 0.17	& 3-6		& \nodata	& & 60	& 0.51	& 600	\\ [-0.3ex]
UV~Cet	& M6V 		&		& 0.23	& 0.12	& 0.16	& 3-6		& \nodata	& & 53	& 0.48	& 560	\\ [1.2ex]
\textit{Sun} & \textit{G2V} & & \textit{$\sim$25} & \textit{1}	& \textit{1}	& \textit{1-3} & \textit{8.5} 	& & \textit{100}	& \textit{0.14}	& \textit{5}	\\
\enddata
\tablecomments{Solar properties are shown for comparison. The hydrostatic equilibrium model, discussed in Section~\ref{section:frequency}, assumes a coronal temperature of 5 MK for the stars and 2 MK for the Sun, and a coronal base density (log $n_e$) of 10.5 for the stars and 8.5 for the Sun, corresponding to a plasma frequency of 1.6 GHz at the base of the stellar coronae and 160 MHz at the base of the solar corona.}
\tablerefs{
\textit{Spectral type:} \cite{henry1994}. \textit{Distance:} SIMBAD database. \textit{Rotation period:} AD Leo, EQ Peg A/B, EV Lac, YZ CMi - \cite{morin2008}. BL Cet, UV Cet - \cite{barnes2017}.
\textit{Mass:} AD Leo, EQ Peg A/B, EV Lac, YZ CMi - photometric mass, \cite{morin2008}. BL Cet, UV Cet - dynamical mass, \cite{kervella2016}.
\textit{Radius:} AD Leo, EQ Peg A/B - radius from spectroscopic effective temperature and bolometric flux, \cite{mann2015}. EV Lac, YZ CMi - theoretical model radius based on photometric mass, \cite{morin2008}. BL Cet, UV Cet - interferometric radius, \cite{kervella2016}.
\textit{Coronal temperature:} Temperature range with peak X-ray emission measure. AD Leo - \cite{gudel2003,maggio2004,wood2018}. EQ Peg A/B - \cite{liefke2008}. EV Lac - \cite{osten2006quiet}. YZ CMi - \cite{raassen2007}. BL Cet, UV Cet - \cite{audard2003}. Sun - \cite{gudel1997,peres2000}.
\textit{Coronal density:} From OVII line ratios. AD Leo, EV Lac, YZ CMi - \cite{ness2004}. EQ Peg A/B - \cite{liefke2008}. Sun - Coronal density models using various types of observations by \cite{allen1947,newkirk1967,saito1970,guhathakurta1999}. OVII line ratios for the quiescent Sun compiled by \cite{ness2001} are near the low-density limit implying log $n_e\lesssim$ 9.5.}
\end{deluxetable*}

\subsection{Choice of Frequency Bands\label{section:frequency}}

A key feature of this survey is its extremely wide fractional bandwidth, defined here as $\nu_{max}/\nu_{min}$, the ratio of highest frequency to lowest frequency observed simultaneously.  Previous detections of stellar radio bursts have achieved continuous frequency coverage of $\nu_{max}/\nu_{min}$ up to 1.4 \citep{osten2008} or have observed widely spaced narrow bands \citep[e.g.,][]{spangler1976}. In contrast, solar spectrographs offer continuous frequency coverage for $\nu_{max}/\nu_{min}$ of 10 to 100, with facilities such as the 10-1000~MHz Green Bank Solar Radio Burst Spectrometer and the 18-1800~MHz Culgoora Solar Radio Spectrograph.  Our survey increases the fractional bandwidth achieved for stellar observations, with continuous frequency coverage over $\nu_{max}/\nu_{min}$ of 6 (1-6 GHz) in 2013 and partial frequency coverage over $\nu_{max}/\nu_{min}$ of 18 (0.22-0.48 and 1-4 GHz) in 2015.  To achieve this wide bandwidth, we divided the VLA into three subarrays to observe three frequency bands simultaneously; the upgraded VLA's wideband feeds and WIDAR correlator enabled up to 2~GHz bandwidth in each band.  The 2013 observations cover L~band (1-2~GHz, 1~MHz channels), S~band (2-4~GHz, 2~MHz channels) and C~band (4-6~GHz, 2~MHz channels), with 9 antennas per subarray.  The 2015 observations cover P~band (224-482~MHz, 125 kHz channels) with 15 antennas to compensate for its lower sensitivity, and L and S~bands with 6 antennas each (the achieved sensitivities are given in Section~\ref{section:bgsub}).  This enables the detection of features in the dynamic spectrum across a much wider bandwidth, with two benefits: 1) increased probability of detecting narrowband bursts; and 2) the potential to track coronal sources as they propagate across 1-2 orders of magnitude in coronal density or magnetic field strength, corresponding to a wide range of heights above the star.

This survey targets relatively high frequencies compared to the distribution of CME-associated solar Type II bursts. Solar Type II radio bursts, which originate from shock fronts in the corona or interplanetary medium, are detected at 100s of MHz and below \citep[start frequency distributions in, e.g.,][]{roberts1959,prakash2010}, but only rarely at the highest frequencies of $\sim$100-600 MHz. The highest-frequency Type II bursts, originating low in the solar corona, are sometimes associated with CME-driven shocks \citep{cho2013,kouloumvakos2015,kumari2017}, but can also be caused without a CME due to coronal shocks driven by blast waves from flare sites \citep{vrsnak2008,magdalenic2012} or by rapidly expanding magnetic structures \citep{su2015}. In contrast, all Type II bursts at low frequencies (<1~MHz) are attributed to CME-driven shocks \citep{cane1987}. However, geoeffective CMEs that cause solar energetic particle (SEP) events tend to form shocks relatively low in the corona, producing Type II bursts that start at high frequencies \citep{prakash2017}.


Active M dwarfs have high coronal densities compared to the Sun, making it plausible to observe coronal plasma emission at higher frequencies. Table~\ref{table:targets} lists coronal electron densities for these stars based on OVII line ratios, which are most sensitive to $\sim$3~MK plasma. The targets with measured densities all have $log n_e\sim10.5$.  This corresponds to a plasma frequency in the low corona of 1.6 GHz, and second harmonic of 3.2 GHz.  However, most solar CMEs do not form shocks low in the corona, accelerating with height as Alfv\'en speed declines with height, enabling them to form shocks at heights of $\sim1.3-2R_*$ \citep{gopalswamy2009,prakash2017}.  For now, to estimate plausible start frequencies for stellar Type II bursts, we consider the case where a shock forms at $r=2R_*$. However, further modeling work is needed in order to predict at what range of heights stellar CMEs can exceed the Alfv\'en speed and form shocks, which may be particularly difficult in the strongly magnetized coronae of active M dwarfs.


To estimate the density at $2R_*$, we assume that density decreases exponentially with height above the base of the corona, following the approach of \cite{crosley2017}, who use Sun-as-a-star data to build a framework for interpretation of stellar Type II bursts.  We use the density scale height for material in hydrostatic equilibrium (HSE) under constant gravity,  $H=k_BT/({\mu}m_Hg)\approx\textrm{(50 Mm)}T_6R_*^2/M_*$, where $R_*$ and $M_*$ are in solar units, and and $\mu=0.6$ is the mean molecular weight of the solar corona.  This constant-gravity HSE model is a simplistic assumption - accounting for the decreased gravity at greater distances from the star would increase the density at $2R_*$ within closed magnetic structures, while the decline in the number of closed magnetic structures and expansion of open magnetic field lines would decrease the average density with height.

Using a coronal temperature of 5~MK for active M dwarfs and 2~MK for the Sun, we obtain $H/R_*$ of 0.3-0.5 for our target stars and 0.14 for the Sun. Assuming a coronal base density $log n_e=10.5$, the plasma frequency at $r=2R_*$ should be in the range of $\sim$300-600 MHz.  In contrast, for the Sun, using $log n_e=8.5$, we obtain a plasma frequency of 5~MHz at 2$R_\odot$, in agreement to order-of-magnitude with the plasma frequency predicted by observational solar coronal density models \citep[e.g.,][]{guhathakurta1999}. Thus, the maximum start frequencies we might expect for CME-associated Type II bursts on these stars, if the CMEs form shocks at about $2R_*$, would be $\sim$0.3-0.6~GHz for first harmonic and 0.6-1.2~GHz for second harmonic emission, within the VLA's P band (0.2-0.5~GHz) and L band (1-2 GHz), much higher than the start frequencies of most solar Type II bursts.  Indeed, in contrast to the Sun, active M dwarfs produce abundant coherent radio bursts at GHz frequencies; AD~Leo, EQ~Peg, and YZ~CMi produce 0.2-0.3 bursts per hour at 1.4 GHz \citep{abadasimon1997}, although so far it is unknown whether any of those bursts are associated with space weather events.

By searching for the high-frequency tail of space weather events, this survey takes advantage of the better sensitivity available at higher frequencies, while complementing lower-frequency observations including targeted stellar observations and wide-area transient searches.  Targeted stellar observations have been conducted with low-frequency radio facilities including LOFAR \citep{crosley2016} and the MWA \citep{lynch2017}. Wide-area low-frequency transient searches include projects using the Long Wavelength Array (LWA) \citep{obenberger2015}, LOFAR \citep{stewart2016,prasad2016}, and the Giant Metrewave Radio Telescope (GMRT) and the MWA \citep{murphy2017}, including a recent MWA detection of two stars at 200~MHz \citep{lenc2018}.

\section{Data Analysis}

\subsection{Calibration\label{section:calibration}}

The VLA data were calibrated and imaged using the CASA software package \citep{mcmullin2007}. The L~band (1-2~GHz), S~band (2-4~GHz), and C~band (4-6~GHz) data were calibrated with the CASA VLA calibration pipeline, using the \citet{perley2013flux} flux density scale, followed by an additional round of automatic flagging with \textit{rflag}.  The P~band (220-480~MHz) data were also calibrated in CASA, following the calibration steps outlined in the CASA Guide ``Basic P-band data reduction-CASA4.6'' and using the \citet{scaife2012} flux density scale with L-band model images of the flux calibrators.  (We bypassed ionospheric TEC correction as it was not supported in CASA 4.6; since it corrects for Faraday rotation of linearly-polarized signals, it does not impact our scientific objectives.) We estimate systematic errors on flux calibration of up to 10\% in P band and 3-5\% at higher frequencies \cite{perley2017}.

Stellar coherent radio bursts can have strong circular polarization.  Measuring Stokes~V provides two benefits: 1) the sense and degree of circular polarization provide information on the emission mechanism and the magnetic field orientation in the source region; and 2) Stokes~V has minimal contamination from background sources, offering better sensitivity to strongly polarized bursts than Stokes~I. At 1 GHz and above, the VLA correlates circularly polarized signals (RR, RL, LR, LL), so Stokes V and Stokes I are both formed from the parallel-hand correlation products, RR and LL.  Polarization calibration is most important for the cross-hand products, so we did not perform any polarization calibration for these frequencies (L, S, and C band).

VLA P~band uses linear feeds, for which Stokes V is formed from the cross-hand products XY and YX, so polarization calibration is required to measure Stokes~V in P~band.  Full-polarization VLA observations at these frequencies had not been fully commissioned at the time of the observations.  We were able to perform certain polarization calibration steps (cross-polarization delay and frequency-dependent leakage) but not another (cross-polarization phase). The consequence of this is that we are able to measure the quantity $\sqrt{U^2+V^2}$ rather than measuring Stokes U and V independently.  Assuming that our sources have no linear polarization, then this gives the absolute value of Stokes~V, allowing us to measure the degree but not the sense (left or right) of circular polarization.

\subsection{Background Source Subtraction and Dynamic Spectroscopy\label{section:bgsub}}

After calibration, we shift the phase center to the position of the target star.  If the star is detected in a Stokes~I or V image of an entire observation, then we use the star's imaged location as the phase center; otherwise we use the star's expected position based on positions and proper motions from Hipparcos and/or VLBA \citep{benz1998}.

It is necessary to model and remove background sources from the visibility data before averaging over all baselines, to avoid the contribution of their sidelobes to the stellar flux density. Each observation was imaged in Stokes~I using CASA's \textit{clean} task with multi-frequency synthesis \citep{rau2011} with 2 Taylor terms, which allows the flux density of clean components to vary linearly with frequency.  These images of background sources were produced using time and frequency ranges chosen to exclude any bright stellar radio bursts.  We created a background model by masking any quiescent stellar emission out of the clean model image, then used CASA tasks \textit{ft} and \textit{uvsub} to subtract the background model from the visibilities. We developed the code \textit{dynspec}\footnote{The code as used is available at \url{https://github.com/jackievilladsen/dynspec}. Consultation is advised before use.} to average the visibilities over all baselines and to extract, analyze, and plot the dynamic spectrum.

The end product of the data reduction is dynamic spectra (such as in Figure~\ref{fig:yzcmi_burst_example}) for all the observations, showing the evolution of stellar emission in the time-frequency plane.  We produce dynamic spectra from interferometric data by averaging the visibility data over all baselines, generating a single complex number per channel per integration.  The real part of this number corresponds to the measured flux density at the phase center at this time and frequency. After successful subtraction of bright sources in the field of view, our dynamic spectra should be thermal noise limited. This is indeed the case for most of our observations. Residual phase errors occasionally led to contamination of dynamic spectra from bright sources, especially for AD~Leo and EV~Lac. Subsequent self-calibration of observations at $>$1~GHz, as needed, allowed us to better model and subtract this background signal. These issues were also prevalent at P band. However, standard self-calibration still resulted in severely dynamic-range-limited images at P~band, due to direction dependent phase errors that occurred since we observed in the VLA's widely-spaced A~configuration, resulting in considerable sidelobe contamination of the dynamic spectrum from bright sources across the wide field of view (FWHM $\sim 2.25 \arcdeg$ in P band). Improving this will require direction dependent self-calibration and source removal \citep[peeling;][]{noordam2004,intema2009}, which is not yet available for wideband VLA data through the CASA software package. We note, therefore, that a future re-analysis of these data with direction-dependent self-calibration can achieve more stringent limits in Stokes~I at low frequencies.

\subsection{Search for Radio Bursts\label{section:burst_search}}

We searched for stellar radio bursts by inspecting the Stokes~I and V dynamic spectra by eye. We searched for bursts in both Stokes~I and V because Stokes~V offered greater sensitivity to circularly polarized bursts for fields with bright background sources, and Stokes~I provided sensitivity to unpolarized and weakly polarized bursts. Searches were initially carried out in the raw unbinned dynamic spectra, with subsequent logarithmic binning in time and frequency to search for events on all timescales probed by our data. Dynamic spectra of all epochs with detected radio bursts are shown in Appendix~\ref{appendix:dynspec}, with each dynamic spectrum binned to a time and frequency resolution chosen to highlight the events in that epoch. While sensitivity depends on resolution, binning to a resolution of 64~MHz and 150~seconds for observations from 1-6~GHz achieves a typical sensitivity of 1~mJy per bin; the thermal sensitivity limit is 0.3-0.6~mJy depending on the band and number of antennas. For 224-482~MHz, binning to 4~MHz and 300~seconds achieves a typical sensitivity of 16~mJy in Stokes~I and 6~mJy in Stokes~V; the thermal sensitivity limit is 4~mJy.

Each pixel of the dynamic spectrum has a complex value, the result of averaging complex-valued visibilities over all baselines.  A point source at the phase center produces a positive signal in the real component of the dynamic spectrum and no signal in the imaginary component.  Bursts were identified visually as a cluster of pixels in the real component of the dynamic spectrum with signal-to-noise ratio (SNR) greater than 3, by making dynamic spectrum plots masked to only show pixels with SNR$>$3, where the noise in each channel was calculated using the standard deviation (along the time axis) of the imaginary component of the dynamic spectrum in that channel.  The other requirement for identifying a burst is the lack of comparable features in the imaginary component of the baseline-averaged visibilities, which would indicate contamination by RFI or background sources. Appendix~\ref{appendix:identification} provides an example that illustrates these conditions.

For each detected radio burst, we measured a set of basic properties (Table~\ref{table:bursts}).  Duration and frequency range are approximate values identified by eye from the dynamic spectrum. The presence of burst components with positive or negative frequency drift was also identified by eye in the dynamic spectrum; while Section~\ref{section:results} discusses general timescales of frequency drift and Section~\ref{section:yzcmi} examines one event, detailed quantitative studies of duration, bandwidth, and frequency drift are left as future work.

The peak flux densities quoted in Table~\ref{table:bursts} depend on the time and frequency resolution of the dynamic spectrum.  To measure peak flux density, rather than impose a single time and frequency resolution on all events, long and short, faint and bright, we clipped each observation to just the time range and frequency of the burst, then binned the dynamic spectrum automatically into roughly 30$\times$30 pixels.  For a few faint events, we manually increased the time bins to improve the SNR.  We then took the 98th percentile of the Stokes~I flux density in the pixels of the dynamic spectrum to represent the peak flux density, choosing this approach to avoid contamination by RFI.  The error given for peak flux density is the standard deviation (across all pixels) of the imaginary component of the baseline-averaged visibilities, which provides an estimate of the noise on an individual pixel in the dynamic spectrum.  Errors are largest for low frequency data due to residuals of background sources.  The reported errors do not include systematic errors on flux calibration with the VLA. In addition, the listed peak flux density is an underestimate in cases where the flux density varies rapidly on timescales shorter than $\sim$3\% of the burst duration.

We also determined the degree of circular polarization of each burst.  To do so, we averaged the dynamic spectrum over the time and frequency range of each burst to obtain an average Stokes~I flux density $<S_I>$ and Stokes~V flux density $<S_V>$, then calculated the degree of circular polarization $r_c = <S_V>/<S_I>$.  We did not subtract quiescent emission (which is a few mJy or less) before doing so, which may result in an underestimate of degree of circular polarization for bursts fainter than $\sim$20~mJy. The quoted error bars for percent circular polarization were calculated using propagation of errors for events above 1~GHz, which have small fractional errors on degree of circular polarization. For low-frequency events, we used $(<S_U>^2+<S_V>^2)^{1/2}$ instead of $<S_V>$, to correct for the lack of cross-hand phase calibration. The low-frequency events had lower SNR than the high-frequency events, leading to asymmetric error bars, so we calculated a 68\% confidence interval or 68\%-confidence lower limit (noted in the table footnotes), using a Bayesian framework assuming Gaussian errors on Stokes I and V and using a flat prior for degree of circular polarization between 0\% and 100\%. In Table~\ref{table:bursts}, a range of values are quoted for certain events, corresponding to burst features at different frequencies. In these cases the errors are not shown but are typically $\sim$1\% for >1~GHz and 10-25\% for 0.2-0.5~GHz.

Table~\ref{table:bursts} also shows the energy of each event, obtained by integrating Stokes~I over the time and frequency range listed in the table. Finally, the table lists the SNR with which the burst was detected in Stokes~I and V. This was calculated as the average flux in the time and frequency range of the burst, divided by the error on that average flux. The error on the average flux was determined by first taking the standard deviation across all pixels of the imaginary component of the baseline-averaged visibilities, divided by the square root of the number of dynamic spectrum pixels. For low-frequency events, we used $(<S_U>^2+<S_V>^2)^{1/2}$ instead of $<S_V>$ in our SNR calculations, to correct for the lack of cross-hand phase calibration. For events on UV~Cet that span all observed frequency bands, we list SNR separately for frequencies below and above 1~GHz to demonstrate that the events are detected with high significance even in the relatively noisy low-frequency band.

\begin{subfigures}
\begin{figure*}[b!]
  \begin{center}
     \includegraphics[width=\textwidth]{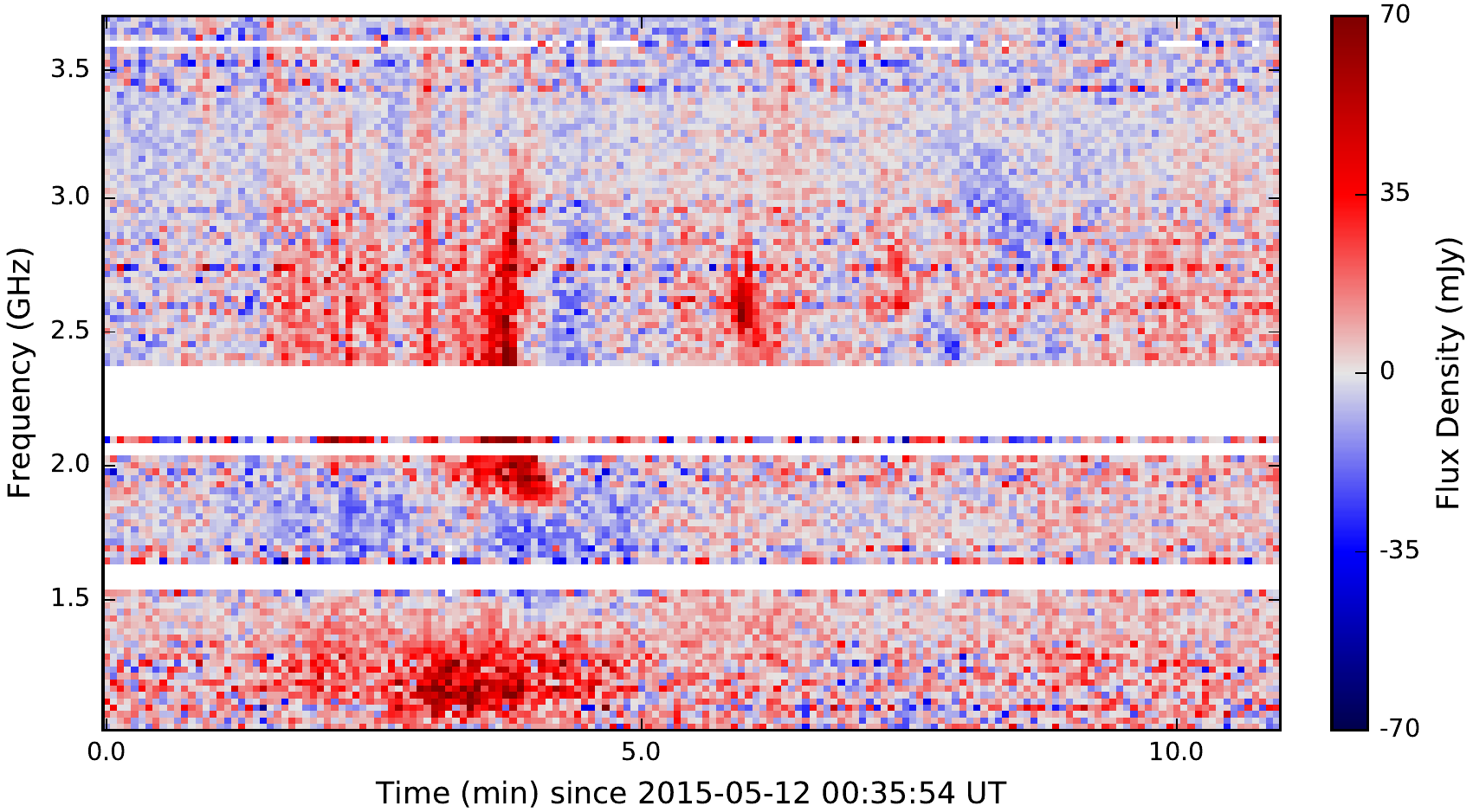}
     \vspace{-2em}
  \end{center}
\caption{Stokes V dynamic spectrum of a short-duration (minutes-long) burst on YZ~CMi, binned to a resolution of 24~MHz and 4~sec. Right circular polarization is red, and left is blue.\label{fig:yzcmi_burst_example}}
\end{figure*}

\begin{figure*}[b!]
  \begin{center}
     \includegraphics[width=\textwidth]{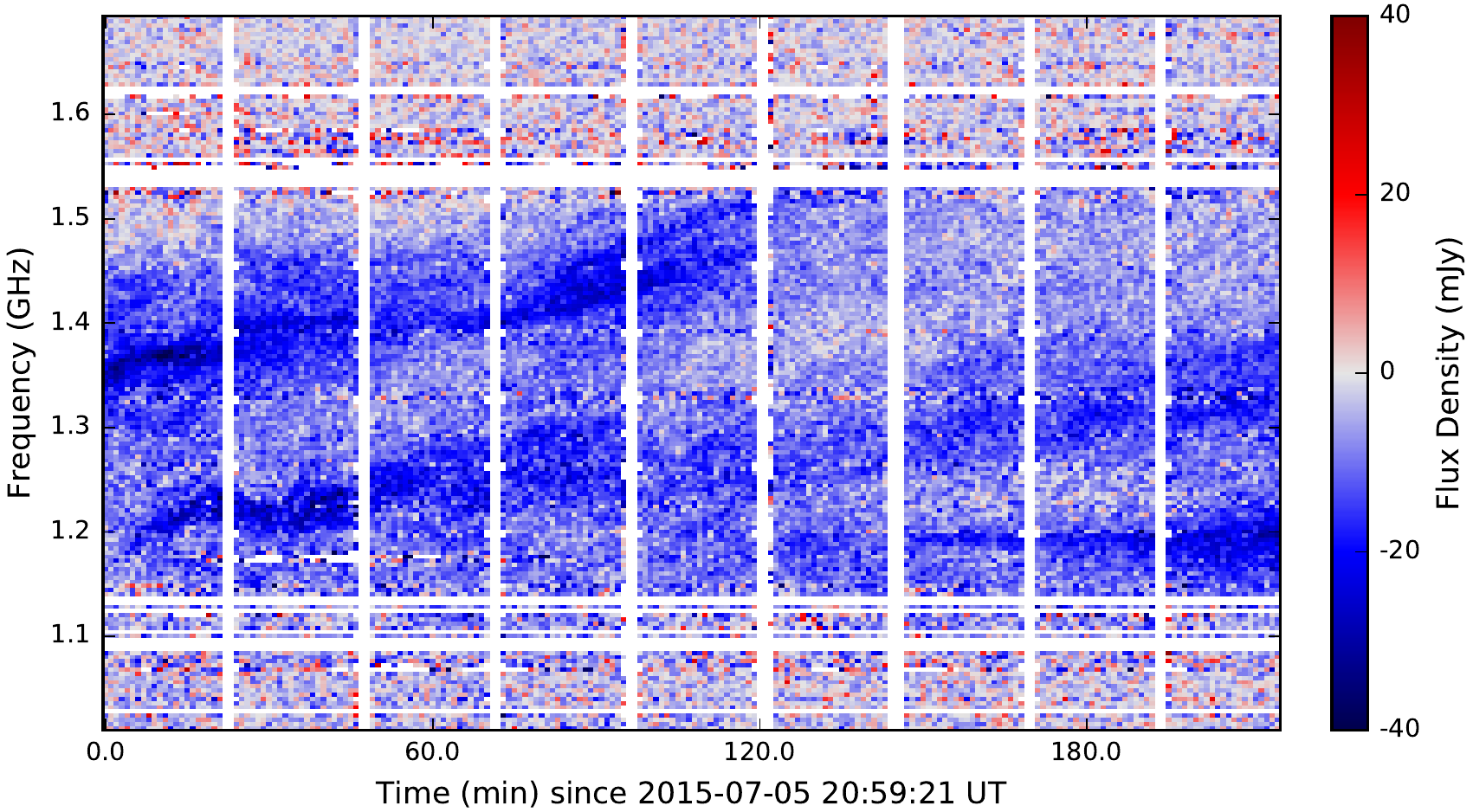}
     \vspace{-2em}
  \end{center}
\caption{Stokes V dynamic spectrum of a long-duration radio burst on AD~Leo, binned to a resolution of 4~MHz and 60~sec. Note that the duration and frequency range are dramatically different from the short-duration event in Figure~\ref{fig:yzcmi_burst_example}.\label{fig:adleo_burst_example}}
\end{figure*}

\begin{figure*}[t!]
  \begin{center}
     \includegraphics[width=\textwidth]{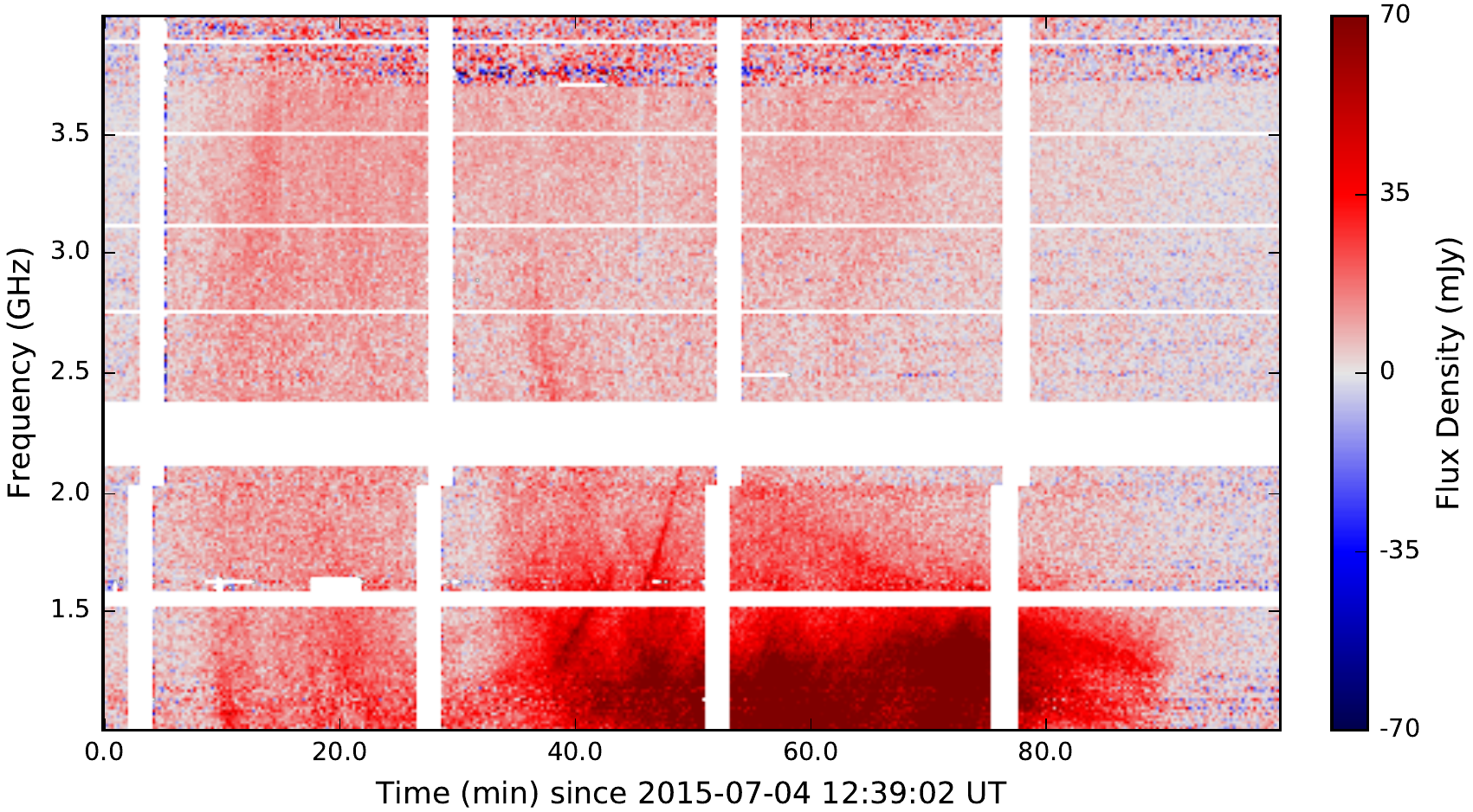}
     \vspace{-2em}
  \end{center}
\caption{Stokes V dynamic spectrum of a long-duration radio burst on UV~Cet, binned to a resolution of 12~MHz and 15~sec.\label{fig:uvcet_burst_example}}
\end{figure*}
\end{subfigures}

\section{Results\label{section:results}}

The VLA survey detected 22 coherent bursts with diverse properties, summarized below and listed in Table~\ref{table:bursts}. We show exemplar dynamic spectra in the text (Figures~\ref{fig:yzcmi_burst_example}-\ref{fig:uvcet_burst_example}) to demonstrate the phenomenology of some of the brightest coherent bursts, and we present a full census of the Stokes I and V dynamic spectra of all epochs with detected radio bursts in Appendix~\ref{appendix:dynspec}. For some of the observations of AD~Leo and UV~Cet, we conducted simultaneous VLBI observing, allowing us to trace dynamics both via the dynamic spectra and via astrometry with the VLBI data.  The VLA data from those sessions are included in this paper, but the joint analysis of the VLA and VLBA data will be presented in subsequent papers. Here we focus on understanding the nature and prevalence of coherent radio bursts on active M dwarfs and their implications for studies of M dwarf stellar activity and widefield transient surveys.  The properties of the population of coherent bursts in the VLA survey are:

\begin{enumerate}
\item \textbf{Rate:} Coherent bursts were detected in 13 out of 23 epochs on 4 of the 5 stars, with most of these bursts detected in multiple bands. As seen in Table~\ref{table:bursts}, several epochs had more than one burst, where bursts were distinguished by separation in the time-frequency plane.  However, the identification of separate bursts within an epoch should be taken as tentative, since the complex substructure seen in many events makes it difficult to decide whether to consider two distinct features in the time-frequency plane as separate events.

\renewcommand{\arraystretch}{1.0} 
\begin{deluxetable*}{cccccccccccc}[htb!]
\rotate
\tablecaption{Detected Stellar Radio Bursts.\label{table:bursts}}
\tablehead{\colhead{Star} & \colhead{Date} & \multicolumn{2}{c}{Duration} & \colhead{Freq Range} & \colhead{Peak Flux} & \multicolumn{2}{c}{Polarization} & \colhead{Frequency drift?} & \colhead{Energy} &  \multicolumn{2}{c}{SNR} \\
& & & & (GHz) & Density (mJy) & & & & ($10^{22}$~erg) & I & V}
\startdata
\\[-1.5ex]
AD~Leo 	& 2015 Jul 5 & long	& >3.5 h \tablenotemark{a} & 1.1-1.55 & 26 $\pm$ 2 & 84 $\pm$ 1 \% & L & positive & 411 $\pm$ 3 & 130 & 280  \\
		& 			& long	& >3.5 h	& 0.28-0.4	& 38 $\pm $15	& $>$ 87 \% \tablenotemark{e} & ? & no & 42 $\pm$ 5 & 8.2 & 11 \\ 
        & 2015 Jul 19 & long & >3.5 h	& 1-1.6		& 64 $\pm$ 2	& 66.5 $\pm$ 0.2 \%	& L & positive & 1413 $\pm$ 4 & 340 & 470 \\ 
        & 			& short	& 30 s		& 1.6-2.2	& 31 $\pm$ 6	& 45 $\pm$ 4 \%		& R & negative? & 2.2 $\pm$ 0.1 & 30 & 13  \\ 
        & 			& short	& 6 m		& 2.8-4		& 9 $\pm$ 2		& 75 $\pm$ 3 \%		& L & negative & 12.5 $\pm$ 0.3 & 36 & 29 \\ 
        & 2015 Sep 5	& long	& >1.5 h	& 1-2.5		& 9 $\pm$ 2		& 65 $\pm$ 2 \%		& L & no & 136 $\pm$ 4 & 31 & 52 \\ 
        & 			& short	& 5 m & 0.29-0.36 & 210 $\pm$ 80 & $>$ 96 \% \tablenotemark{e} & ? & negative? & 2.0 $\pm$ 0.6 & 3.2 & 27 \\ [1.5ex]
UV~Cet	& 2013 May 26 & long & >25 m \tablenotemark{b} & 1-6 & 52 $\pm$ 1 & 77.2 $\pm$ 0.3 \% & R & positive & 1386 $\pm$ 5 & 300 & 350 \\
		& 		   	& short	 & 10 m		& 1-6		& 9 $\pm$ 1		& 52 $\pm$ 2 \%		& R & positive & 54 $\pm$ 1 & 55 & 34 \\
        & 2013 Jun 2 	& long 	& 1 h 		& 1-4 		& 31 $\pm$ 4	& 65 $\pm$ 1 \%		& R & positive & 148 $\pm$ 2 & 74 & 33 \\
        & 2015 May 15 & short & 8 m 		& 1.2-2 \tablenotemark{c} & 41 $\pm$ 7 & 78 $\pm$ 3 \% & R & negative & 94 $\pm$ 2 & 38 & 40 \\
        & 2015 May 16 & long 	& >20 m \tablenotemark{b} & 0.3-3.7 & 79 $\pm$ 2 & 67 - 70\% 	& R & positive & 1810 $\pm$ 10 & 150 (L,S), 6.5 (P) \tablenotemark{f}	 & 180 (L,S), 14 (P)  \\
        %
        & 2015 Jul 4 	& long 	& 1.5 h 	& 0.3-8.5 \tablenotemark{d} & 107 $\pm$ 1 & 43 - 68 \% & R & both & 5610 $\pm$ 20 & 420 (L,S), 17 (P) & 400 (L,S), 17 (P) \\
        & 2015 Jul 18 & long 	& >30 m & 0.35-8.5 \tablenotemark{d} & 83 $\pm$ 1 & 53 - 58 \% & R & both & 2470 $\pm$ 10 & 430 (L,S), 14 (P) & 320 (L,S), 14 (P)  \\
        & 			& short	& 25 m		& 1-3.7		& 20 $\pm$ 3	& 62 $\pm$ 2 \%		& R & negative & 140 $\pm$ 2 & 70 & 49 \\
        & 			& short	& 10 m		& 0.28-0.37	& 120 $\pm$ 30	& [82,95] \% \tablenotemark{e}	& ? & no & 25 $\pm$ 2 & 12 & 28 \\
        & 2015 Sep 6	& long	& 1.5 h		& 0.25-8.5 \tablenotemark{d} & 111 $\pm$ 1 & 50 - 59 \% & R & both & 7950 $\pm$ 20 & 770 (L,S), 17 (P) & 770 (L,S), 18 (P) \\ [1.5ex]
EQ~Peg	& 2015 May 16 & long 	& >30 m 	& 1-3 		& 14 $\pm$ 2	& 59 $\pm$ 2 \%		& R & negative? & 58 $\pm$ 2 & 35 & 46 \\
		&			& long	& >35 m		& 0.35-0.385 & 270 $\pm$ 60 & [70,94] \% \tablenotemark{e}	& ? & no & 9.2 $\pm$ 1.5 & 6.1 & 7.2 \\ [1.5ex]
YZ~CMi  & 2015 May 11 & long 	& >1.5 h 	& 3-3.7 		& 15 $\pm$ 1	& 53 $\pm$ 1 \%		& L & negative & 44.4 $\pm$ 0.5 & 95 & 53 \\
		& 			& short & 10 m 		& 1-3.7 		& 42 $\pm$ 2	& \multicolumn{2}{r}{33\% L - 90\% R} & unresolved? & 75.6 $\pm$ 0.5 & 137 & 40 \\
        & 2015 May 13 & long 	& >45 m 		& 2-2.7 	& 13 $\pm$ 2	& 84 $\pm$ 2 \%		& L & no & 30.6 $\pm$ 0.6 & 53 & 52 \\[1.5ex]
\enddata
\tablenotetext{a}{A > symbol on duration indicates that the burst started before or ended after the observations.}
\tablenotetext{b}{We classify these three events on UV~Cet with the long-duration events, which are $\gtrsim$30~m, due to their morphological similarity to long events and their continuation beyond the end of the observations.}
\tablenotetext{c}{Only 1-2 GHz data were available for this observation.}
\tablenotetext{d}{8.2-8.5 GHz from simultaneous VLBA observations.}
\tablenotetext{e}{68\% confidence interval or 68\% confidence lower limit on degree of circular polarization given for cases with asymmetric errors.}
\tablenotetext{f}{For UV~Cet's bursts that span all observed frequency bands, we quote SNR separately for $>$1 GHz (L, S band) and $<$1 GHz (P band).}
\end{deluxetable*}

In spite of the challenges in determining the number of distinct events and thus the rate of bursts, it is worth doing so to inform future observational programs.  We estimate the rate of bursts by counting events within one frequency band at a time, considering events that overlap in time as a single event, thus discounting some of the events in Table~\ref{table:bursts} that are morphologically distinct in the dynamic spectrum but that overlap in time.  In P~band (220-480~MHz), we detected 8 events in 42 hours, a rate of one per 5.3 hours.  In L~band (1-2~GHz), we detected 15 events in 56 hours, or one per 3.7 hours.  In S~band (2-4~GHz), we detected 13 events in 56 hours, or one per 4.3 hours.  And in C~band (4-6~GHz), we detected 2 events in 14 hours, or one per 7 hours. As discussed in Section~\ref{section:burst_search}, these burst rates are for a $3\sigma$ detection threshold of $\sim$2-3~mJy in Stokes I and V for 1-6 GHz (for a time and frequency resolution of 64~MHz and 150~seconds), and a $3\sigma$ detection threshold of $\sim$50~mJy for Stokes I and $\sim$20~mJy for Stokes V for 220-480~MHz (for a time and frequency resolution of 4~MHz and 300~s). For the most distant star in our survey (EQ~Peg at 6.2~pc), a detection threshold of 3~mJy (for the 1-6 GHz events) corresponds to a spectral luminosity of 1.4$\times10^{14}$~erg\,s$^{-1}$\,Hz${-1}$; all but one of the events (the short event on 2013~May~26) fall above this spectral luminosity threshold.

\item \textbf{Brightness temperature:} We do not have measurements of source size for these bursts.  Coherent sources can have millisecond-timescale variation \citep[e.g.,][]{osten2008} indicating small source size and high brightness temperatures.  Our time resolution of 1 second is longer than the light-crossing time for the stellar disks, providing very little constraint on source size, so we instead calculate the stellar disk-averaged brightness temperature.  For the five stars in our sample, the implied disk-averaged brightness temperature is $\sim10^{11}$~K for a typical flux density of 20~mJy at 1 GHz.  The detected low-frequency events, with flux densities of 100-300~mJy, imply the highest brightness temperatures or largest source sizes: 0.5-1$\times10^{13}$~K for 200~mJy at 0.35~GHz (assuming a source the size of the stellar disk).  However, these may well be significant underestimates of the true brightness temperature of a much smaller source.

\item \textbf{Morphology:} The morphology of bursts in the time-frequency plane provides rich information about the phenomena underlying solar radio bursts. We find that the morphology of stellar radio bursts often departs from the solar classification scheme but shows a similarly rich phenomenology. As illustrated by Figures~\ref{fig:yzcmi_burst_example} to~\ref{fig:uvcet_burst_example}, the bursts have diverse morphology in the time/frequency plane, reflecting the diversity in the underlying physical processes and conditions responsible for the bursts. A number of bursts show a sharp cutoff in the frequency spectrum above or below a certain frequency, as exemplified by the burst on AD~Leo in Figure~\ref{fig:adleo_burst_example}, which drops off rapidly below 1.1~GHz and above 1.6~GHz. Many of the bursts, such as the UV~Cet burst in Figure~\ref{fig:uvcet_burst_example}, show complex substructure without a clear analog in the classification system of solar radio bursts.  We discuss next two aspects of morphology that are of particular significance for the identification and interpretation of solar radio bursts: frequency drift and harmonic structure.

\item \textbf{Frequency drift:} Table~\ref{table:bursts} notes whether bursts show positive or negative frequency drift, or both, or possible ``unresolved'' drift for cases where vertical features in the dynamic spectrum are suggestive of rapid frequency drift unresolved by our 1-second integration times. These identifications have been determined by visual inspection of the dynamic spectrum at different time and frequency resolutions. While some bursts show a downwards frequency drift, including on the timescale of seconds to minutes expected for space weather events, no bursts have both frequency drift on these timescales and emission that continues beyond the lower edge of the observed bands, which would be a signature of a disturbance continuing to move outwards from the star. In many cases, the frequency drift is seen in substructure of a very complex burst (as in Figure~\ref{fig:uvcet_burst_example}) that does not have an analog in solar radio bursts.  While the complex substructure could be attributed to a very dynamic and complex set of source motions, it may also be that the patterns in the time-frequency plane should not be interpreted within the solar paradigm (Section~\ref{section:freq_drift_interpretation}). Based on the lack of features that persist below the minimum frequency observed, as well as the complex structure of many bursts, we have identified no candidate Type~II-like events in this survey.

\item \textbf{Harmonic structure:} The majority of metric solar Type~II bursts show harmonic structure, with emission at similar flux levels at both the fundamental plasma frequency and its second harmonic \citep{roberts1959}.  In spite of the wide frequency coverage of our observations, we detect no bursts with harmonic structure. There is only one possible exception, which we conclude is not likely to be true harmonic structure (Figure~\ref{fig:yzcmi_burst_example}; Section~\ref{section:yzcmi}).

\item \textbf{Duration:} Burst durations range from seconds to longer than a 4-hour observation.  We choose to divide the bursts into two categories: long-duration ($\gtrsim$30~min) and short-duration (seconds to minutes).  This is a departure from the terminology used to describe solar radio bursts, where ``long-duration'' is used to describe the minutes-long Type II bursts associated with CMEs, events which here would fall in the ``short-duration'' category. The motivation for adopting this categorization scheme is: high-speed electron beams (responsible for most plasma emission) in the solar corona decay on a timescale of milliseconds to seconds due to collisional damping and other mechanisms \citep{gudel1990time}, so bursts longer than seconds imply ongoing electron acceleration throughout the duration of the burst (at our observed frequencies; Type~III bursts from high-speed electron beams can persist longer as they propagate to very low frequencies).  Events of seconds to minutes correspond to a typical duration for the impulsive phase of a flare or for a coronal mass ejection to move a distance comparable to the stellar radius.  Thus, our short-duration category includes the range of durations we expect for radio bursts associated with space weather events.  In contrast, the long-duration bursts require a process that produces ongoing electron acceleration over a timescale of hours. Such long-duration acceleration processes are responsible for some low-frequency radio emission from the Sun (solar noise storms) and for auroral processes on planets.

We categorize all of our events according to this scheme in Table~\ref{table:bursts}.    The YZ~CMi event in Figure~\ref{fig:yzcmi_burst_example} falls in the short-duration category, while the AD~Leo and UV~Cet bursts in Figures~\ref{fig:adleo_burst_example} and \ref{fig:uvcet_burst_example} are classified as long-duration.  Long-duration bursts are more common and typically more luminous than short-duration bursts.  In the whole survey, we detect 12-13 long-duration bursts (depending on whether simultaneous but distinct spectral features on AD~Leo on 2015 July 5 are related) and 8 short-duration bursts.

\item \textbf{Polarization:} Polarization is the one property in which the bursts show a lack of diversity, with all bursts having a high degree of circular polarization. Almost all of the bursts fall in the 50-100\% range for degree of circular polarization, with some of the 0.2-0.5~GHz events as low as 40\%. It should be noted that for low frequencies, the higher noise due to dynamic range limits in Stokes I causes a bias towards detecting strongly polarized events. This bias towards strong polarization should not affect 2-6~GHz at all, can have a minor effect for 1-2~GHz for the fields with bright background sources (AD~Leo and EV~Lac), and is significant for 220-480~MHz for all targets.  220-480 MHz is also the frequency range (out of the bands observed) in which we most expect the stars to produce CME-associated Type~II-like bursts. In contrast to the detected stellar bursts, solar Type~II bursts generally have much less than 50\% circular polarization \citep{komesaroff1958} with occasional exceptions \citep{zlobec1993,thejappa2003}.

Table~\ref{table:bursts} lists the degree of polarization for all the bursts, and the sense of polarization for all bursts above 1~GHz (sense of polarization could not be determined below 1~GHz; Section~\ref{section:calibration}).  Most of the targets produced bursts of both polarizations, although UV~Cet shows a clear preference for right-hand polarization.  However, if we consider only the long-duration bursts, a pattern starts to emerge. For the three stars with more than one burst, the long-duration bursts tend to have a consistent sense of circular polarization: left-hand for AD~Leo and YZ~CMi, and right-hand for UV~Cet.  This tentative trend, of consistent sense of polarization in the long-duration bursts from a given star, is explored further in Section~\ref{section:pol_ZDI} in the context of magnetic field measurements of these stars.  In contrast, the few short-duration bursts, such as that in Figure~\ref{fig:yzcmi_burst_example}, appear to show arbitrary polarization.

\end{enumerate}

\section{Interpretation}

In the previous section, we classified our events as short- or long-duration, motivated by the expectation that flares and CMEs should produce bursts that are seconds to minutes long, whereas a different electron acceleration process is required to produce hours-long bursts. For the purposes of interpretation, we first highlight some physical principles that apply to all events, then consider the short- and long-duration events as separate populations.

Although the separation of events by duration is physically motivated, we caution that the distinction may be purely artificial, with all events originating from the same underlying process.  There may also be multiple phenomena responsible for either the short- or long-duration bursts, as is the case for the complex phenomenology of solar radio bursts.

\subsection{Physical Processes Underlying Stellar Radio Bursts\label{section:mechanisms}}

Interpretation of the dynamic spectra of stellar coherent radio bursts hinges on two interrelated questions: 1) What is the emission mechanism: plasma radiation or electron cyclotron maser emission?  2) Is frequency drift in the dynamic spectrum due to source motion or to rotational modulation of highly-beamed emission?
This section discusses these questions in light of the observed bursts.

\subsubsection{Emission Mechanism}

The strong circular polarization observed in most bursts, combined in many cases with narrowband features in the frequency spectrum, favors a coherent emission mechanism. The two emission mechanisms that may be responsible for stellar coherent radio bursts are plasma emission and electron cyclotron maser emission (ECM). Identification of the emission mechanism is necessary in order to use the emission frequency to determine either plasma density or magnetic field strength in the source region, which maps to an approximate source height and potentially a source velocity.  The question of emission mechanism is also a question of which paradigm in which to interpret the coherent bursts of active M~dwarfs: Do active M~dwarfs resemble the Sun, whose low-frequency bursts (including the space weather-associated Type II and III bursts) are mostly due to plasma emission, with ECM responsible for some high-frequency bursts originating in strongly-magnetized active regions?  Or do active M~dwarfs resemble auroral emitters, including planets \citep{zarka1998}, brown dwarfs \citep{hallinan2007}, and magnetic massive stars \citep{trigilio2011,das2018}, which produce periodic pulses and/or occasional isolated radio bursts due to ECM?

There are a number of properties that can help identify the emission mechanism: brightness temperature \citep{melrose1991}, fine structures in the dynamic spectrum \citep{treumann2006,osten2008}, knowledge of plasma density or magnetic field strength in the source \citep[e.g.][]{morosan2016}, constraints due to absorption \citep{dulk1985,osten2006,benz1992}, and degree and sense of circular polarization.  However, most of these properties do not help discriminate between emission mechanisms for our observations, because we do not have the time resolution needed to detect the ultra-high brightness temperatures or fine structures in the dynamic spectrum associated with ECM, and because measurements of coronal density and magnetic field strength for stars provide only global or large-scale information, allowing a wide range of possible properties for small-scale coronal radio sources.  The wide range of possible coronal properties means that the observed emission frequencies (0.28-6~GHz) could be due to either emission mechanism: the targets have X-ray measurements of average coronal plasma density ($\sim10^{10.5}$~cm$^{-3}$; Table~\ref{table:targets}) and Zeeman broadening measurements of magnetic field strength \citep[a few kG;][]{reiners2007}, corresponding to typical plasma frequencies of $\sim$1.6~GHz (but with possible large variation around this average) and cyclotron frequencies of $\lesssim$10~GHz in the low corona.  And while both emission mechanisms face potentially prohibitive levels of absorption at GHz frequencies, due to free-free opacity for plasma emission and gyroresonant opacity for ECM, the absorption issues can be avoided with plausible coronal conditions. GHz-frequency plasma emission may escape the corona when emitted along a steep density gradient \citep{benz1992}, which occurs due to the fibrous nature of the solar corona \citep{chen2013}, while GHz-frequency ECM (especially at the second harmonic rather than the fundamental) can escape the corona if emitted in a low-density cavity \citep{melrose1982}.

For our data set, the most promising observational indicator of emission mechanism is the degree and sense of circular polarization. Plasma emission at the fundamental is expected to be in the ordinary mode (o-mode) since it is below the minimum frequency at which the x-mode can propagate; the second harmonic is most often unpolarized but sometimes observed with weak o-mode polarization that is predicted not to exceed roughly 50\% \citep{melrose1978}. Therefore, the strong polarization observed in almost all events indicates that if it is plasma emission, it should be at the fundamental frequency and in the o-mode. ECM, in the low-density limit \citep[ratio of plasma frequency to cyclotron frequency < 0.3;][]{melrose1984}, should be dominated by emission at the fundamental frequency in the extraordinary mode (x-mode), observed to have up to 100\% circular polarization.  Propagating outwards from a north magnetic pole, o-mode corresponds to left circular polarization, and x-mode to right circular polarization. Since we observed the long-duration radio bursts to have a consistent sense of circular polarization on a given star, we can compare this sense of polarization to stellar magnetic field measurements to assess whether the emission is in the x-mode or the o-mode and thereby constrain the emission mechanism for the long-duration events (Section~\ref{section:pol_ZDI}). For the short-duration events, which do not have a consistent sense of polarization, we are unable to differentiate between fundamental plasma emission or electron cyclotron maser emission.

\subsubsection{Cause of Frequency Drift\label{section:freq_drift_interpretation}}
Identification of the emission mechanism helps determine whether it is appropriate to use frequency drift to measure source velocity. Frequency drift cannot always be used to measure source velocity because there is more than one potential cause of frequency drift in coherent bursts: source motion, rotational modulation, frequency-dependent propagation effects, and evolving plasma conditions at a fixed location. Here we consider the two most commonly identified causes of frequency drift in solar and planetary radio bursts, which are source motion and rotational modulation.
\begin{enumerate}
\item \textbf{Source motion.} Both plasma emission and ECM can show frequency drift caused by source motion, occurring on timescales of milliseconds to minutes. Of note are solar Type II and III bursts, which are plasma emission originating from coronal shock fronts and electron beams, respectively. Measuring frequency drift rate, combined with a model of the spatial variation of plasma density or magnetic field strength with height $h$, enables estimation of an apparent source velocity \citep[more detailed treatments can be found in, e.g.,][]{crosley2017}. If a source is moving outwards radially at constant speed $V=dh/dt$, the frequency drift rate $\dot{\nu}$ is:
\begin{equation}
\dot{\nu} = \frac{d\nu}{dh} \frac{dh}{dt} = V \frac{d\nu}{dh} .
\end{equation}
We define a scale height for emission frequency, $H_\nu = \nu/(d\nu/dh)$, which depends on the emission mechanism and the coronal structure.  For plasma emission, $H_\nu$ is twice the density scale height, and for cyclotron emission, $H_\nu$ is the magnetic scale length $B/(dB/dr)$. We can then write an expression for velocity:
\begin{equation}
V = H_\nu \left( \frac{\dot{\nu}}{\nu} \right)
\label{eq:drift_speed}
\end{equation}
We apply this equation to an event on YZ~CMi in Section~\ref{section:yzcmi}.

The timescale for a radio burst caused by a moving source is approximately $H_\nu/V$. The longest duration for a burst (or burst substructure) that is caused by a moving source will be for the largest expected $H_\nu$ and smallest expected velocity.  Generously taking $H_\nu \sim R_*$ (larger than the expected hydrostatic equilibrium scale height; see Table~\ref{table:targets}) for our largest star AD~Leo, and using a velocity of 1000 km\,s$^{-1}$ that is comparable to that of a shock-forming solar CME, we see that the timescale should be of order 5 minutes or less.  Thus, while our short-duration events may plausibly originate from moving sources, the long-duration events either require some other explanation, or they must be the combined signal from many different moving sources.

\item \textbf{Rotational modulation.} Coherent emission can be highly beamed, particularly ECM, which is beamed into the thin surface of a wide cone.  (Propagation effects in the stellar atmosphere can modify the opening angle of the cone.) As the surface of the cone emitted from a particular region passes through our line of sight, we briefly detect emission from that region.  Frequency drift over time can be produced as a purely geometric effect, as the beamed emission from different source regions rotates into the line of sight.  This effect has been modeled for Jupiter's decametric emission, using beam widths of order $1\degree$ as observed by \citet{kaiser2000}, to explain arc shapes in the dynamic spectrum with frequency structure varying over minutes to hours \citep{hess2008}, a timescale comparable to our long-duration bursts.  Plasma emission can also have narrow angular beaming, as measured for solar Type~III bursts at the fundamental plasma frequency \citep{thejappa2012}, raising the possibility that long-duration plasma emission may also show frequency drift due to rotational modulation.
\end{enumerate}

In order to detect bulk plasma motion in stellar coronae, it is important to distinguish between true source motion and frequency drift due to rotational modulation or propagation effects.  A smoking gun that rules out true source motion is repetition of the emission feature once per stellar rotation period, such as seen on brown dwarfs \citep{hallinan2007} and CU~Virginis \citep{trigilio2011}.  Since our 2- and 4-hour blocks are shorter than the stars' rotation periods (Table~\ref{table:targets}), we cannot test periodicity of radio bursts on the stars in this sample, but can rely on other clues.  Notably, the UV~Cet burst in Figure~\ref{fig:uvcet_burst_example} resembles bursts on UV~Cet in other epochs with very similar complex structure in the time-frequency plane.  This repetition suggests that the patterns of frequency drift in these event are due to rotational modulation, since it would be unlikely for such a complex set of source motions to recur in the same pattern in multiple epochs.

\subsection{Short-Duration Events}

To identify stellar space weather events, we look for radio bursts with frequency drift caused by outwards source motion, analogous to solar Type II and III bursts. The eight short-duration events detected in this survey occur on timescales of seconds to minutes, consistent with the timescales for electron acceleration expected for space weather events such as flares and CMEs.  However, the observed stellar short-duration bursts have a number of properties that diverge from those of solar radio bursts associated with geoeffective space weather events. In this section, we first use the YZ~CMi event from Figure~\ref{fig:yzcmi_burst_example} to highlight similarities to and differences from the population of solar radio bursts, then we discuss the properties and physical origins of the ensemble of short-duration stellar bursts.

\subsubsection{Short-Duration Radio Burst on YZ~CMi\label{section:yzcmi}}

A luminous burst on YZ~CMi, shown in Figure~\ref{fig:yzcmi_burst_example}, demonstrates an intriguing resemblance to solar radio bursts associated with source motion, and yet also shows properties dramatically different from most solar radio bursts. The 10-minute burst, which spans 1-4~GHz, has a number of distinct features in the time-frequency plane, with a range of frequency drift rates from fast to slow.  This bears a morphological resemblance to some solar radio bursts, where features with a range of negative drift rates (such as Type II and III bursts) may appear together due to sources moving outwards at different speeds that are associated with one flare/eruptive event. However, in contrast to solar Type II and Type III radio bursts, this event shows a high degree of circular polarization whose sense (left or right) varies with frequency and time.

\textbf{Frequency drift rate.} This event consists of features with a range of frequency drift rates.  One such feature, providing potential evidence of rapid frequency drift, is a collection of right-polarized vertical spikes in the dynamic spectrum (t$\sim$2-4~min in Figure~\ref{fig:yzcmi_burst_example}).  For these features, no frequency drift is resolved with our 1-second integration time, but they bear a morphological similarity to solar radio bursts caused by high-speed electron beams.  By analogy, the bandwidth of the vertical spikes may be due to unresolved rapid frequency drift caused by high source speeds.  For example, one such feature in the dynamic spectrum, at t=5.0 minutes, extends from at least 2.4~GHz to 3.5~GHz, or $\Delta\nu$=1.1~GHz. The range of frequencies covered within one integration time $t_\textrm{int}$ is determined by the frequency drift rate $\dot{\nu}$, the spike duration $\Delta{t}$ (which is $\leq t_\textrm{int}$), and the emission's instantaneous fractional bandwidth $f_\textrm{BW}$: $\Delta\nu = |\dot{\nu}|\Delta{t} + f_\textrm{BW}\nu$.  This implies a lower limit on frequency drift rate of:
\begin{equation}
|\dot{\nu}| > \frac{\Delta\nu - f_\textrm{BW}\nu}{t_\textrm{int}}.
\end{equation}
Observations of fast-drift bursts on AD~Leo with milli-second time resolution \citep{osten2006,osten2008} measured a median instantaneous bandwidth of 5\% \citep[range from $<$0.5-15\%][]{osten2006,osten2008}. While we cannot rule out the possibility that the instantaneous bandwidth in this case is larger, if we assume $f_\textrm{BW}\sim$0.15, we obtain 
$|\dot{\nu}|>0.66$~GHz\,s$^{-1}$ or $|\dot{\nu}/\nu|> 0.22$~s$^{-1}$.

Converting frequency drift rate to speed depends on how rapidly emission frequency varies with height in the source region, encapsulated by scale length $H_\nu$ in Equation~\ref{eq:drift_speed}.  $H_\nu$ is challenging to estimate without spatially resolved plasma density (for plasma emission) or magnetic field measurements (for ECM).  We consider a range of scale heights from 0.1$R_*$ to 1$R_*$, which brackets the hydrostatic equilibrium density scale height of 0.34$R_*$ (Table 3), so this is a plausible range of scale heights for plasma emission.  This also covers a range of possible length scales for decline in magnetic field near small-scale features ($H_\nu\sim 0.1R_*$) or in a region dominated by the global dipole field ($H_\nu\sim R_*$), which contains a significant fraction of the total magnetic flux \citep{morin2008}.  For $H_\nu\sim 0.1-1R_*$, Equation~\ref{eq:drift_speed} yields speeds of 1.5\% to 15\% of the speed of light.

Different types of solar radio bursts often appear together, such as the fast-drift Type III and the slow-drift Type II bursts.  Similarly, this event's vertical features, which may be due to unresolved fast frequency drift, are followed within a few minutes by other burst features, some amorphous and some with gradual negative frequency drift.  However, none of the features bear a clear morphological resemblance to a Type~II burst, and the frequency drift is too gradual to be caused by a coronal shock front.  For instance, a left polarized feature at t$\sim$10~minutes in Figure~\ref{fig:yzcmi_burst_example} moves from 3.0 to 2.6~GHz in the course of roughly 30~seconds, so $\dot{\nu}/\nu\sim-4.7\times10^{-3}$~sec$^{-1}$. Using $H_\nu\sim0.1-1R_*$, we estimate an apparent outward speed of 100-1000~km\,s$^{-1}$.  However, the Alfv\'en speed in the strongly magnetized coronae of active M~dwarfs is likely of order a few thousand km\,s$^{-1}$, so that a disturbance moving at a few hundred~km\,s$^{-1}$ could not form a shock front. Therefore, we prefer an alternative explanation for the slow frequency drift, such as evolving conditions in a source region or modulation by angular beaming.

While this burst shows possible evidence of coronal source motion, including unresolved vertical spikes that may be due to rapidly moving sources, it is significant that these features do not continue down to low frequencies. The burst is not detected in the P~band dynamic spectrum nor in a cleaned image of the P~band data (entire bandwidth) at the times of the higher frequency burst peak, 2015 May 12 0:38-0:40~UT.  The non-detection in this image puts a 3$\sigma$ upper limit of 10.8~mJy in Stokes~I and 5.4~mJy in Stokes~V on the P-band flux density of the burst, lower than the detected 1-2~GHz flux density of 29.4~mJy (I) and 8.2~mJy (V) and 2-4~GHz flux density of 23.1~mJy (I) and 7.6~mJy (V), integrated over the same time range. (For reference, 1~mJy on this star is 4.3$\times10^{13}$~erg\,s$^{-1}$\,Hz${-1}$.)  In contrast, solar Type~III bursts often continue down to very low (kHz) frequencies, which is interpreted as evidence of electrons escaping the corona along open field lines (likely accompanied by accelerated protons that can impact planetary atmospheres).  The non-detection of this event at low frequencies means that in this case, there is no evidence that accelerated particles are escaping the corona.

\begin{figure}[h]
  \begin{center}
     \includegraphics[width=\columnwidth]{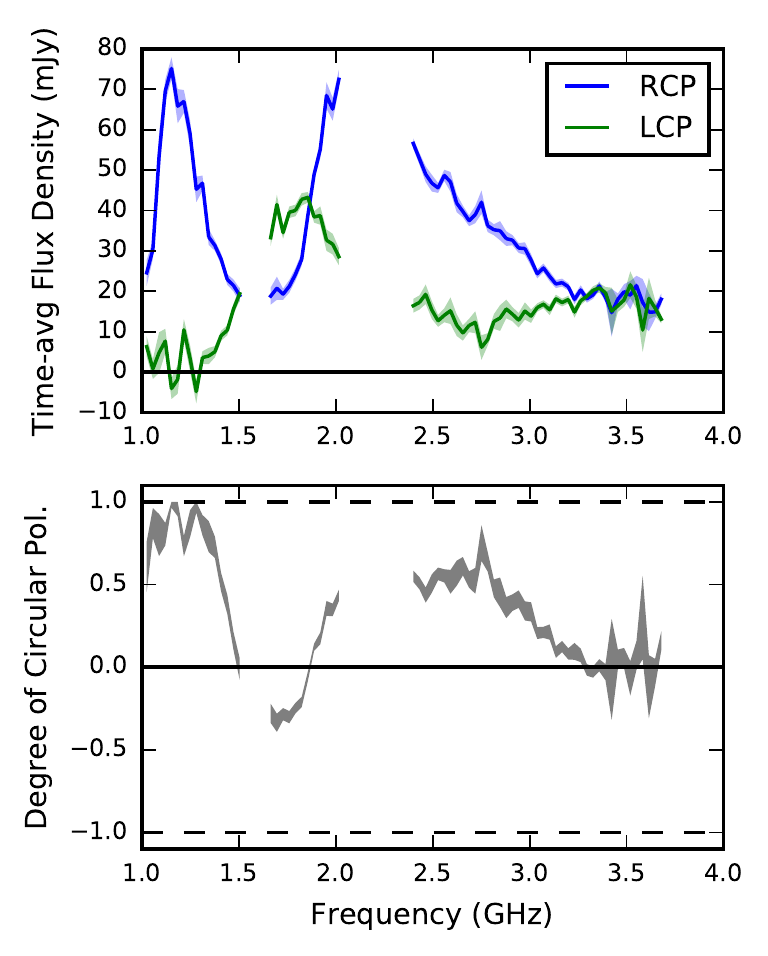}
     \vspace{-2em}
  \end{center}
\caption{Burst peak spectrum for the YZ~CMi event from Figure~\ref{fig:yzcmi_burst_example}, spanning 2015 May 12 0:38-0:40 UT. \textit{(top)} Right circular polarization (RCP) and left circular polarization (LCP) flux density spectra, where polarized flux density is defined using the convention that I=(RCP+LCP)/2. The 1-$\sigma$ statistical error in each channel is shown as an envelope around each spectrum. \textit{(bottom)} 68\% confidence interval for degree of circular polarization versus frequency; +1 is 100\% right polarized and -1 is 100\% left polarized.\label{fig:yzcmi_pol_spec}}
\end{figure}

\begin{figure}[h]
  \begin{center}
     \includegraphics[width=\columnwidth]{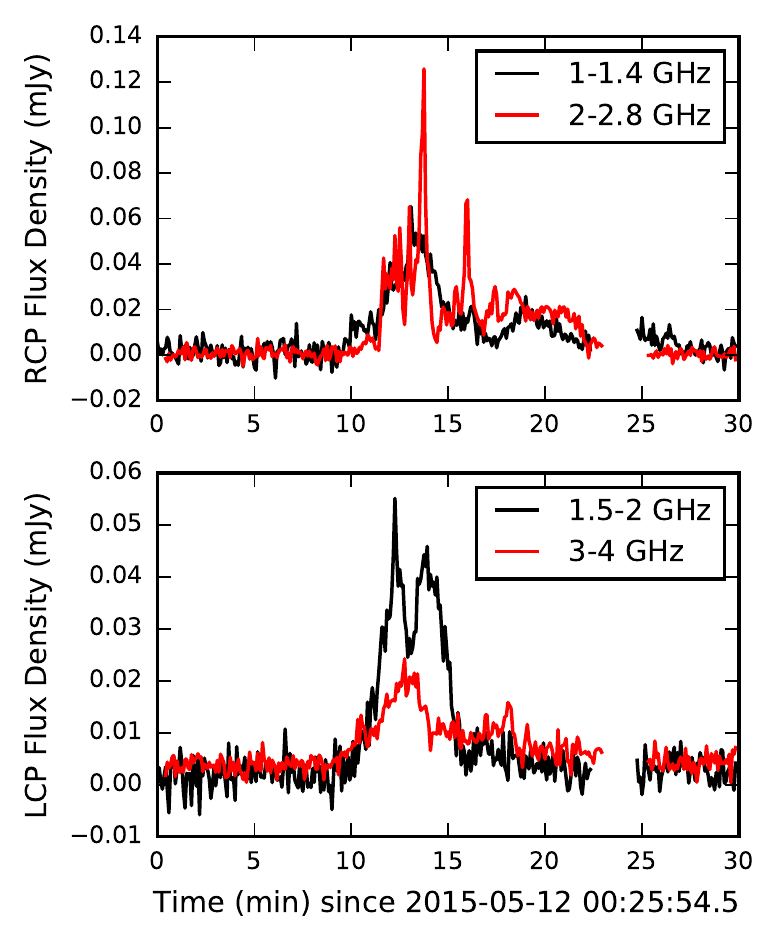}
     \vspace{-2em}
  \end{center}
\caption{Time series of flux density of the YZ~CMi burst from Figure~\ref{fig:yzcmi_burst_example}. The right polarized burst features \textit{(top)} and left polarized burst features \textit{(bottom)} are integrated over frequency ranges corresponding to distinct spectral features. While all frequencies show flare activity at the same time, the features at $\nu$ (black line) and $2\nu$ (red line) with shared polarization do not have similar time variation within the burst.\label{fig:yzcmi_tseries}}
\end{figure}

\textbf{Polarization structure.} Figure~\ref{fig:yzcmi_pol_spec} shows the frequency dependence of sense of polarization during the burst peak from 0:38-0:40~UT. During this time, the mapping between frequency and sense of polarization is, roughly: RCP for 1-1.5~GHz, LCP for 1.5-1.9~GHz, and RCP above 1.9~GHz, declining to near zero polarization above 3.2~GHz. The presence of right polarized burst features simultaneously at frequencies $\nu$ (1-1.5 GHz) and $2\nu$ ($\sim$2-3~GHz) is reminiscent of harmonic structure.  However, time series of the emission during the burst peak (Figure~\ref{fig:yzcmi_tseries}) do not show correlated behavior at $\nu$ and 2$\nu$.  In addition, it it challenging to explain this burst's strong polarization at both $\nu$ and 2$\nu$ in the same sense of polarization using harmonic structure, for either plasma emission \citep[which should have a weakly polarized second harmonic;][]{melrose1978} or ECM \citep[for which the fundamental and harmonic should have opposite polarization if they appear together; e.g.,][]{winglee1985,mellott1986}.

Instead of harmonic structure, there are a few other possible explanations for the multiple alternations in sense of polarization as a function of frequency. The reversals in polarization may be attributed to: spatial variations in the ratio of plasma frequency to cyclotron frequency \citep{melrose1984,winglee1985} or in the energy distribution of the exciting electrons \citep{melrose1984,zhao2016}; different emission mechanisms (plasma and ECM) contributing x-mode and o-mode; or multiple interconnected source regions with opposite magnetic polarity.
None of the scenarios provide a straightforward explanation for why there is more than one polarization reversal with increasing frequency, and we find no definitive way of differentiating between them with the available data.  However, the last scenario, two separate but connected source regions, would permit independent time variation in the two polarizations (due to a disturbance affecting both regions, but causing different electron acceleration time profiles in the two source regions), which would also help explain why the right and left polarized features do not line up perfectly in time, and certain features show up in only one polarization, such as the vertical spikes seen in RCP at 2.4 to 3.5~GHz.

\subsubsection{Physical Origins of Short-Duration Events}

The YZ CMi event demonstrates attributes characteristic of the population of short-duration stellar coherent bursts. The short-duration bursts all have strong circular polarization (50-100\%); this is true for most short-duration bursts in the literature as well, with only a few cases of unpolarized or weakly polarized bursts \citep{kellett2002}.  The short-duration bursts in this survey do not show evidence of harmonic structure, with the possible exception of some features in the YZ~CMi event. Significantly, while multiple events show frequency drift, the frequency drift is either too slow to be associated with a coronal shock, or too fast to be resolved (if it is indeed frequency drift) and therefore much too fast for a coronal shock.  In addition, no short-duration events continue to drift to frequencies below the edge of our observing band.

The lack of harmonic structure and strong degree of polarization of the stellar short-duration coherent bursts are both characteristic of the electron cyclotron maser, but may also be explained by fundamental plasma emission.  Indeed, these properties are characteristic of solar Type~I noise storms \citep{komesaroff1958,jaeggi1982}, which are attributed to fundamental plasma emission \citep{melrose1980}, but which tend to be long-duration events. In contrast, the short-duration plasma-emission solar Type~II bursts (which trace CMEs) and Type~III bursts (which trace fast electron beams) most often show weak to moderate circular polarization that only occasionally exceeds $\sim$50\% \citep{zlobec1993,dulk1980}, and the fundamental and second harmonic are often detected together \citep{roberts1959,smerd1976}, although fundamental emission may be suppressed above $\sim$100~MHz \citep{dulk1980}. Short-duration solar bursts can show a high degree of circular polarization \citep[millisecond spike bursts;][]{benz1982,nonino1986,stahli1986}, but such bursts are generally attributed to ECM \citep{melrose1994}.

The sense of circular polarization of the short-duration events provides an additional clue to their origins: multiple events on the same star may show different polarizations, or a single event may have both right and left polarized features (Figure~\ref{fig:yzcmi_burst_example}).  This arbitrary polarization implies that the bursts originate in regions with apparently random magnetic polarity, as expected for the small-scale magnetic features that should dominate most of the stellar surface \citep[since only $\sim$14\% of magnetic flux is in the large-scale field for active mid-M~dwarfs;][]{reiners2009topology}. This is consistent with the expectation for coronal disturbances (such as flares) driven by reconnection processes associated with the evolution of small-scale magnetic structures.

In contrast to many solar Type~II and Type~III bursts, which can continue down to kHz frequencies, none of the short-duration events appear to continue to frequencies below the edge of our observing band. As discussed above for the YZ~CMi event, this means that whatever coronal disturbances are responsible for the short-duration bursts, they likely do not propagate freely away from the star into interplanetary space.  However, another possibility is that disturbances further from the star do not fulfill the conditions for coherent emission or they produce directional coherent emission that is consistently beamed away from us. It is possible that such events are still caused by the same phenomena as solar bursts.  For example, short-duration bursts could be caused by a shock front that only forms briefly when a coronal disturbance (such as a CME) passes through a weakly magnetized region where it exceeds the Alfv\'en speed.  However, it is also possible that the bandwidth and duration of these short events are determined by an angular beaming pattern, rather than by source motion.

The population of short-duration coherent bursts do not have clear analogs within the solar classification system, in terms of polarization, harmonic structure, and frequency range and drift rate, preventing an immediate interpretation of these bursts in terms of space weather. However, it is still possible that these short-duration bursts are associated with coronal disturbances that drive space weather, but that such coronal disturbances result in different types of radio bursts in the extreme coronae of active M~dwarfs than on the Sun. If so, we expect that the radio bursts should be associated with flares at optical/X-ray/UV wavelengths.  Past multi-wavelength studies have found that radio bursts on active M~dwarfs are only sometimes associated with flares at other wavelengths \citep{kahler1982,nelson1986,kundu1988,gudel1996,vandenoord1996,gagne1998,osten2005}, but these studies used radio time series. With the additional information from wide-bandwidth dynamic spectroscopy, multi-wavelength campaigns can help sort short-duration coherent radio bursts into different populations based on flare association, identifying which radio bursts may be associated with space weather events.

\subsubsection{Non-Detection of Radio-Loud CMEs}
Further work is needed to characterize the population of stellar short-duration coherent bursts and determine their physical origins.  However, none of the short-duration bursts detected by the survey are a clear analog to solar Type~II bursts produced by coronal shock fronts (which are often driven by super-Alfv\'enic CMEs). For this reason, we calculate our flux density detection threshold for radio-loud coronal shock fronts around these stars, and scale it to a distance of 1~AU to contrast it to the flux densities observed for solar Type~II bursts. We also consider whether our non-detection may be due to the rate of shock-forming CMEs expected on these stars based on their energetic flare rates.

We concentrate here on the 220-480~MHz band, since, as discussed in Section~\ref{section:frequency}, it is more likely for stellar Type~II-like bursts to occur in this band than at higher frequencies.  To put an upper limit on the possible flux density of Type~II-like bursts during our observations, it is helpful to predict the bandwidth and duration of such events.  The bandwidth is determined by the range of plasma densities in the emitting region, difficult to predict a priori, so we assume a typical solar value for instantaneous fractional bandwidth of $\Delta \nu / \nu \sim 0.2-0.4$ \citep{mann1995}, where 0.2 corresponds to a burst bandwidth of 70~MHz at an observing frequency of 350~MHz.  The duration is then the time for a burst with drift rate $\dot{\nu}$ to cross the burst bandwidth: $\Delta t = \Delta \nu / \dot{\nu}$.  Using a CME speed of 2000~km\,s$^{-1}$ (see discussion of Alfv\'en speed later in this section) and a density scale height of 75~Mm (middle of the range in Table~\ref{table:targets}), we use Equation~\ref{eq:drift_speed} to obtain $\dot{\nu}/\nu = V/(2H) \sim$ 1/(70~s) and $\Delta t \sim$15~sec. In a dynamic spectrum pixel of 70~MHz and 15~sec, for the 220-480~MHz band, we obtain a typical Stokes~I sensitivity of 17~mJy. We consider only the Stokes~I sensitivity, and not Stokes~V, since solar Type~II bursts tend to have fairly weak polarization. Our 3$\sigma$ upper limit on Stokes~I flux density of possible Type~II bursts during our 42~hours of 220-480~MHz observations is roughly 50~mJy.

We scale our survey's detection threshold to 1~AU to compare to the flux densities of observed solar Type~II bursts to evaluate whether sensitivity may cause the lack of Type~II detections. If these stars were at 1~AU, the 50~mJy upper limit on flux density would correspond to an upper limit of 1.6-8.2$\times10^6$~solar flux units (SFU), where $10^6$~SFU at 1~AU corresponds to a spectral luminosity of 2.8$\times10^{14}$~erg/s/Hz. There is limited information available on solar flux densities at metric wavelengths.  The first observed solar Type~II burst \citep{payne-scott1947} peaked at $10^7$~SFU at 60~MHz. \cite{nelson1985book} report that the brightest metric bursts observed by the Culgoora radioheliograph are of order $10^5$~SFU, and flux densities can be as low as 1~SFU.  Thus, the sensitivity of our survey would be enough to detect the stellar equivalent of only the most exceptionally bright solar Type~II bursts, so it is plausible that radio-loud CMEs on these stars may be occurring below the detection limit. \cite{crosley2018vla} assess detectability of stellar Type~II bursts for their observations of EQ~Peg with all 27 VLA antennas, finding that a CME shock with the maximum observed brightness temperatures for solar Type~II bursts of $10^{14}$~K \citep{benz1988}, and source size comparable to the stellar disk, would be barely observable, but they note that if stellar Type~II bursts have higher brightness temperatures of $10^{18}$~K, consistent with the high brightness temperatures observed in some stellar coherent bursts \citep{osten2008}, these events should be easily detectable by the VLA.

Given the uncertainty in expected brightness temperature for stellar Type~II bursts, the non-detection may be caused by sensitivity or by the rate of CME shocks around the observed stars. We can compare our non-detection of radio-loud CMEs during this survey to the rate of energetic flares at other wavelengths to test our expectations based on solar flare-CME correlations. We compare to flare rate in two ways:
\begin{enumerate}
\item \textit{Rate of flares expected to be associated with CMEs.} The most energetic solar flares show an almost 100\% association rate with CMEs \citep{yashiro2009}. This is true for solar flares with GOES X-ray fluence above 0.2~J\,m$^{-2}$, which corresponds to a U-band stellar flare energy of $6\times 10^{30}$~erg \citep[using the flare energy scaling relationships of][]{osten2015}.  Based on the flare frequency distributions of \cite{lacy1976}, a flare above this energy is expected to occur every: 7 hours on AD~Leo, 1 hour on EQ~Peg \citep[consistent with the prediction of][]{crosley2018multiwavelength}, 4 hours on EV~Lac, 3 hours on YZ~CMi, 15 hours on UV~Cet.  Based on our survey's duration on each target, we expect that during the survey about 11 optical flares occurred at an energy high enough that such a flare would almost certainly be associated with a CME if it occurred on the Sun.  Therefore, the solar flare-CME association rate alone cannot be used to predict the rate of radio-loud CMEs at 220~MHz and above for our sensitivity threshold.

\item \textit{Rate of flares expected to be associated with fast CMEs that can form shocks.} In the solar corona, more energetic flares tend to be associated with faster CMEs \citep{yashiro2009} that are more likely to form shocks and produce radio bursts.  To form a shock, a CME must exceed the local fast magnetosonic speed, which is roughly the Alfv\'en speed in the magnetically-dominated corona. The Alfv\'en speed is
\begin{equation}
v_A=\left(1940~\textrm{km\,s$^{-1}$}\right)
	\left(\frac{B}{50~\textrm{G}} \right)
    \left( \frac{n_e}{10^{9.5}~\textrm{cm}^{-3}} \right)^{-1/2}.
\end{equation}
To assess what CME speed is required to form a shock, we estimate the Alfv\'en speed at $2R_*$ for our targets.  To do so, we use the density at $2R_*$ that corresponds to the plasma frequency estimated in Table~\ref{table:targets}.  The large-scale magnetic field of the target stars is mostly poloidal, with a strong dipole component of order a few hundred Gauss~\citep{morin2008,kochukhov2017}.  We therefore estimate that a typical field strength at $2R_*$ is of order $(400~G)/2^3=50~G$.  We then obtain Alfv\'en speeds at $2R_*$ of 1600 to 3500~km\,s$^{-1}$ for our targets, where the later-type stars have lower Alfv\'en speeds.  This threshold velocity for CMEs to form radio-loud shocks is significantly higher for flare stars than for the Sun: the solar Alfv\'en speed is of order 500 km\,s$^{-1}$ at heights above 1.2$R_\odot$ \citep{gopalswamy2001,mann2003}.

We next estimate the typical energy of a stellar flare associated with such a fast CME, based on solar correlations between CME properties and flare energy. \cite{crosley2016} combine the correlations between solar flares and CME properties from \cite{aarnio2012} and \cite{osten2015} to derive an equation predicting stellar U-band flare energy from CME velocity: $E_U = (9.8\times 10^{30}~\textrm{erg}) v^{5.4}$ (from their Equation~7). Our Alfv\'en speed estimates correspond to U-band flare energies of $10^{32}$ to $10^{34}$~erg for our targets, and the flare frequency distributions of \cite{lacy1976} predict one such energetic flare per: 200 hours on EV~Lac or YZ~CMi, 400 hours on AD~Leo and UV~Cet, and 1800 hours on EQ~Peg. (In contrast, \cite{crosley2018vla} find roughly the same threshold U-band flare energy for EQ~Peg of $10^{34}$~erg, but predict that such a flare occurs once per 27 hours.) This corresponds to only a 15\% chance that during our survey a sufficiently energetic flare occurred to produce a fast shock-forming (and thus radio-loud) CME, if stellar flares follow the same flare-CME correlations as solar flares.

This estimation process is extremely uncertain due to the order-of-magnitude scatter in scaling relationships, the steep dependence of predicted U-band flare energy on CME speed, and ignorance of conditions in the stellar corona \citep[whose strong magnetic fields may suppress or slow CMEs;][]{alvarado-gomez2018}.  This calculation does serve, however, to demonstrate that the lack of detections of stellar Type~II-like radio bursts does not necessarily imply a lack of CMEs; it could also be caused by the rarity of CMEs fast enough to exceed the high Alfv\'en speed in the strongly magnetized coronae of active M~dwarfs.  It remains a possibility too that stellar Type~II-like bursts are occurring in the observed frequency range, but that our observations have not reached the requisite sensitivity.

\end{enumerate}

Due to the range of factors that may cause a non-detection of stellar Type~II bursts, our observations do not constrain the occurrence rate of stellar coronal mass ejections. However, these observations demonstrate that the observed population of coherent radio bursts on active M~dwarfs is not caused by the mechanism of plasma emission from a CME shock front, and therefore that the current-generation VLA does not provide a systematic method to detect and characterize stellar CMEs on active M~dwarfs.

\subsection{Long-duration Events - Solar Storms or Ultracool Dwarf Aurorae?}

The long-duration coherent bursts observed in this survey share the strong circular polarization of short-duration events, indicating that they may originate from the same emission process.  However, we can derive additional clues about the origins of long-duration events from their persistence over hours, their frequency dependence, and their consistent sense of polarization. These properties may be considered within the framework of solar radio bursts or in the framework of periodic radio aurorae produced by brown dwarfs and planets, or they may belong to a population of radio bursts unique to active M~dwarfs.


\subsubsection{Hours-Long Duration\label{section:acceleration}}

Stellar coherent emission processes derive energy from a positive gradient in the electron velocity distribution (such as an excess of high-speed electrons). The observed long-duration bursts require ongoing electron acceleration to supply these high-speed electrons throughout their hours-long duration. The presence of such an ongoing acceleration process is perhaps not surprising, given that these stars continuously produce non-thermal incoherent quiescent emission at GHz frequencies \citep{gudel1994}. The electron acceleration process responsible for the incoherent quiescent emission is an ongoing subject of debate \citep{benz1994,airapetian1998,kellett2002}, and it is unclear whether the incoherent and coherent emission have shared or distinct acceleration mechanisms. Solar and planetary processes both provide examples of possible acceleration processes responsible for the long-duration coherent emission.  Possible explanations within the paradigm of solar and stellar activity include: ongoing magnetic reconnection caused by emerging magnetic flux \citep[thought to be responsible for Type~I solar storms; e.g.,][]{stewart1986,bentley2005}; quasi-steady acceleration due to frequent small flares \citep[discussed in the context of active binaries by][]{lefevre1994}; gradual energy release during a flare decay phase \citep[which may explain long M~dwarf flare decay times;][]{kowalski2013}, possibly due to post-eruptive magnetic reconfiguration \citep{katsova1999}; or reconnection in an equatorial current sheet formed by the stellar wind \citep[proposed to explain X-ray and radio emission from magnetic massive stars;][]{havnes1984,linsky1992}. Possible explanations within the paradigm of planetary and brown dwarf auroral radio emissions \citep{zarka1998,hallinan2015} include: interaction of the large-scale magnetosphere with plasma distant from the star (co-rotation breakdown, analogous to Jupiter's main auroral oval) or currents on a magnetic field line connected to a close-in satellite (analogous to the Jupiter-Io interaction).  The Sun is not known to produce such auroral emission, but the strong large-scale magnetic field of these stars (0.1-1~kG in the dipole field, in contrast to about 1~G for the Sun) means that they are better able to generate powerful currents through magnetospheric processes distant from the star.

The AD~Leo and UV~Cet bursts in Figures~\ref{fig:adleo_burst_example} and \ref{fig:uvcet_burst_example} provide evidence that the acceleration mechanism operates over a timescale of weeks to months.  The AD~Leo burst is narrowband, with sharply defined lower and upper cutoff frequencies at 1.1 and 1.6~GHz. An observation 2 weeks later detects nearly identical emission (Figure~\ref{fig:2015_ADLeo_4} in the appendix) throughout that entire 4-hour observation as well, suggesting that this event might be ongoing for 2 weeks or more.  With its narrow bandwidth, strong circular polarization, and potentially weeks-long duration, this event bears a resemblance to solar Type~I noise storms \citep{elgaroy1977}.  Type~I noise storms are attributed to fundamental plasma emission from a closed magnetic field structure, where the lowest density within the magnetic field structure determines the low frequency cutoff of the burst.  Type~I noise storms can last for days, associated with the emergence of new magnetic flux through the photosphere \citep{stewart1986}.  However, the strongly polarized emission in Type~I storms consists of short sub-second spikes of emission, whereas the AD~Leo burst flux density varies smoothly (to within our sensitivity for our 1-second integration time).  In addition, the sense of polarization of the AD~Leo bursts is consistent with the x-mode (Section~\ref{section:pol_ZDI}), whereas solar Type~I storms are polarized in the sense of the o-mode. Nevertheless, the narrow bandwidth of the AD~Leo burst (and of the long-duration bursts observed on YZ~CMi; Figure~\ref{fig:2015_YZCMi_2} in the appendix) may be caused by emission from within a closed magnetic structure, analogous to Type~I storms. Alternatively, the narrow bandwidth may be attributed to an angular beaming pattern for which only certain frequencies are observable at Earth.  The AD~Leo burst has narrowband substructure with gradual positive frequency drift, which may also be due to modulation by an angular beaming pattern.

As discussed in Section~\ref{section:freq_drift_interpretation}, the UV~Cet burst in Figure~\ref{fig:uvcet_burst_example} strongly resembles events detected in other epochs separated by months.  Due to the complexity and repeated nature of this burst, it is most easily explained as periodic radio bursts analogous to those seen on planets and brown dwarfs, where the complex pattern in the time-frequency plane is due to rotational modulation of angularly beamed emission.

\subsubsection{Frequency Dependence}

Within this survey, luminous long-duration bursts are most abundant at 1-1.4~GHz compared to lower and higher frequencies (Table~\ref{table:bursts}; Section~\ref{section:subbands}).  While the peak frequency may vary somewhat from epoch to epoch or from star to star, this population of bursts stands in contrast to the Sun, on which luminous bursts are an order of magnitude more frequent below 1~GHz than in bands above 1~GHz \citep{nita2002}. The abundance of >1~GHz coherent radio bursts compared to the Sun can be attributed to the stronger magnetic field strengths (and cyclotron frequency) and higher plasma densities (and plasma frequency) in the coronae of active M dwarfs.  The decline of burst occurrence rate below 1~GHz observed on these stars indicates that the population of bursts observed by this survey are dominated by disturbances that are restricted to the low corona, where magnetic field strengths are >300~G and/or plasma densities are >$10^{10}$~cm$^{-3}$. The presence of this population of coherent bursts peaking at 1-1.4~GHz, which has no analog on the Sun, suggests that a non-solar process may be responsible for these bursts.

\begin{deluxetable*}{cccccccc}
\tablecaption{Magnetic Polarity and Long-Duration Radio Burst Polarization.\label{table:ZDI}}
\tablehead{\colhead{Star} & \colhead{Visible Magnetic Pole} & \colhead{Epoch} & \colhead{Reference} & \colhead{Long-duration Radio} & \colhead{Number} & \colhead{Epoch} & \colhead{Reference} \\[-1ex]
& & & & \colhead{Burst Polarization} & \colhead{of Events}}
\startdata
\multirow{3}{*}{AD~Leo}	& \multirow{3}{*}{S} & \multirow{3}{*}{2007, 2008} & \multirow{3}{*}{1} & LCP (1.4 GHz) & 1 & 1985 & 3 \\
			& 		& 				& 	& LCP (>1 GHz)	& 3	& 2015	& * \\
			& 		& 				& 	& RCP (<1 GHz)	& 2	& 2015	& * \\
\hline
\multirow{2}{*}{EQ~Peg A/B}	& \multirow{2}{*}{N/N\,\tablenotemark{a}}	& \multirow{2}{*}{2006}	& \multirow{2}{*}{1}	& RCP (1.5 GHz) & 1 & 1985 & 4   \\
	&   	& &  	& RCP (1-3 GHz)	& 1	& 2015 & *	\\
\hline
EV~Lac		& N, S\,\tablenotemark{b} 	& 2006, 2007	& 1	& no bursts		&	& 		\\
\hline
\multirow{6}{*}{YZ~CMi} & \multirow{6}{*}{S} & \multirow{6}{*}{2007, 2008} & \multirow{6}{*}{1} & RCP? (0.41 GHz) \tablenotemark{c}	& 2	& 1977 & 5 \\
            &		&				&	& LCP (4.9 GHz)\,\tablenotemark{d}	& 1 & 1983 & 6 \\
            &		&				&	& LCP (1.5 GHz) & 1 & 1984 & 6 \\
            &		&				&	& LCP (1.5 GHz) & 1 & 1985 & 4 \\
            &		&				&	& LCP (1.5 GHz) & 1 & 1987 & 7 \\
            &		&				&	& LCP (1.8-4 GHz) & 2 & 2015 & * \\
\hline
\multirow{3}{*}{BL~Cet/UV~Cet} & \multirow{3}{*}{N/N\,\tablenotemark{a}} & \multirow{3}{*}{2013} & \multirow{3}{*}{2} & RCP (1.5 GHz) & 1 & 1985 & 4 \\
		& 	& 		&  & RCP (0.2-8.5 GHz)\,\tablenotemark{e}	& 6	& 2013, 2015 & * \\
        &	&		&  & RCP, LCP (0.15 GHz)		& 4 & 2015 & 8 \\
\enddata
\tablecomments{Long-duration radio bursts are included from this work and the literature. See also the compilation in \cite{kellett2002}.}
\tablenotetext{a}{Both binary components have the north magnetic pole visible.}
\tablenotetext{b}{The sign of the net longitudinal field on EV~Lac varies with rotational phase, indicating that both magnetic poles rotate in and out of view over the course of the rotation period.}
\tablenotetext{c}{Observations had only right-polarized feeds, so burst may have had an LCP component.}
\tablenotetext{d}{Classified by the authors as slowly-variable quiescent emission, but included here due to its nearly 100\% circular polarization.}
\tablenotetext{e}{Two bursts on UV~Cet have a faint LCP signal that appears before the start of the RCP emission.}
\tablerefs{
(*) This work. (1) \cite{morin2008}. (2) \cite{kochukhov2017}.
(3) \cite{white1986}. (4) \cite{kundu1988}. (5) \cite{davis1978}.
(6) \cite{lang1986}. (7) \cite{lang1988}. (8) \cite{lynch2017}.
}
\end{deluxetable*}

\subsubsection{Consistent Sense of Polarization\label{section:pol_ZDI}}

On a given star, the long-duration bursts above 1~GHz show a consistent sense of circular polarization. 
Our targets all have magnetic field measurements from optical/IR spectropolarimetry \citep{morin2008,kochukhov2017}, providing the opportunity to compare the dominant radio polarization to the polarity of the large-scale photospheric magnetic field.

The majority of active mid M~dwarfs have a strong large-scale magnetic field that is nearly rotationally symmetric and poloidal \citep{morin2008}. For all of our targets except EV~Lac, the measured longitudinal field (the line-of-sight magnetic field averaged across the visible hemisphere of the star) has the same sign at all rotational phases, implying that the same magnetic pole is visible throughout the entire rotation period.  Table~\ref{table:ZDI} lists the observed polarity of the large-scale field of our targets and the sense of polarization of long-duration radio bursts from this survey and from the literature. For long-duration bursts above 1~GHz, the radio emission is LCP for stars with the magnetic south pole visible (AD~Leo and YZ~CMi) and RCP for stars with the magnetic north pole visible (EQ~Peg and UV~Cet).  This is consistent with emission in the extraordinary mode (x-mode) relative to the polarity of the large-scale field.  

The observed x-mode polarization and the high degree of circular polarization of the long-duration coherent bursts are characteristic of electron cyclotron maser emission.  A low-density environment ($\nu_p/\nu_c\lesssim0.3$) is required to explain the the x-mode polarization \citep{melrose1984} and the escape of the radiation in spite of gyroresonant opacity \citep{melrose1982,zaitsev2005}.  For 1~GHz emission, this implies electron densities of $\lesssim10^9$~cm$^{-3}$ in the source region, significantly lower than the average coronal densities of order $10^{10.5}$~cm$^{-3}$ (Section~\ref{section:frequency}). However, if any such hot, diffuse cavities exist in the corona, the emission can escape to greater heights via multiple reflections along the cavity, a process seen in Earth's auroral kilometric radiation \citep[e.g.,][]{zarka1998}. The long-duration bursts at 1-6~GHz originate in regions with field strengths of 0.36-2.1~kG (if fundamental emission) or 0.18-1.1~kG (if second harmonic).  We expect such field strengths to occur in the low corona, relatively close to the star, so that we only observe the polarization corresponding to the visible magnetic pole.

In contrast, low frequencies (<1~GHz) correspond to greater heights in the stellar atmosphere, which are less likely to be obscured by the star.  We therefore expect that low frequency bursts of either polarization may be observed.  Indeed, bursts observed below 0.5~GHz on UV~Cet \citep{lynch2017} and YZ~CMi \citep{davis1978} sometimes show opposite polarization compared to higher frequencies.

The agreement between radio polarization and polarity of the large-scale magnetic field indicates consistent magnetic polarity in the source region, suggestive of an electron acceleration process that populates the large-scale field. In a solar-like corona, flares should not selectively occur in regions that match the polarity of the large-scale field. Instead, the electron acceleration is likely occurring at a distance from the star where the large-scale field dominates, then electrons propagate along field lines down to the low corona where they produce radio emission.  This is analogous to the auroral current systems on planets and brown dwarfs.  However, we note that one other possible explanation (instead of source regions sharing the polarity of the large-scale field) is weak-mode coupling, which would cause all x-mode emission to match the handedness (R or L) of the x-mode in the large-scale field regardless of the polarity of the source region \citep[e.g.,][]{benz_book}.

While the polarization in the x-mode relative to the large-scale field is suggestive of auroral processes, the long-duration coherent bursts observed on active M~dwarfs in this survey are not a perfect analog to the periodic pulsed radio emission from Jupiter and brown dwarfs. Other than for the set of similar bursts on UV~Cet, there is no evidence that most of these long-duration M dwarf bursts are periodic. Depending on the geometry (shape of emitting region, inclination of the rotation axis, magnetic dipole axis, beam opening angle, and propagation effects), auroral radio emission from planets and brown dwarfs can have features of both polarizations, and the duration of individual pulses may be quite short or long. Ultracool dwarf radio pulses sometimes vary in sense of polarization between epochs, perhaps indicative of magnetic reversals \citep{route2016}, whereas active M~dwarfs seem to consistently prefer one radio polarization over decades \citep[Table~\ref{table:ZDI}; and][]{kellett2002}. There is only one pulsing ultracool dwarf with a spectropolarimetric measurement of the magnetic field, and the relationship between the polarization of radio pulses and magnetic field orientation is ambiguous: M8.5 dwarf LSR~J1835+3259 produced LCP pulses in 2006 \citep{hallinan2008} and RCP in 2011 \citep{hallinan2015}, and \cite{berdyugina2017} in 2012 detected a longitudinal magnetic field oriented away from the observer (corresponding to a southern magnetic pole).

\subsubsection{Time Evolution of Magnetic Activity}

In the Sun, various magnetic activity properties, including the rate of flares and coronal mass ejections, are linked to the evolution of the Sun's global magnetic field throughout the solar cycle, as well as to small-scale evolution of the magnetic field as active regions emerge and decay. The Sun's large-scale magnetic field reverses polarity every 11 years. It is unknown whether such magnetic polarity reversals will also occur on active mid-M~dwarfs, whose magnetic dynamo operates under different conditions than on the Sun \citep[rapid rotation and fully convective interiors, similar to the ultracool dwarf observations in][]{route2016}.

In contrast to changes in brown dwarf radio burst polarization that \cite{route2016} interpret as potential evidence of magnetic polarity reversals, the consistent sense of polarization of radio bursts in our survey, in agreement with non-contemporaneous magnetic measurements, suggests a lack of magnetic reversals between the spectropolarimetric observations in 2007-2012 and the radio observations in 2013-2015.  \cite{kellett2002} compiled observations of UV~Cet and YZ~CMi from 1983-1996, finding that the preponderance of radio bursts have polarizations that agree with the long-duration events in our survey, indicating that these stars may not have undergone a polarity reversal in over 30 years. The lack of magnetic reversals is consistent with the findings of \cite{vida2016}, who perform Doppler imaging of active mid-M dwarf V374~Peg, revealing a stable spot pattern on the star over 15~years. This survey adds to the evidence that active M~dwarfs may go decades without polarity reversals, but with so few objects studied so far (in this survey and in the literature), this pattern cannot be generalized to all active M~dwarfs.

While the large-scale magnetic field of these stars appears to be stable, the majority of magnetic flux is still in small-scale magnetic features \citep{reiners2009topology} that evolve on timescales of weeks to years \citep{giles2017}.  This may drive time variation in the observed levels of magnetic activity, including the rate of coherent bursts.  Within our survey, we note a higher burst detection rate on AD~Leo in 2015 (3 out of 5 epochs) compared to 2013 (0 out of 4 epochs).  This difference is largely due to long-duration bursts (with two nearly identical long-duration bursts detected two weeks apart in 2015), consistent with an acceleration mechanism for long-duration bursts that evolves on timescales of weeks to years.

Some past observations at $\sim$1.5~GHz have found relatively low rates of coherent bursts compared to our survey.  While we detected 1 burst per 3.7~hours at 1-2~GHz, \cite{osten2006} report 1 per 8~hours on AD~Leo at 1.1-1.6~GHz with Arecibo (16 hours total), and \cite{abadasimon1997} report 1 per 7~hours on AD~Leo, EQ~Peg, and YZ~CMi at 1.4~GHz with Arecibo (49.4 hours total, counting distinct events by requiring >5 minutes separation). This difference can be largely attributed to their lack of detection of long-duration bursts; our rate of detection of short-duration bursts at 1-2~GHz is only 1 per  11.2~hours (5 in 56 hours). The lack of detection of long-duration bursts in these surveys may be due to the use of a single dish telescope, which requires subtraction of smooth time variations to remove gain variations and noise. While there have been other reports of long-duration bursts on active M~dwarfs \citep{davis1978,lang1986,lang1988,white1986,kundu1988,slee2003}, these observations have all used interferometers. Some long-duration coherent bursts may also be classified by other authors as slowly-variable quiescent emission \citep[e.g., Figure 1 in][]{kundu1988}. This survey's broad bandwidth may also have helped improve the detection rate of narrowband coherent emission.

\begin{figure*}[b]
  \begin{center}
     \includegraphics[width=0.95\textwidth]{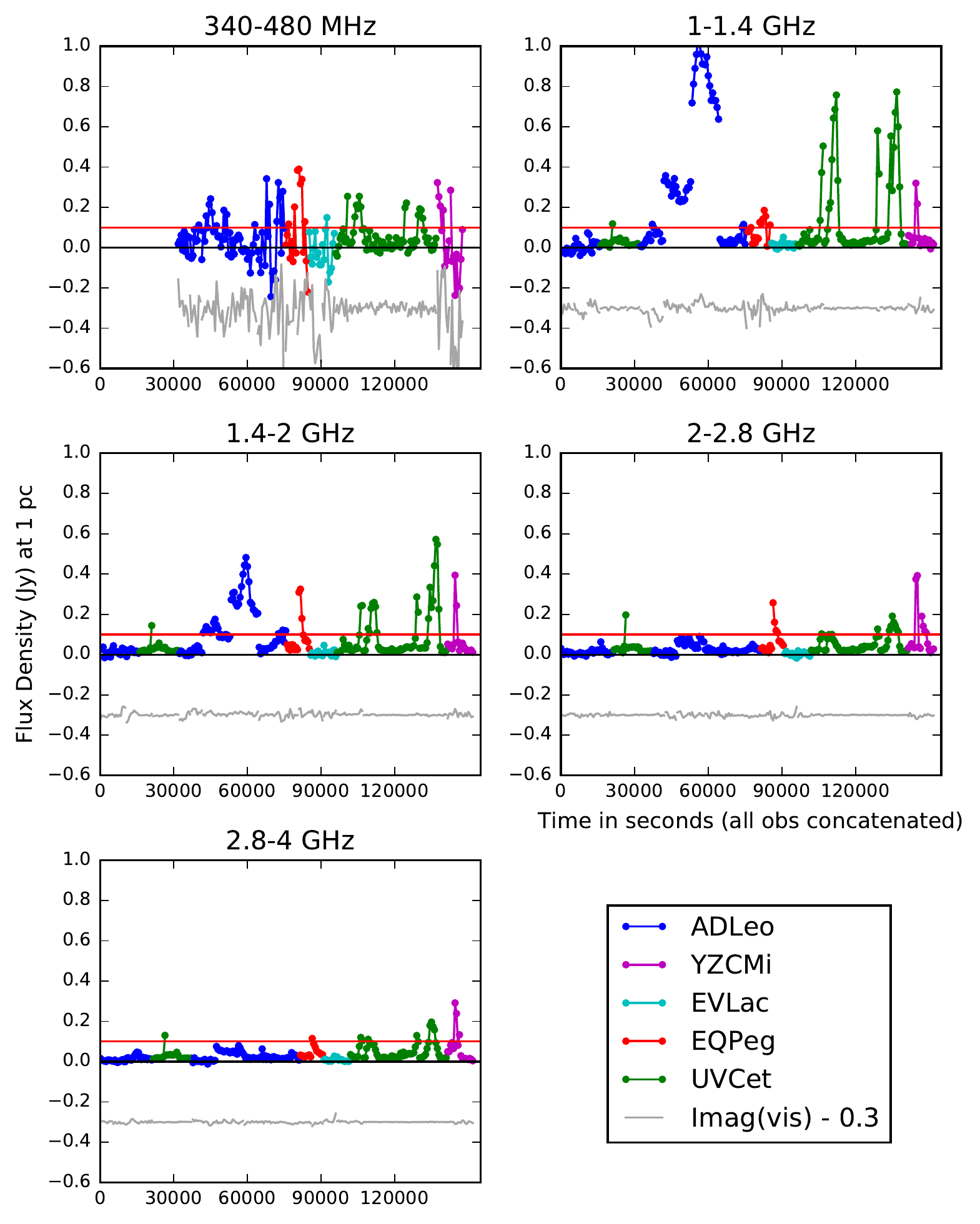}
     \vspace{-2em}
  \end{center}
  \caption{Time series of all observations, scaled to a distance of 1 pc.  The Stokes~I flux density is averaged to 10-minute integrations. The frequency bands were chosen to all have the same fractional bandwidth of $\nu_\textrm{max}/\nu_{min}\approx\sqrt{2}$. The observations are concatenated, removing times without data, so the x-axis shows the total time on source of our survey.  The data are colorized by source, with multiple observations of a source in a given year shown as a continuous block.  The time series is the real component of the baseline-averaged visibilities; shown offset by -0.3~Jy (gray line) is the imaginary component of the same quantity, which should not contain stellar signal and thus provides an estimate of the noise in the time series. The red horizontal line shows $S_\textrm{1pc}=$100~mJy, which is used as the minimum threshold flux density for calculating expected transient rates due to stellar coherent radio bursts.\label{fig:halfband_tseries}}
\end{figure*}

\section{Implications for Transient Surveys \label{section:transient_rate_estimation}}

Stellar flares from active M~dwarfs and close binaries are expected to be among the most common types of radio transients at decimetric wavelengths \citep{osten2008proc,williams2013,mooley2016}.  This expectation applies to ongoing and upcoming transient surveys by telescopes including ASKAP, MeerKAT, and the VLA.  Long-duration coherent bursts should dominate the population of flare star transients at low frequencies (below a few~GHz).  However, due to the narrowband nature of coherent emission, estimation of the transient rate is complicated because the frequency coverage of a given transient survey is not an exact match to that of past stellar observations. The wideband nature of our observations makes it possible to estimate the active M~dwarf transient rate adjusted for the frequency coverage of individual surveys.

To characterize the dependence of transient density on frequency, we first divide our data into sub-bands with equal fractional bandwidth, $\nu_\textrm{max}/\nu_\textrm{min}=\sqrt{2}$, and calculate transient density for each sub-band (Section~\ref{section:subbands}).
We then apply the same process to the frequency bands of three ongoing or upcoming transient surveys (Section~\ref{section:transient_surveys}): the ASKAP VAST survey, the MeerKAT ThunderKAT survey, and the VLA Sky Survey (VLASS).

\subsection{Transient Density Dependence on Frequency \label{section:subbands}}

To calculate transient density, we take three steps: 1) For a given frequency band, calculate the burst duty cycle (fraction of time spent bursting) as a function of luminosity. 2) Estimate the volume density of active M~dwarfs.  3) For a given frequency band, calculate the expected transient density, which is the number of bursts (above a given flux density) per area on the sky at an instant in time.

\subsubsection{Coherent Burst Duty Cycle\label{section:luminosity_dist}}

We divide our data into five sub-bands with equal fractional bandwidth: 340-480~MHz, 1-1.4~GHz, 1.4-2~GHz, 2-2.8~GHz, and 2.8-4~GHz. We use constant fractional bandwidth
$\nu_\textrm{max}/\nu_\textrm{min}$ so that, for a simple exponential coronal model (as in Section~\ref{section:frequency} and Equation~\ref{eq:drift_speed}), each sub-band will correspond to roughly the same interval in physical height $\Delta{h}=h\left(\nu_\textrm{min}\right)-h\left(\nu_\textrm{max}\right)=H_\nu\ln\left(\nu_\textrm{max}/\nu_\textrm{min}\right)$.
We will identify bursts as times when the Stokes I luminosity time series for a sub-band is above a certain threshold, using the same luminosity threshold for all sub-bands. We originally covered almost the entire observed range of frequencies by including two additional sub-bands, 240-340~MHz and 4-5.6~GHz, but have removed them from the analysis because the Stokes~I noise levels in 240-340~MHz were too high to yield meaningful results, and the coherent bursts at 4-5.6~GHz were too faint to be detected using our luminosity threshold (which was set by the noise level at 340-480~MHz).

To measure the coherent burst duty cycle for a given sub-band, we must first compute a luminosity time series for all of our data in that sub-band.  For each star, we first convert flux density to a proxy for luminosity, $S_\textrm{1pc}$, which is the flux density of the star scaled to a distance of 1~parsec. We then average the dynamic spectra along the frequency axis to produce time series, weighting each frequency channel by its inverse variance. The inverse variance is measured from the imaginary component of the dynamic spectrum; this downweights RFI to maximize sensitivity of the resulting time series. We compute time series of the imaginary component in the same fashion, to serve as a control time series that should contain no stellar bursts.  Figure~\ref{fig:halfband_tseries} shows the resulting time series of all the survey data.

\begin{figure*}[t!]
  \includegraphics[width=\textwidth]{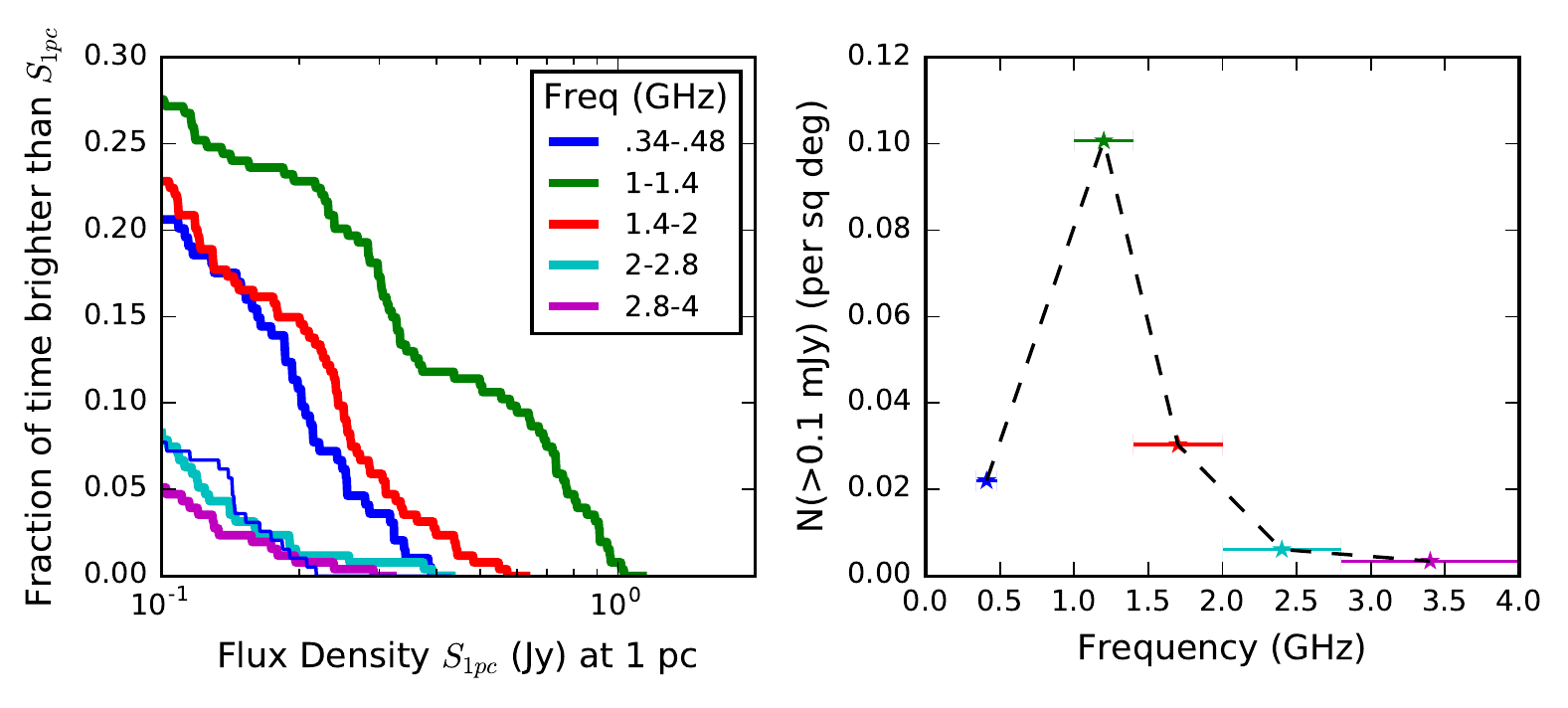}
  \caption{Frequency dependence of duty cycle and transient density of coherent bursts on active M~dwarfs. \textit{(left)} Fraction of time that a highly active M dwarf, viewed from 1~pc, will be brighter than a given flux density in the listed bands.  The thin lines near the bottom show the false alarm rate for each band due to noise, which is near zero above 1.4~GHz.  \textit{(right)} The expected instantaneous density of transients with flux density greater than 0.1~mJy due to stellar coherent radio bursts, as a function of frequency. The width of the horizontal bars shows the frequency range of each band used to calculate transient density. We estimate a factor of 2 systematic uncertainty in transient density (not shown), affecting all frequency bands the same way, due to uncertainty in the volume density of radio-active M~dwarfs. \label{fig:halfband_burstrate}}
\end{figure*}

From the time series, we calculate the burst duty cycle, the fraction of time that stars spend bursting above a certain luminosity. We perform this calculation for the time series of the stellar emission, and for the control time series to estimate a false alarm rate due to time-variable contamination by sidelobes, RFI, and thermal noise.  The left panel of Figure~\ref{fig:halfband_burstrate} shows the burst duty cycle and false alarm rate for bursts with $S_\textrm{1pc}>100$~mJy, corresponding to a luminosity of $1.2\times10^{14}$~erg\,s$^{-1}$\,Hz$^{-1}$ (assuming isotropic radiation). We restrict this calculation to a minimum 1-pc flux density of 100~mJy to keep the P~band false alarm rate low; this restriction does cause some fainter coherent bursts to be neglected, especially in S~band, but we found that using lower threshold values did not have a major impact on predicted transient density, implying that the most luminous events dominate the predicted transient population.

We find that the coherent burst duty cycle is highest in the lower half of L~band, 1-1.4~GHz, with the target stars producing coherent bursts with $S_\textrm{1pc}>$100~mJy ($1.2\times10^{14}$~erg\,s$^{-1}$\,Hz$^{-1}$) more than 25\% of the time observed.  Referring to Table~\ref{table:bursts}, we find that the duty cycle is highest in this frequency range because bursts are more common, more luminous, and longer duration compared to other bands. This implies that coherent bursts on active M~dwarfs will be most significant as a transient source for L~band surveys (at least compared to P, S, and C~bands, the other bands considered in this calculation).

\subsubsection{Volume Density of Radio-Active M Dwarfs}

To predict transient areal density, we need to know the volume density of active M~dwarfs, or more specifically, those responsible for the majority of M~dwarf radio transients. To do so, we make a few assumptions:

\begin{enumerate}

\item \textit{The volume density of stellar systems like those in our sample is constant with distance.} The approximation of constant spatial density will break down for estimating rates of very faint events, because the density of active M dwarfs declines with height above the galactic plane, due to a decline in overall stellar density and an increase in average stellar age. Increased stellar age means lower activity: the fraction of M4 dwarfs with magnetic activity declines by about a factor of 2 at a height of 100 pc above the Galactic plane \citep{west2008}.  Since our brightest events have $S_\textrm{1pc}\sim$1~Jy, and the assumption of constant volume density of active mid M dwarfs is good to within a factor of two at 100~pc, we expect that this assumption is not the dominant source of error on our estimates of transient density for flux densities above 0.1~mJy.

The assumption of constant source density as a function of distance means that our predicted transient densities  $N(>S)$, as a function of flux density $S$, will follow a Euclidean relationship of ${N(>S)\propto S^{\,-3/2}}$.

\item \textit{Our sample is representative of the active M~dwarf stellar systems responsible for most radio transients.} Our sample consists of single or wide-binary mid-M~dwarfs with strong magnetic activity, selected in part based on their previously-known radio variability. Active mid-M dwarfs are more prolific sources of radio bursts compared to active early-M~dwarfs \citep[e.g.,][]{abadasimon1997} and are also a more common type of star than earlier-type active M~dwarfs, so stars like the ones in our sample should dominate the transient population originating from active M~dwarfs. However, our burst rate may be skewed by the fact that the majority of our survey time is spent on only two stars, AD~Leo and UV~Cet, in only two years; this limits our ability to account for the variation in activity properties over time and between stars.

\item \textit{The volume density of M~dwarfs producing radio transients can be estimated from the number of known nearby radio-bright M~dwarfs.} One of the most uncertain parts of estimating radio transient rates due to active M~dwarfs is the source volume density. \cite{henry2018} estimate there are $\sim$300 stellar systems within 10~pc, in which 75\% of stars are M~dwarfs \footnote{\url{http://www.recons.org/census.posted.htm}, accessed on 2018 Sep 14.}; $\sim$30\% of mid-M~dwarfs show magnetic activity in H$\alpha$ \citep[with a strong dependence on spectral type, 15-50\% for M4 to M6;][]{west2011}, implying a volume density of stellar systems containing magnetically active M~dwarfs of $\sim$0.016~pc$^{-3}$.

However, the five systems targeted in this survey were chosen for their track record of strong magnetic activity in radio, a more stringent requirement than H$\alpha$ emission. For purposes of comparison, we calculate a lower limit for the volume density of highly radio-active stellar systems based only on our sample:
\begin{eqnarray}
n_\star &=& \frac{\textrm{number of star systems in our sample}}{\textrm{volume of our sample}} \nonumber \\
&=& \frac{5}{(4/3) \pi (6.2\textrm{ pc})^3} = 0.005\textrm{~pc}^{-3}.
\end{eqnarray}
There are a handful of other nearby, fairly well-known M-dwarf radio emitters that we have not observed, such as Proxima~Cen \citep{lim1996proxcen,slee2003}, CN~Leo and Wolf~630 \citep[detected as radio stars in e.g.,][]{gudel1993}.

For a third estimate counting more radio-active stars beyond those in our sample, we turn to \cite{bower2009}, who conducted a 5-GHz survey of 104 M~dwarfs within 10~pc, detecting 29 at levels of 130~$\mu$Jy or more. This is roughly 45\% of the total census of M~dwarf stellar systems from \cite{henry2018}, but many of the Bower stars were X-ray-selected so that the sample likely contains more than 45\% of the active M dwarfs within 10~pc.  Assuming a total of 30-65 (65=30/0.45) radio-active M~dwarfs within 10~pc yields a source density of 0.007-0.016~pc$^{-3}$.

Our three estimates of source density range from 0.005-0.016~pc$^{-3}$. For the following calculations we use a source density of 0.01~pc$^{-3}$, with an estimated factor of 2 uncertainty in transient rate caused by uncertainty in source density.
\end{enumerate}

\begin{figure*}[bth!]
  \includegraphics[width=\textwidth]{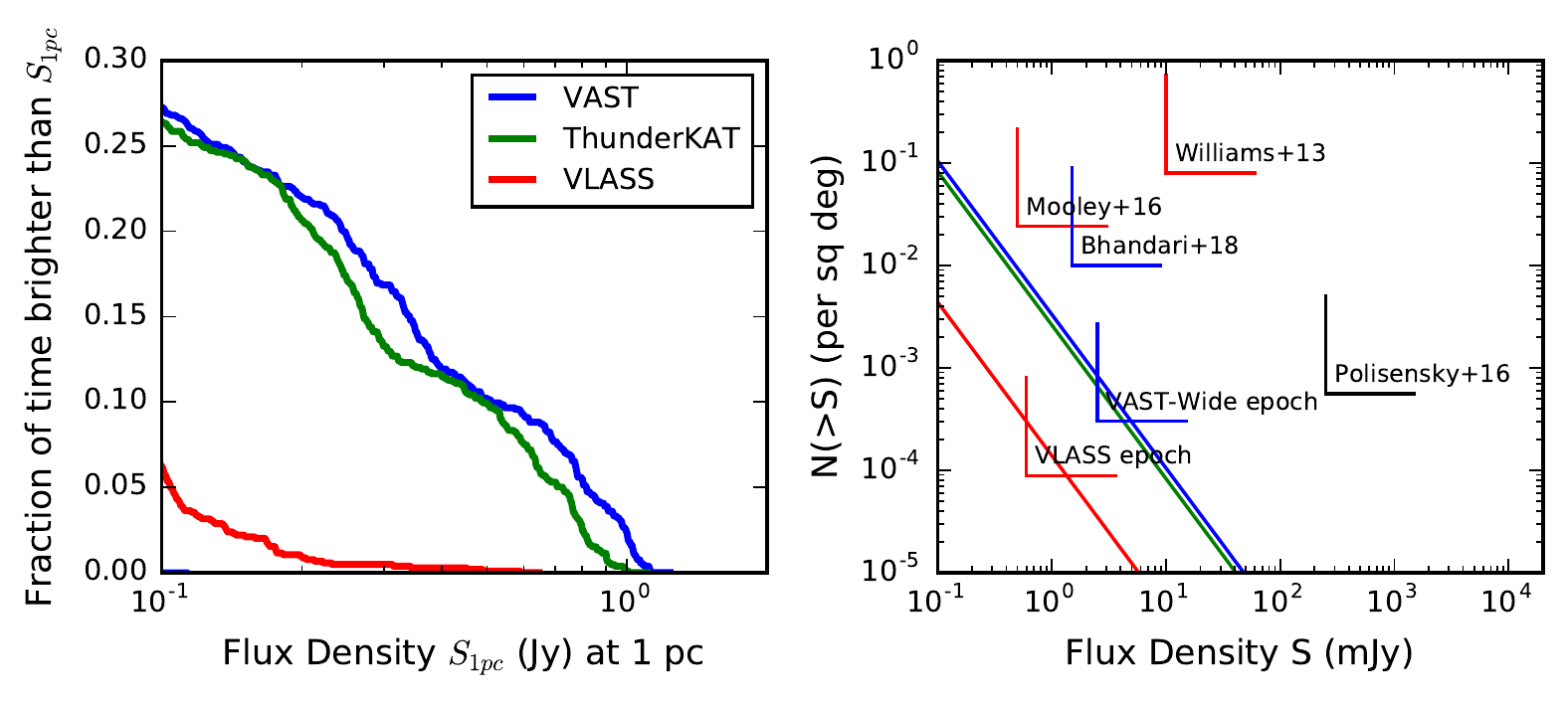}
\caption{Predicted rate of stellar coherent radio bursts in transient surveys. \textit{(left)} Coherent burst duty cycle, in the style of Figure~\ref{fig:halfband_burstrate}, with the exception that the duty cycle is calculated using time series with a shorter integration time of 150~sec. \textit{(right)} Expected instantaneous transient areal density due to stellar coherent radio bursts (line colors same as left panel). We assume a constant source density as a function of distance, so the predicted transient rates follow a Euclidean relationship of $N(>S)\propto S^{-3/2}$. The wedges show a representative flux density detection threshold and measured/predicted 95\%-confidence upper limit on transient density for completed/ongoing surveys, colorized by frequency band (black: 0.34~GHz; blue: 1-2~GHz; red: 2-4~GHz): \cite{williams2013,mooley2016,polisensky2016,bhandari2018}; and single epochs of VAST-Wide and VLASS.\label{fig:survey_burstrate}}
\end{figure*}

\subsubsection{Transient Density}

We combine the duty cycle of coherent bursts as a function of luminosity (left panel of Figure~\ref{fig:halfband_burstrate}) with the estimated volume density of active M~dwarfs to predict a transient density due to active M~dwarfs.  The transient density is the area density (number per square degree on the sky) of sources producing coherent bursts above a certain flux density at an instant in time.  The right panel of Figure~\ref{fig:halfband_burstrate} shows, for each sub-band, the expected instantaneous transient density $N(>S)$ above $S=0.1$~mJy due to active M~dwarfs (which can reasonably be extrapolated to higher flux density thresholds up to $\sim$10~mJy using $N(>S)\propto S^{-3/2}$). As expected based on the high duty cycle of coherent bursts at 1-1.4~GHz, the transient density peaks at that frequency, with a density of one >100-$\mu$Jy transient per 10 square degrees. The transient density declines at lower and higher frequencies, dropping to only one >100-$\mu$Jy transient per 300 square degrees for 2.8-4~GHz.  The rate of stellar flares expected for a radio transient survey is therefore highly dependent on the survey's frequency band.

Since the majority of our survey's time focused on two stars, AD~Leo and UV~Cet, this introduces uncertainty in our predictions of transient density versus radio frequency from the entire population of active M~dwarfs. The coronae of M~dwarfs with strong magnetic activity have fairly similar coronal densities (Table~\ref{table:targets}) and magnetic field strengths \citep{morin2008,reiners2009saturation}, due to coronal saturation \citep[e.g.,][]{wright2011,jeffries2011}, implying similar coronal cyclotron and plasma frequencies. However, coronal densities and magnetic field strengths are inhomogeneous, so bursts may vary in frequency depending where in the corona they occur in different epochs and on different stars, potentially leading to a broadening of the frequency dependence of M~dwarf coherent bursts shown in Figure~\ref{fig:halfband_burstrate}.


\subsection{Application to Current Transient Surveys \label{section:transient_surveys}}

We next calculated the coherent burst duty cycle and expected transient density for the frequency bands of three ongoing or upcoming transient surveys (Figure~\ref{fig:survey_burstrate}): VAST \citep[1.13-1.43~GHz;][]{murphy2013}, ThunderKAT L-band (0.9-1.67~GHz; but we used 1-1.67~GHz, the part of the band that overlaps with our frequency coverage), and VLASS (2-4~GHz).  We generated these predictions following the same procedure described in the previous section, but using a shorter integration time of 150~sec, enabled by the better sensitivity of these bands compared to the 340-480~MHz band considered in the previous section.

For a given flux density threshold, a survey's frequency coverage can have a dramatic effect on instantaneous transient density: for a minimum flux density of 0.1~mJy, we predict a transient density of roughly one per 10~deg$^2$ for the VAST and ThunderKAT frequency bands, and one per 250~deg$^2$ for the VLASS frequency band. The L~band surveys, VAST and ThunderKAT, will detect a higher density of stellar coherent radio bursts above a given flux density than the S~band survey VLASS because, based on the targets and epochs in this paper, the duty cycle of luminous coherent bursts peaks in L~band.

For comparison, \cite{mooley2016} use their observations and the literature to predict a $\sim$1-6~GHz flare star transient density of one per 50~deg$^2$ at >0.3~mJy, or one per 10~deg$^2$ at >0.1~mJy, a rate which is higher than their predictions for any other type of galactic transient. Their rate, which encompasses all types of flare stars including active binaries, is comparable to our transient density predictions for the L~band surveys. Thus, we expect that coherent radio bursts from active M~dwarfs will be among the dominant sources of galactic transients in the 1-2~GHz band, but not at higher frequencies, due to their intrinsically narrowband emission process.

The total number of transients detected by these surveys will depend on their single-epoch detection threshold, the area covered per epoch, and the number of epochs. This information is available for VAST \citep{murphy2013} and VLASS, so we predict their transient yield due to active M~dwarfs.  The VAST survey has multiple sub-surveys with different strategies. Here we consider VAST-Wide, which will cover 10,000~deg$^2$ once per day for 2 years with a per-epoch 5$\sigma$ detection threshold of 2.5~mJy.  At this sensitivity, we expect a transient density at 1.13-1.43~GHz of 1 per 1100 deg$^2$, so VAST-Wide should detect $\sim$9 coherent radio bursts from active M~dwarfs per day.  The brightest coherent bursts in our survey ($S_\textrm{1~pc}=1$~Jy) can be detected with a flux density threshold of 2.5~mJy out to a distance of 20~pc.  Using our estimated source density of 0.01~pc$^{-3}$, we can expect about 340 active M~dwarf systems to be responsible for these bursts. Assuming a 5$\sigma$ requirement for VLASS as well gives a per-epoch detection threshold of 0.6~mJy; VLASS covers 34,000~deg$^2$ in each of 3 epochs.  We therefore expect coherent radio bursts on active M~dwarfs to be responsible for $\sim$10 transients per VLASS epoch. We note again that these predictions are biased by the small number of stars in our sample, but we know of no published surveys of transients or radio stars at these low frequencies that have sufficient sensitivity to overcome the bias introduced by small number statistics.

The right panel of Figure~\ref{fig:survey_burstrate} shows our predicted transient areal density as a function of frequency for the frequency bands of these three surveys. The figure also shows 95\% confidence upper limits from a selection of published surveys at frequencies from 0.3-4~GHz \citep{williams2013,mooley2016,polisensky2016,bhandari2018}, and predicted upper limits for VAST-Wide and VLASS single epochs, \textit{assuming no detections were made}. Another relevant transient search, \cite{thyagarajan2011}, used data from the VLA FIRST survey at 1.4 GHz to search for transient/variable sources, detecting 5 radio-variable stars, but we do not plot that detection since they do not distinguish between types of stars. Figure~\ref{fig:survey_burstrate} demonstrates that VLASS and VAST-Wide are among the first radio surveys expected to systematically detect stellar coherent radio bursts as a transient population.

Transient surveys will provide an unbiased sample of the population of stellar coherent bursts.  Detection of a single event offers the potential to trace source motion in a stellar corona, if the survey has the capability to extract a dynamic spectrum once stellar events are identified; identification and interpretation of such an event is strengthened by the capability to measure circular polarization. The scientific return of such detections can be maximized by rapid follow-up observations, including VLBI (to search for extended structures in the post-flare radio corona) and lower-frequency radio observations (to search for a low-frequency continuation of space weather events and for planetary radio emission triggered by these events). When surveys reach enough coverage to detect a population of events, rather than 1-2 single events, they will reveal whether the stars in this paper are ``oddballs'' or if their behavior is typical for active mid-M~dwarfs. Detection of a population of events will also make it possible to compare the coherent emission properties of different spectral types, illuminating the question of what types of star can sustain the large-scale magnetospheric current systems responsible for long-duration coherent bursts.  Such detections will identify new targets and categories of star for future targeted stellar observations, guiding the search for extrasolar space weather events.

\section{Conclusions}

We have conducted a 58-hour survey of 5 nearby active M dwarfs with the VLA, in order to characterize the frequent coherent radio bursts on these stars and search for events caused by energetic CMEs with radio-loud shock fronts. This survey includes observations from 2013 with continuous coverage of 1-6~GHz, and from 2015 with coverage of 0.22-0.48 and 1-4~GHz.  This wide bandwidth, enabled by the VLA upgrade, greatly extends the maximum fractional bandwidth of stellar dynamic spectroscopy, opening the possibility of tracking coronal sources across regions varying by 1-2 orders of magnitude in density or magnetic field strength.

We detect frequent coherent bursts in all bands observed, with a total of 22 events occurring across 13 of the 23 epochs.  The events can be very bright, with 9 of the epochs having a peak flux density above 50~mJy, enabling detailed characterization in the time-frequency plane. The properties of the bursts (duration, frequency range, and morphology in the time-frequency plane) are very diverse, reflective of the wide range of dynamic processes occurring in stellar coronae. The bursts show uniformity only in their high degree of circular polarization, in most cases 40-100\%, consistent with the preponderance of active M~dwarf coherent bursts in the literature but in contrast to the weak to moderate polarization observed in many classes of solar coherent bursts. This strong circular polarization requires the emission to either be cyclotron maser emission or fundamental plasma emission, and indicates that the depolarization mechanisms operating in the Sun do not strongly affect the coronae of active M~dwarfs. We divide the bursts into two classes based on duration: short-duration (seconds to minutes) and long-duration (30~min or more).

The timescale of the short-duration events is consistent with the duration of a ``space weather'' event, such as the impulsive phase of a flare or a CME crossing the corona. We do not detect any clear analogs to solar Type~II radio bursts (associated with CME shock fronts), which may be due to: insufficient sensitivity, a CME rate much lower than expected based on the stellar flare rate, or CMEs failing to form radio-loud shocks due to the high Alfv\'en speed in active M~dwarf coronae. In spite of the lack of Type~II-like bursts, frequency drift (which can be caused by moving coronal sources) occurs in many bursts on a variety of timescales. In an exceptional event on YZ~CMi, vertical spikes (if due to unresolved frequency drift) are consistent with source motion at 1.5-15\% of the speed of light, with morphological resemblance to solar bursts caused by beams of high-speed electrons. The sense of polarization varies with frequency in the YZ~CMi event, and in general varies between short-duration events on a given star, pointing to emission from regions with different magnetic polarity, or emission in both x-mode and o-mode.  If due to source regions with different magnetic polarity, this indicates that the short-duration bursts originate from local phenomena in regions dominated by the small-scale magnetic field (such as flaring in an active region) rather than from a global magnetospheric process. The short-duration events do not have clear analogs in the solar classification system, preventing immediate identification of the underlying coronal processes and their potential impact on space weather. To identify these phenomena will require multi-wavelength data, combined with the diagnostic power of wide-bandwidth radio dynamic spectroscopy. Even without a full understanding of the underlying processes, it is significant that we do not detect any short-duration bursts that continue beyond our lowest observed frequency. This means that we find no evidence of coronal disturbances propagating freely away from active M~dwarfs to distances where they could impact (hypothetical) planets.  Observations with deeper sensitivity and/or at lower frequencies can refute or strengthen this conclusion.

The long-duration bursts, which last hours and perhaps weeks to months between epochs, require an ongoing electron acceleration mechanism during the burst.  Candidate acceleration processes are found within the paradigm of solar radio bursts and within the paradigm of periodic radio aurorae produced by ultracool dwarfs and planets.  Some long-duration events on AD~Leo and YZ~CMi show narrow bandwidth with a sharp lower frequency cutoff, which may be caused by emission from a closed magnetic structure, analogous to solar noise storms.  However, complex patterns of frequency drift repeated between epochs on UV~Cet point to a periodic radio aurora, where the frequency drift is caused by rotational modulation of angularly beamed emission.  Angular beaming may also explain the frequency cutoffs and slow frequency drift observed on AD~Leo.  The duration and occurrence rate of long-duration bursts peak at 1-1.4~GHz compared to both lower and higher frequencies, in contrast to solar radio bursts which become more common below 1~GHz, indicating that the majority of observed stellar long-duration coherent emission originates in the low corona.  The sense of polarization of long-duration bursts above 1~GHz is consistent on a given star, both in our survey and in the literature from 1977-2015 (albeit with sparse coverage), indicating that these stars have not undergone magnetic reversals on a timescale of decades.  Comparison of the long-duration radio burst polarization with the orientation of the large-scale magnetic field (measured by optical spectropolarimetry) reveals that long-duration bursts are all in the x-mode relative to the large-scale field.  Due to their strong x-mode polarization, we favor the electron cyclotron maser as the emission mechanism for long-duration coherent bursts on active M~dwarfs.  The dependence of polarization on the large-scale field either indicates that the long-duration bursts originate from a population of accelerated electrons in the large-scale magnetosphere, analogous to auroral processes on very low mass objects, or it can be caused by weak mode coupling.  Weak mode coupling would also help explain the consistent high degree of circular polarization of active M~dwarf radio bursts compared to solar radio bursts, but it faces a challenge in the varying sense of polarization of the short-duration bursts, so we favor instead the interpretation that the long-duration bursts are caused by a global magnetospheric process.

We find that the coherent burst duty cycle is highest in the lower half of L~band, 1-1.4~GHz, with the target stars producing coherent bursts with $S_\textrm{1pc}>$100~mJy ($1.2\times10^{14}$~erg\,s$^{-1}$\,Hz$^{-1}$) more than 25\% of the time observed. 

Long-duration coherent bursts will dominate the population of GHz-frequency transients from active M~dwarfs. Due to the intrinsically narrowband nature of coherent emission, the predicted transient density due to coherent bursts depends strongly on frequency, peaking at 1-1.4~GHz. For this band, the duty cycle of coherent bursts more luminous than $1.2\times10^{14}$~erg\,s$^{-1}$\,Hz$^{-1}$ is just over 25\%. We predict a 1-1.4~GHz transient density of one per 10 deg$^2$, making active M~dwarfs the most prolific expected source of galactic transients at these frequencies. We use our wideband data to predict transient densities for the specific frequency bands of upcoming or ongoing transient surveys, ASKAP's VAST, MeerKAT's ThunderKAT, and the VLA Sky Survey. VAST and ASKAP's observations in the 1-2~GHz band should find a high rate of stellar radio bursts from active M dwarfs; we predict an active M~dwarf transient density for these surveys more than an order of magnitude higher than for VLASS.  Such surveys will identify new targets for study of the global magnetospheric processes probed by long-duration bursts.

\acknowledgments
The authors thank Stephen Bourke for writing the initial version of the \textit{tbavg} code for producing dynamic spectra in CASA and for providing guidance on background source subtraction, and Michael Eastwood for testing direction-dependent calibration of our P~band data. J.R.V. thanks Tim Bastian for discussion of stellar radio bursts and feedback on this paper, Urvashi Rao for imaging advice, and Rick Perley and Frank Schinzel for help with P~band polarization calibration. The authors thank the referee for thoughtful and constructive feedback.

This material is based upon work supported by the National Science Foundation under Grant No.~AST-1311098. J.R.V. additionally thanks the Troesh family and the PEO International Scholar Award program for financial support of her graduate research.

The National Radio Astronomy Observatory is a facility of the National Science Foundation operated under cooperative agreement by Associated Universities, Inc. This research has made use of the SIMBAD database, operated at CDS, Strasbourg, France. This research has made us of NASA's Astrophysics Data System. This research made use of Astropy, a community-developed core Python package for Astronomy. The acknowledgements were compiled using the Astronomy Acknowledgement Generator. 

\facilities{VLA} 
\software{CASA, Astropy \citep{astropy_ref}, Anaconda, SIMBAD, ADS, NumPy \citep{van2011numpy}, SciPy \citep{Jones_scipy_2001}, Matplotlib \citep{Hunter:2007}}

\bibliographystyle{yahapj}
\bibliography{references}

\appendix

\section{Distinguishing Stellar Bursts and Artifacts in the Dynamic Spectrum\label{appendix:identification}}

\begin{figure*}[h]
  \begin{center}
     \includegraphics[width=\textwidth]{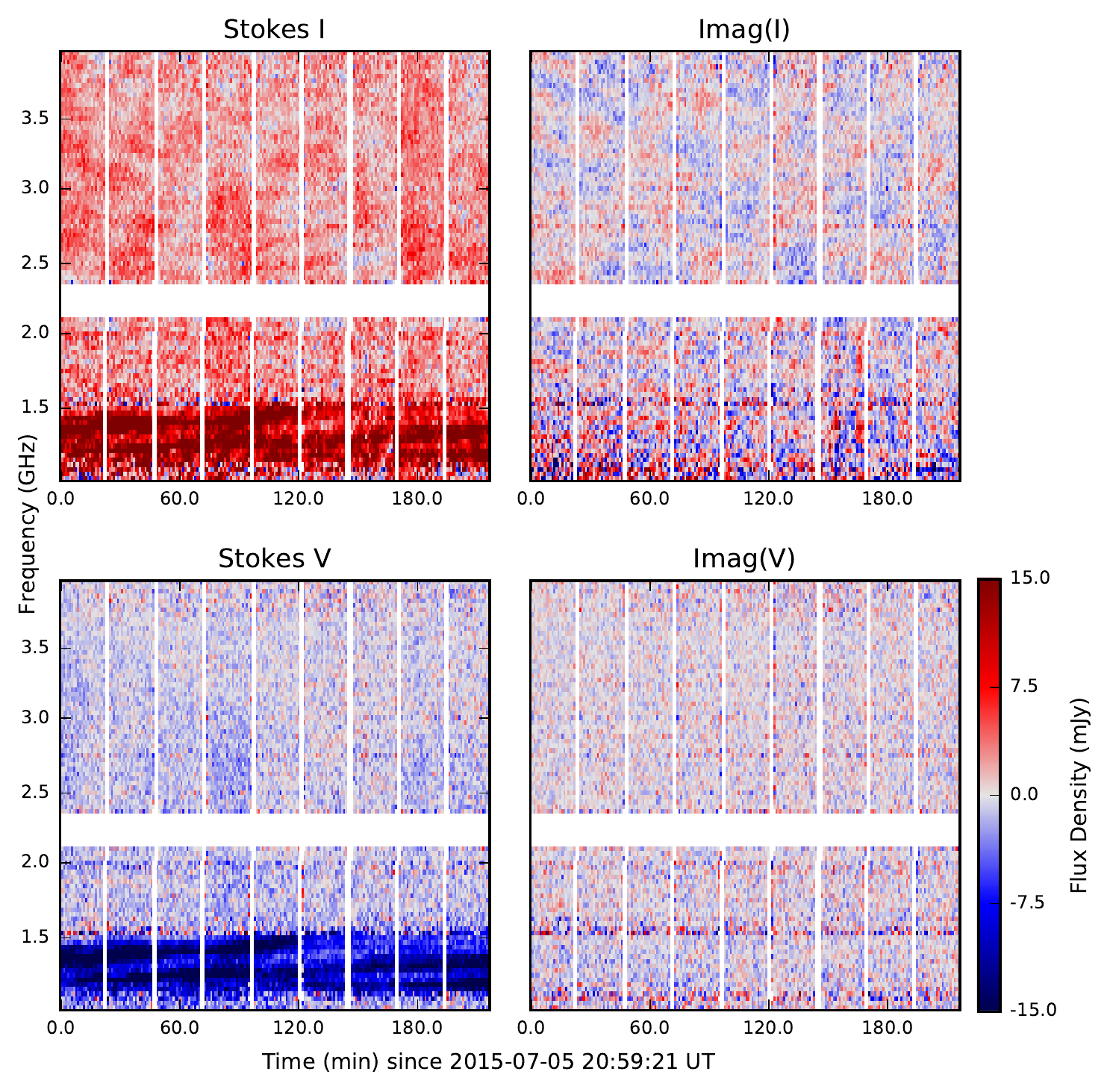}
  \end{center}
  \caption{Example of data products used to identify radio bursts. All panels are 1-4~GHz dynamic spectra of a 4-hour observation of AD~Leo, binned to a resolution of 32~MHz and 60~sec. \textit{(top left)} Stokes I dynamic spectrum of the target, generated by averaging the Stokes I visibilities over all baselines and taking the real component. All stellar Stokes I flux density should be positive, or red. \textit{(top right)} Imaginary component of the baseline-averaged Stokes I visibilities.  No signal from the target star at the phase center appears in the imaginary component, so this spectrum provides an estimate of noise levels due to thermal noise, RFI, and residuals of background sources. \textit{(bottom left)} Stokes V dynamic spectrum of the target. Right circular polarization is positive/red, and left is negative/blue. \textit{(bottom right)} Imaginary component of the baseline-averaged Stokes~V visibilities. Stokes V has lower noise levels than Stokes I, due to the absence of circularly polarized background sources.\label{fig:adleo_ds_example}}
\end{figure*}

Figure~\ref{fig:adleo_ds_example} shows an example of the different data products used to identify radio bursts and characterize noise levels in the dynamic spectrum.  The dynamic spectrum is the real component of the baseline-averaged visibilities (broken down into separate polarizations, e.g., Stokes I and V), which contains the emission from the star at the phase center.  Sources of noise and artifacts include: thermal noise, RFI, and residual sidelobes from imperfectly-subtracted background sources.  None of the sources of noise and artifacts are located at the phase center, with the result that they produce striation patterns in the dynamic spectrum (alternating between positive/red and negative/blue), and they also produce a signal in the imaginary component of the baseline-averaged visibilities. In contrast, the target star (a point source at the phase center) only produces emission in the real component of the baseline-averaged visibilities, with no contribution to the imaginary component.  In Appendix~\ref{appendix:dynspec}, we present dynamic spectra of detected bursts, showing only the real component of the baseline-averaged visibilities, but the imaginary component of the baseline-averaged visibilities has been inspected as well for each event; these imaginary-component dynamic spectra are included as figures, identical in format to those in Appendix~\ref{appendix:dynspec}, in this paper's online supplementary files. The stellar origin of a number of bursts has been confirmed by imaging.

In the absence of noise, stellar emission in the Stokes I dynamic spectrum is always positive (red), and in Stokes~V the stellar emission consists of continuous patches of right-polarized emission (positive values; a clump of red pixels) and/or left-polarized emission (negative values; a clump of blue pixels). \textbf{In Figure~\ref{fig:adleo_ds_example}, a left-polarized stellar burst is present throughout the entire duration in the Stokes I and V dynamic spectra (1st and 3rd panels from left) at 1.1-1.6~GHz.  (There is also fainter broadband left-polarized stellar emission present throughout the entire frequency range and duration.)} Horizontal white gaps indicate frequency ranges that have been entirely flagged due to RFI; vertical white gaps are time intervals with no data due to calibrator observations.  Other features in the figure provide examples of different types of noise/artifacts:
\begin{itemize}
\item \textit{Thermal noise.} The imaginary component of the baseline-averaged visibilities for Stokes~V (bottom right panel) is dominated by thermal noise at most frequencies.  Thermal noise appears in the dynamic spectrum as random variation that is not correlated across adjacent pixels, with values centered around zero (so thermal noise does not introduce a systematic offset in the average flux density).
\item \textit{RFI.} The same panel shows the effect of RFI at $\sim$1.5~GHz. The heightened intensity at this frequency is due to 2 effects: 1) RFI signals (which can produce noise that is correlated across multiple pixels), and 2) increased thermal noise levels because certain baselines with strong RFI have been flagged at these frequencies.  The fact that this feature is seen in the imaginary component of the baseline-averaged visibilities (unlike stellar emission), along with the alternation between positive and negative values, demonstrates that it is not at the phase center and not of stellar origin (which can be confirmed by imaging the affected time and frequency range). We did not completely flag all frequency ranges affected by RFI since it would remove a significant fraction of our observing band, and because certain stellar bursts are detected in these RFI-contaminated ranges (in these cases, the stellar origin of the burst has been confirmed by imaging).
\item \textit{Residual sidelobes.} Residuals after background source subtraction affect Stokes~I but do not significantly affect Stokes~V, due to the absence of circularly polarized background sources.  This can be seen in the imaginary component of the baseline-averaged Stokes~I visibilities (top right panel), as alternating positive and negative ``wiggles'' in the time-frequency plane.  These wiggles are most intense at the lowest frequencies, which have numerous bright background sources and a wide field of view. The effect of background sources in the Stokes~I dynamic spectrum (top left panel) is to introduce low level variations in the structure of the stellar emission across the dynamic spectrum.  These variations are not seen in the Stokes~V dynamic spectrum, so at low frequencies Stokes~V provides a more accurate picture of fine structure in stellar radio bursts.
\end{itemize}
To avoid interpreting artifacts as stellar bursts, the imaginary component of the baseline-averaged visibilities was inspected for each event while interpreting the dynamic spectrum.  The noise levels in each frequency channel were measured using the standard deviation across time of the imaginary component in that channel, a quantity that accounts for the effects of all 3 types of artifacts (thermal noise, RFI, and residual sidelobes). The approach of measuring noise levels across multiple pixels of the imaginary component of the baseline-averaged visibilities is used throughout this work, used as a way to assess the significance of bursts (Section~\ref{section:burst_search}), to conduct a weighted average of dynamic spectra to produce time series (Section~\ref{section:transient_rate_estimation}), and to determine the errors on peak flux density, degree of circular polarization, and energy quoted in Table~\ref{table:bursts}.

\newpage

\section{Dynamic Spectra of All Radio Bursts\label{appendix:dynspec}}

Figures~\ref{fig:2015_ADLeo_3}-\ref{fig:2015_UVCet_5} show the dynamic spectra from all epochs with detected radio bursts. All epochs are shown with the same scale in time and frequency (i.e., hours per inch and GHz per inch) but with different binning and color scales to highlight the features in each epoch.

\begin{subfigures}

\begin{figure}[h]
  \centering
  \includegraphics[angle=0]{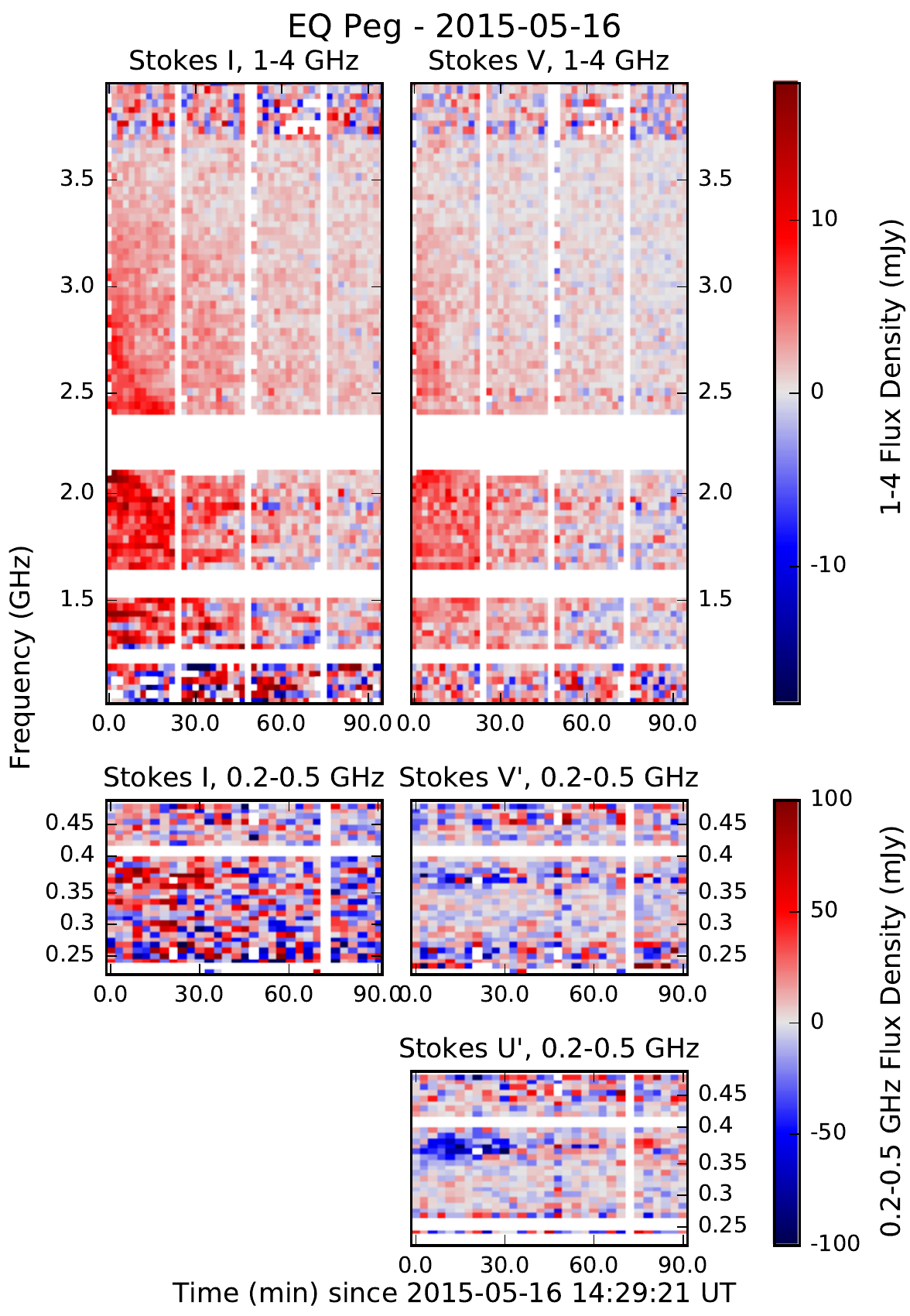}
  \caption{Dynamic spectra of the 2-hour observation of EQ~Peg on 2015 May 16. The left column shows the Stokes~I dynamic spectrum, with different color scales and binning for 1-4~GHz (top) and 0.2-0.5~GHz (bottom).  The right column shows the Stokes~V dynamic spectrum.  For 0.2-0.5~GHz, we show both Stokes V' and U', using the ' symbol to denote that cross-hand phase has not been corrected for this frequency band, such that the true Stokes V has undergone an arbitrary rotation between V' and U'.  In this epoch, we identify two coherent bursts, distinct spectral features that overlap in time: a broadband right-polarized event at 1-3 GHz for the first 30+ minutes, and a narrowband strongly polarized event at 0.35-0.385~GHz for the first 35 minutes. Since the low-frequency event occurs at RFI-contaminated frequencies, we verified its stellar origin by imaging the star during the burst, detecting it with flux density 31.3$\pm$6.4 (I), -25.9$\pm$3.9 (U'), and -12.1$\pm$4.3 (V').\label{fig:2015_EQPeg_2}}
\end{figure}

\begin{figure}[h]
  \centering
  \includegraphics[angle=0]{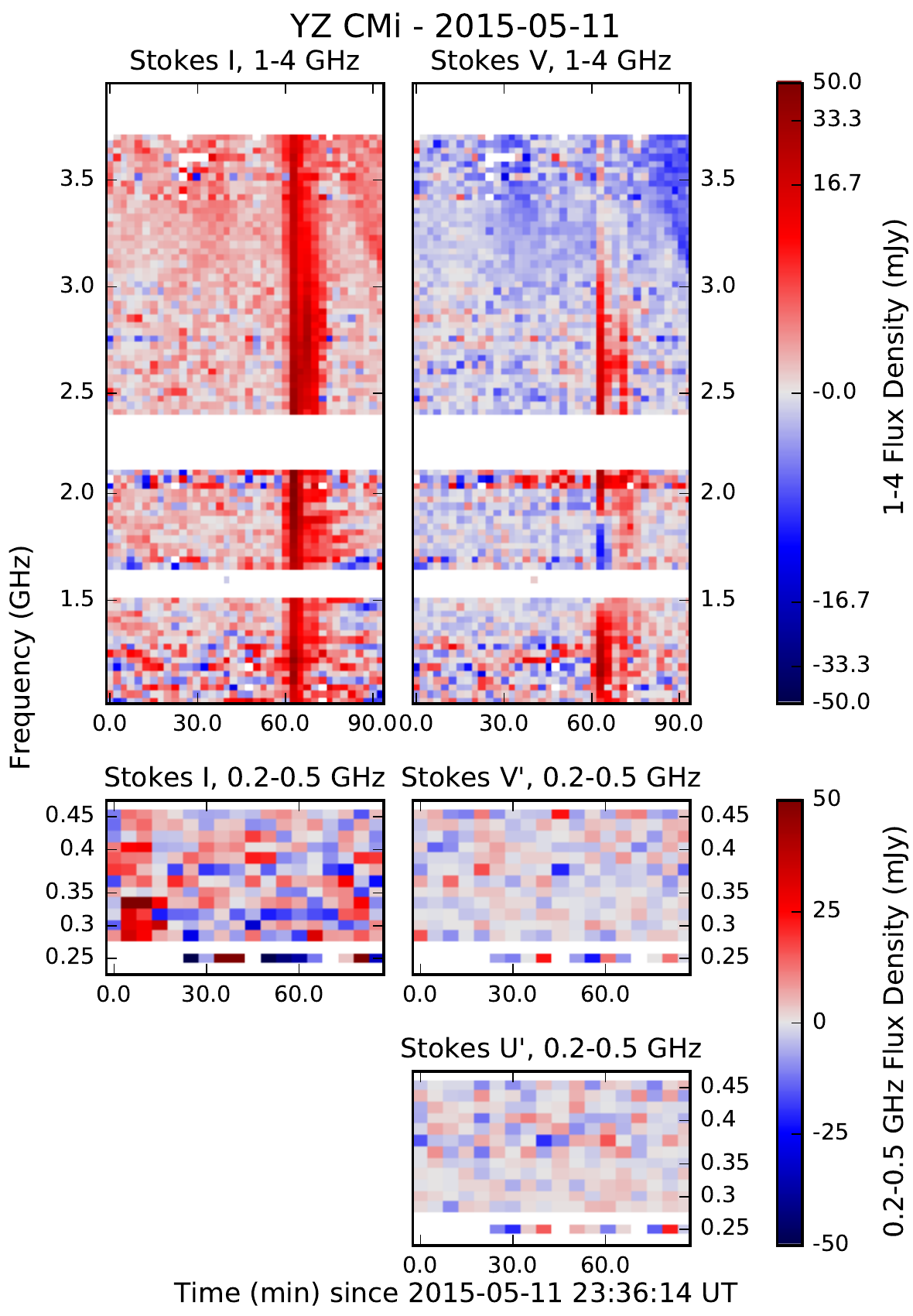}
  \caption{Dynamic spectra of the 2-hour observation of YZ~CMi on 2015 May 11, in the style of Figure~\ref{fig:2015_EQPeg_2}.  In this epoch, we identify two coherent bursts, which overlap in time but are distinct spectral features: a short-duration event with complex polarization structure at 1-3.7~GHz at t$\sim$60-70~min, and long-duration left-polarized emission throughout the observation at 3-3.7 GHz. This epoch is shown using matplotlib's ``symlog'' scale (linear near the origin, logarithmic at large flux densities) in order to highlight events at a range of intensities. The color scale and binning are chosen to make the long-duration emission visible, and the short-duration event is shown in greater detail in Figure~\ref{fig:yzcmi_burst_example}. \label{fig:2015_YZCMi_1}}
\end{figure}

\begin{figure}[h]
  \centering
  \includegraphics[angle=0]{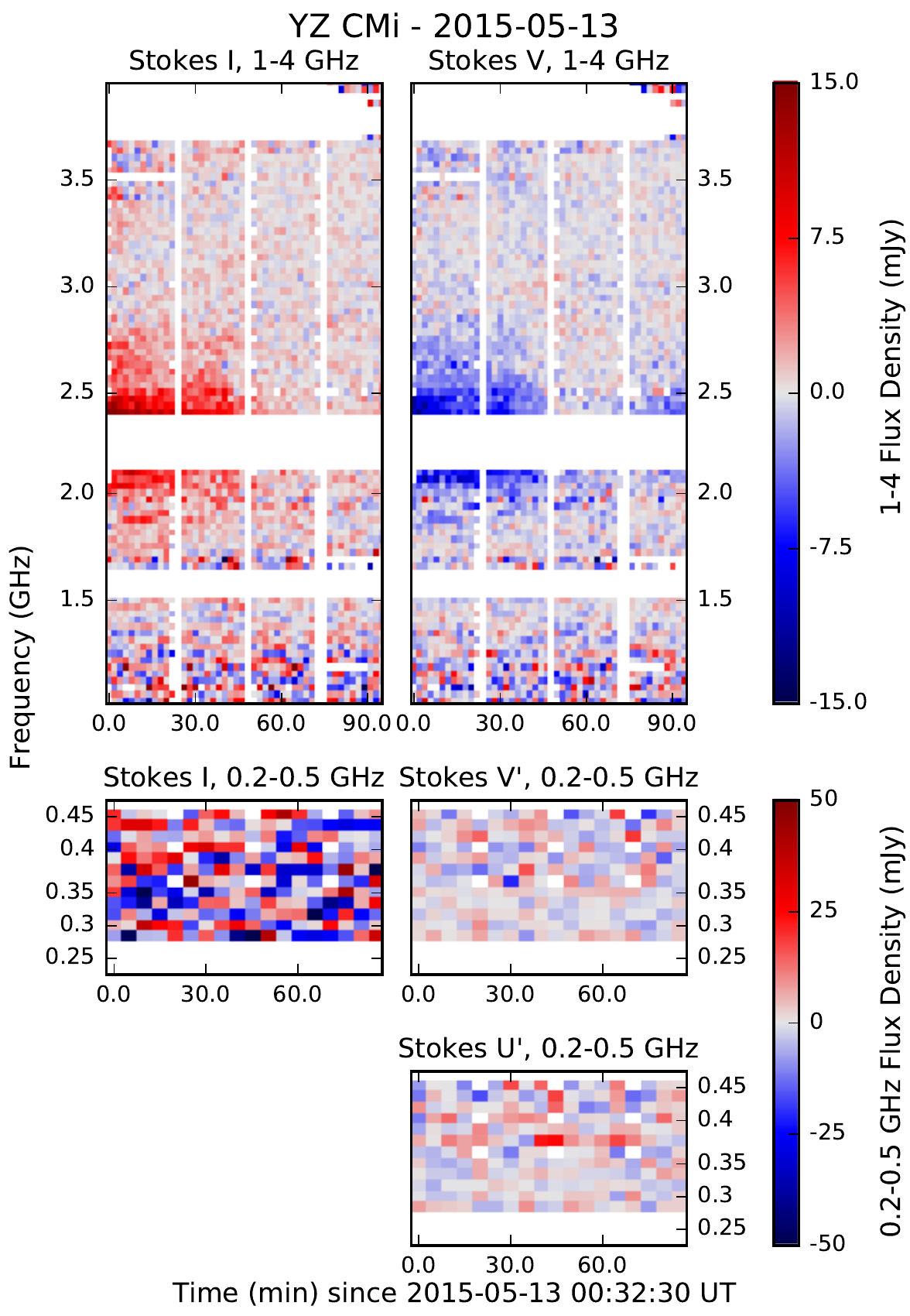}
  \caption{Dynamic spectra of the 2-hour observation of YZ~CMi on 2015 May 13, in the style of Figure~\ref{fig:2015_EQPeg_2}.  In this epoch, we identify one coherent burst, a long-duration event from 2-2.7~GHz lasting at least 45~minutes.\label{fig:2015_YZCMi_2}}
\end{figure}

\begin{figure}[h]
  \centering
  \includegraphics[angle=270]{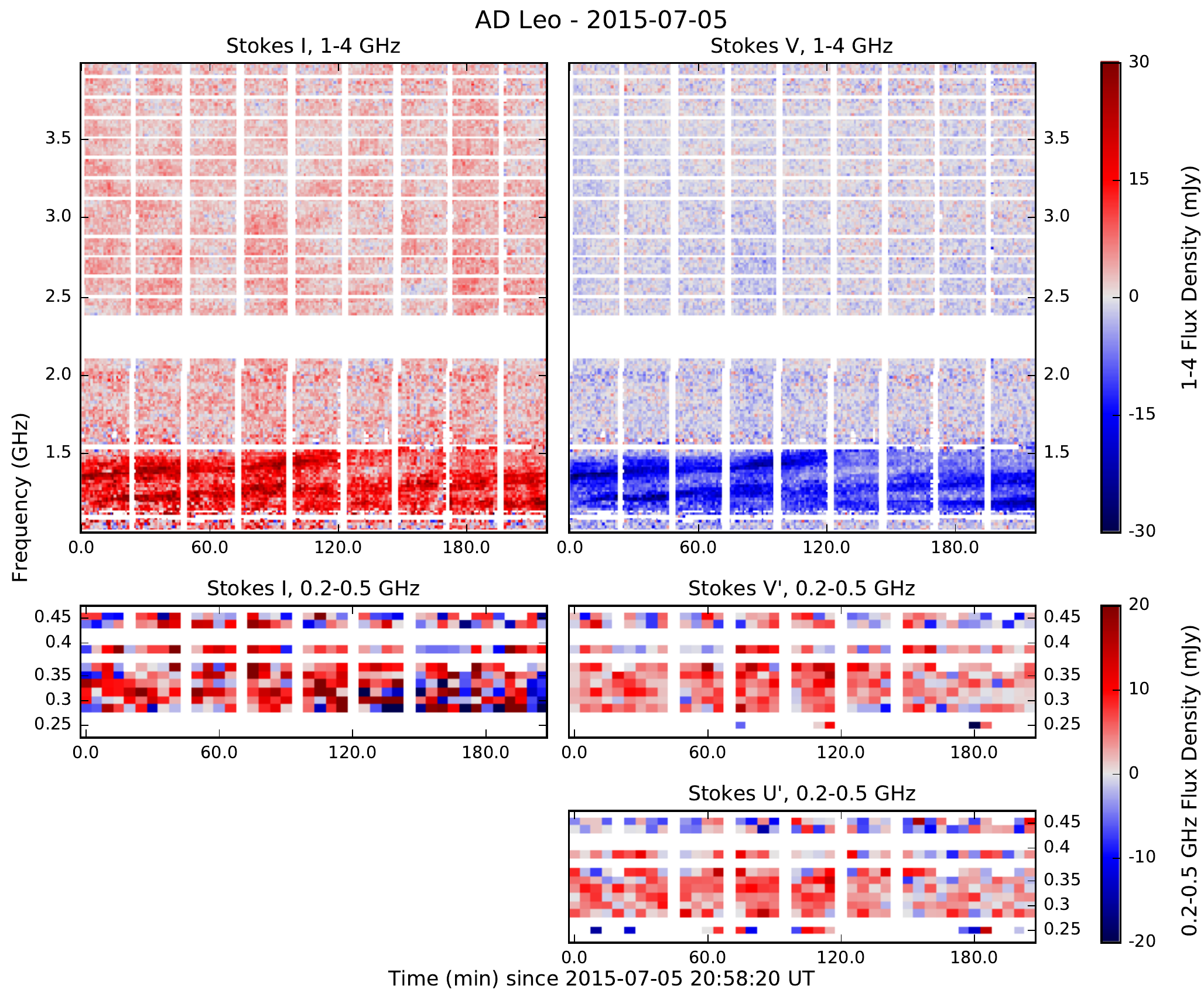}
  \caption{Dynamic spectra of the 4-hour observation of AD~Leo on 2015 July 5, in the style of Figure~\ref{fig:2015_EQPeg_2}.  In this epoch, we identify two coherent bursts: both long-duration events lasting throughout the observation, one at 1.1-1.6~GHz, and one at 0.28-0.4 GHz. While these events occur at the same time, we identify them as separate events since they are distinct spectral features and because the 1.1-1.6~GHz emission recurs without the lower-frequency component on 2015 July 19 (Figure~\ref{fig:2015_ADLeo_4}).\label{fig:2015_ADLeo_3}}
\end{figure}

\begin{figure}[h]
  \centering
  \includegraphics[angle=270]{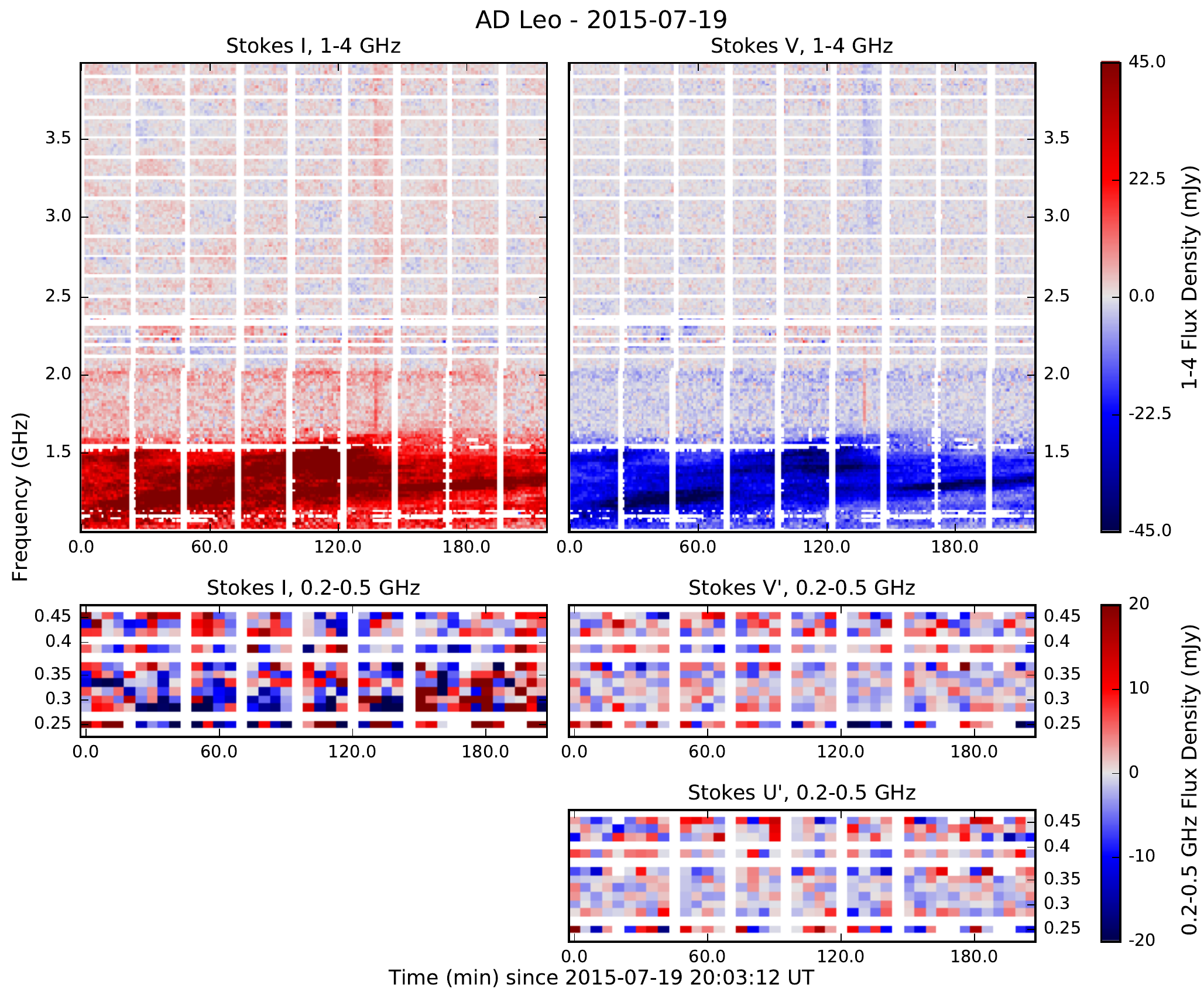}
  \caption{Dynamic spectra of the 4-hour observation of AD~Leo on 2015 July 19, in the style of Figure~\ref{fig:2015_EQPeg_2}.  In this epoch, we identify two coherent bursts: long-duration emission at 1-1.6~GHz throughout the observation, similar to the event on 2015~July~4 (Figure~\ref{fig:2015_ADLeo_3}), and two short-duration events that may be related, both at t$\sim$140~min: a 1.6-2.2~GHz right-polarized burst lasting 30~sec, and a 2.8-4~GHz burst lasting 6~min.\label{fig:2015_ADLeo_4}}
\end{figure}

\begin{figure}[h]
  \centering
  \includegraphics[angle=270]{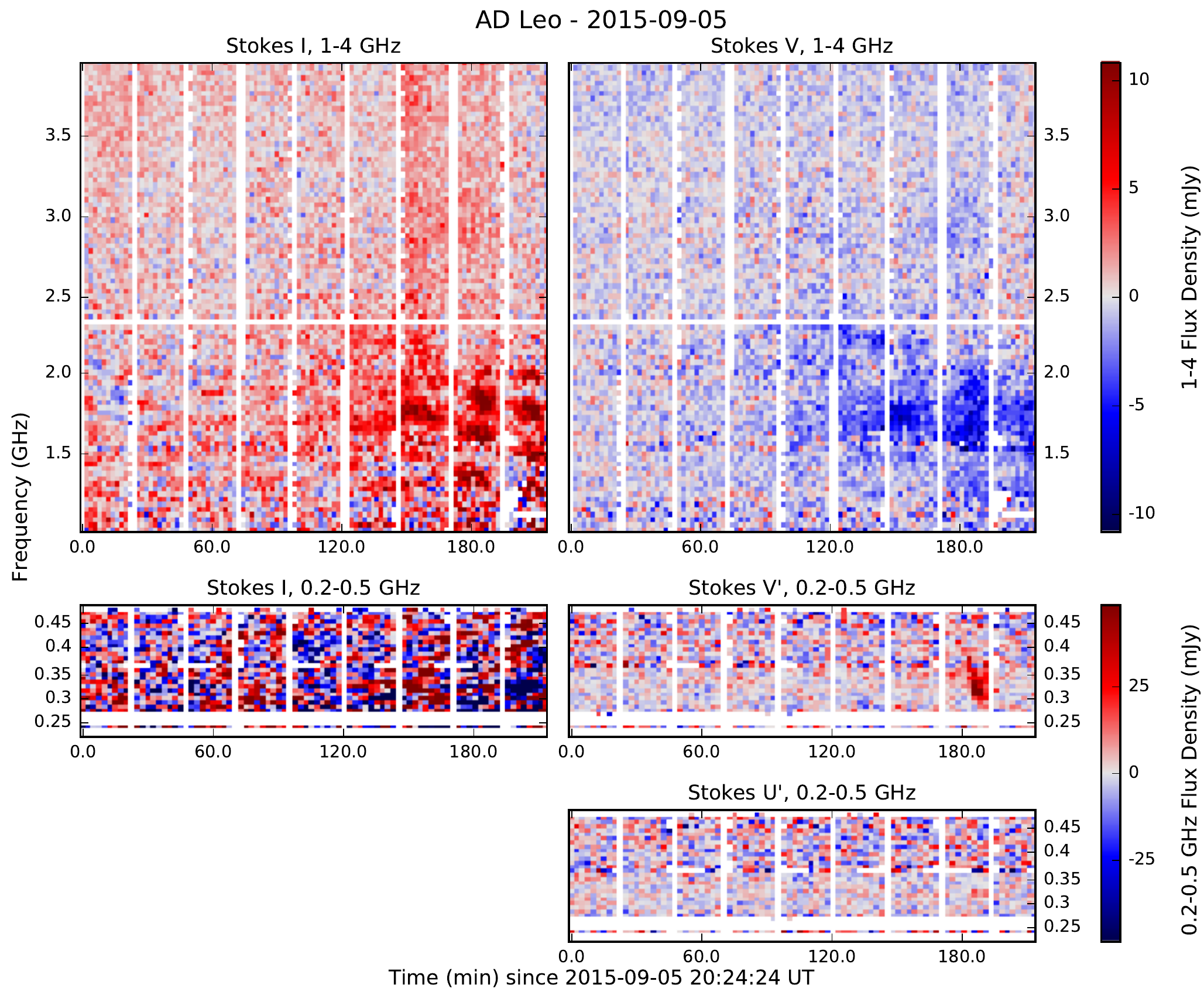}
  \caption{Dynamic spectra of the 4-hour observation of AD~Leo on 2015 September 5, in the style of Figure~\ref{fig:2015_EQPeg_2}.  In this epoch, we identify two coherent bursts: long-duration broadband emission at 1-2.5~GHz, lasting $>$1.5~hours, and a strongly-polarized short-duration event at 0.29-0.36~GHz at t$\sim$190~min.\label{fig:2015_ADLeo_5}}
\end{figure}

\begin{figure}[h]
  \centering
  \includegraphics{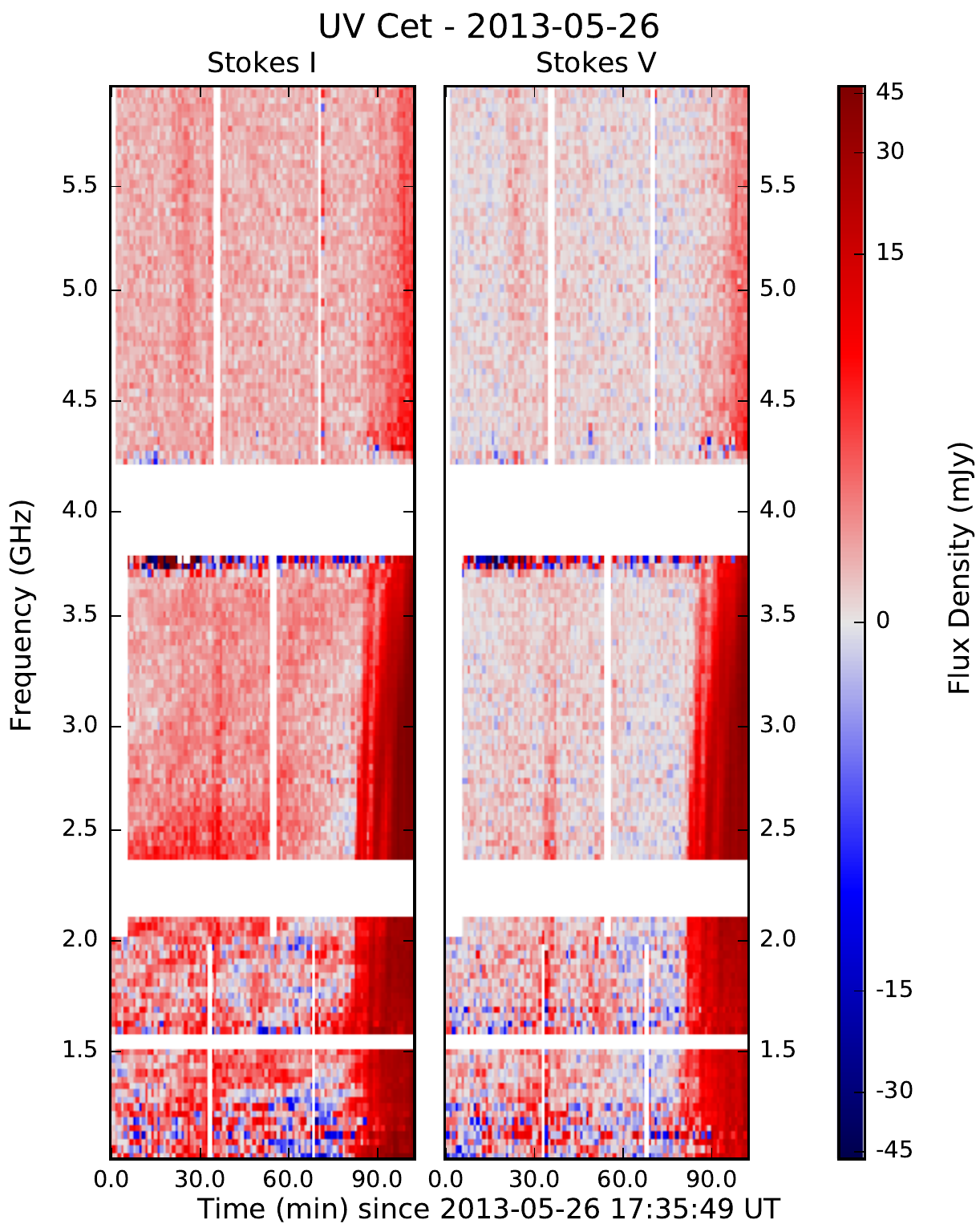}
  \caption{Dynamic spectra of the 2-hour observation of UV~Cet on 2013 May 26, in the style of Figure~\ref{fig:2015_EQPeg_2}, but covering 1-6~GHz.  This epoch is shown using matplotlib's ``symlog'' scale (linear near the origin, logarithmic at large flux densities) in order to highlight events at a range of intensities.  This epoch contained a short-duration event (two right-polarized sub-features at t$\sim$30 min, together spanning at least 1-6~GHz), and a long-duration event (1-6 GHz, last 25 minutes of observation).  \label{fig:2013_UVCet_1}}
\end{figure}

\begin{figure}[h]
  \centering
  \includegraphics[angle=270,trim={0 0 0.6cm 0},clip]{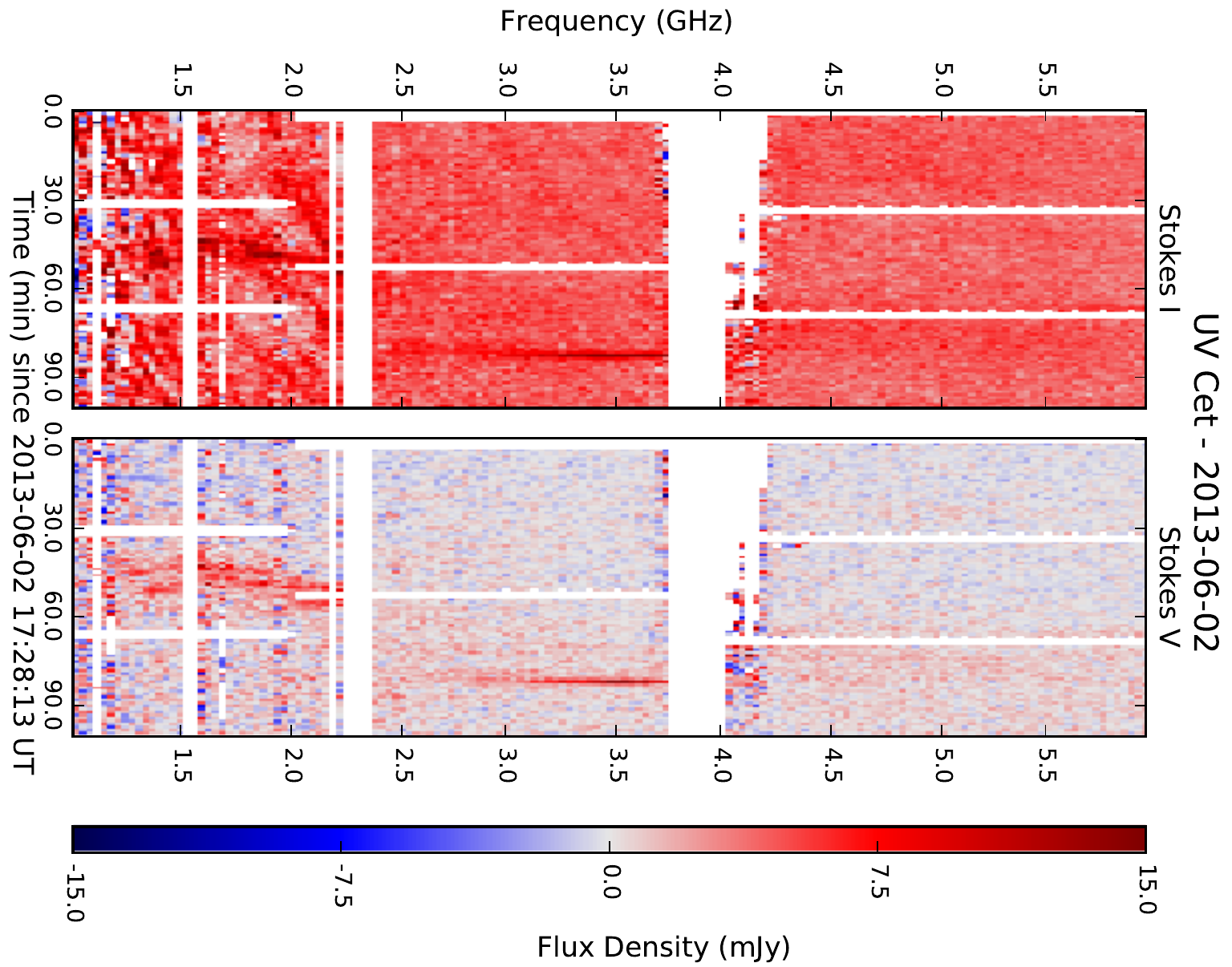}
  \includegraphics[angle=270,trim={0.5cm 0 0 0},clip]{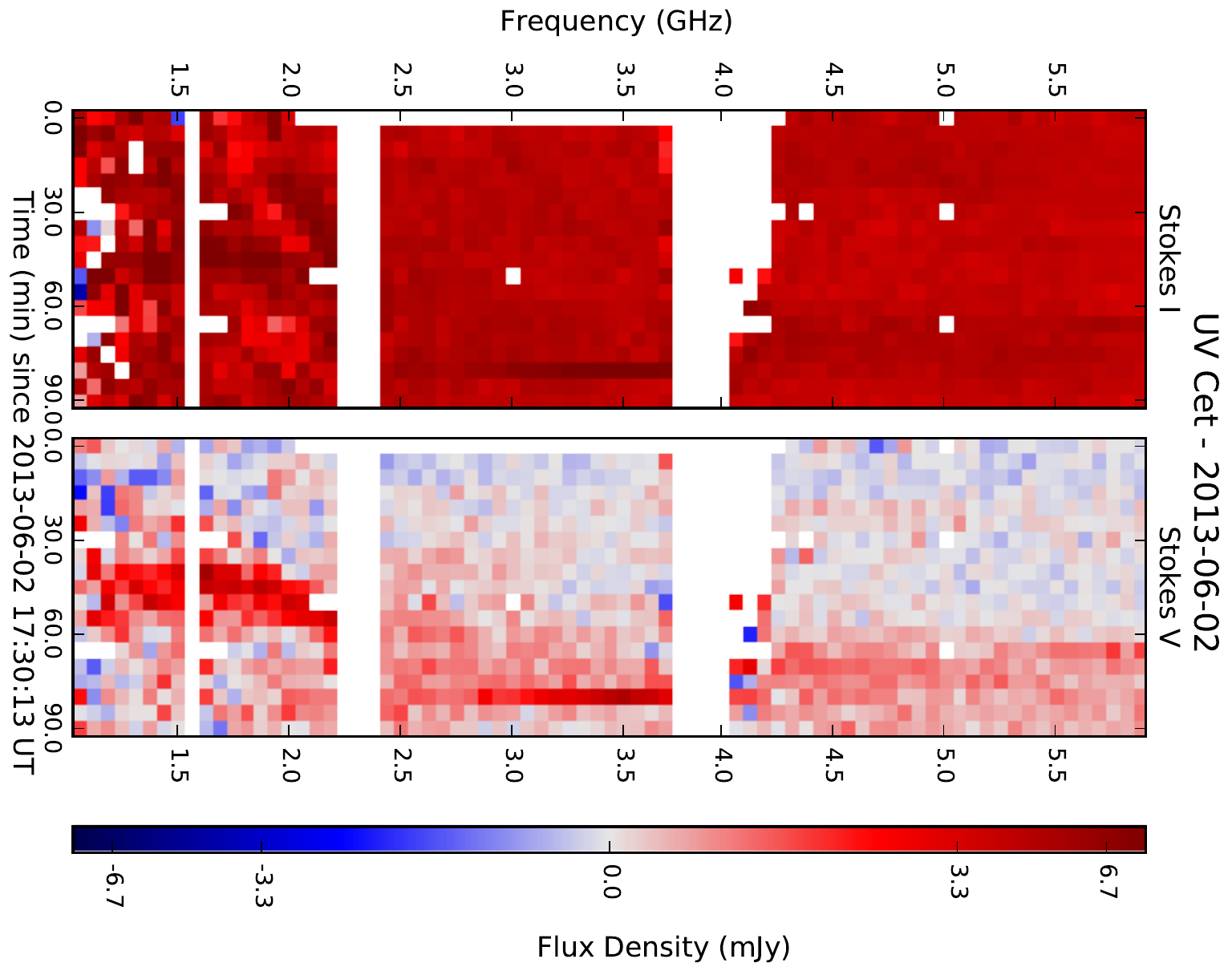}
  \caption{Dynamic spectra of the 2-hour observation of UV~Cet on 2013 June 2, in the style of Figure~\ref{fig:2015_EQPeg_2}, shown twice with different binning and color scales. There is a long-duration event spanning 1-4~GHz and lasting roughly 1 hour, with multiple bright short-duration features embedded within it. \label{fig:2013_UVCet_2}}
\end{figure}

\begin{figure}[h]
  \centering
  \includegraphics[angle=0]{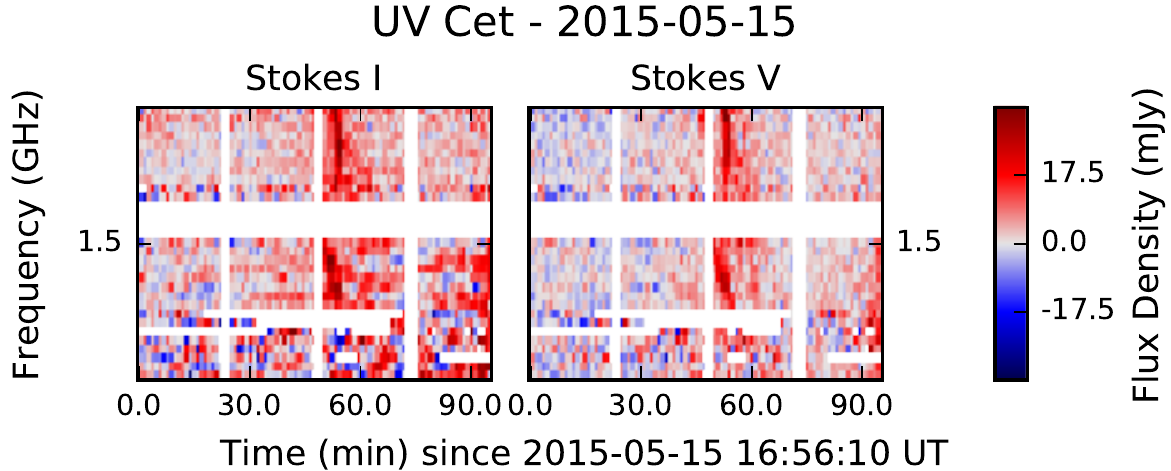}
  \caption{Dynamic spectra of the 2-hour observation of UV~Cet on 2015 May 15, in the style of Figure~\ref{fig:2015_EQPeg_2}. Only 1-2~GHz was observed on this date. There is a short-duration burst spanning 1.2-2~GHz at t$\sim$60~min.\label{fig:2015_UVCet_1}}
\end{figure}

\begin{figure}[h]
  \centering
  \includegraphics[angle=0]{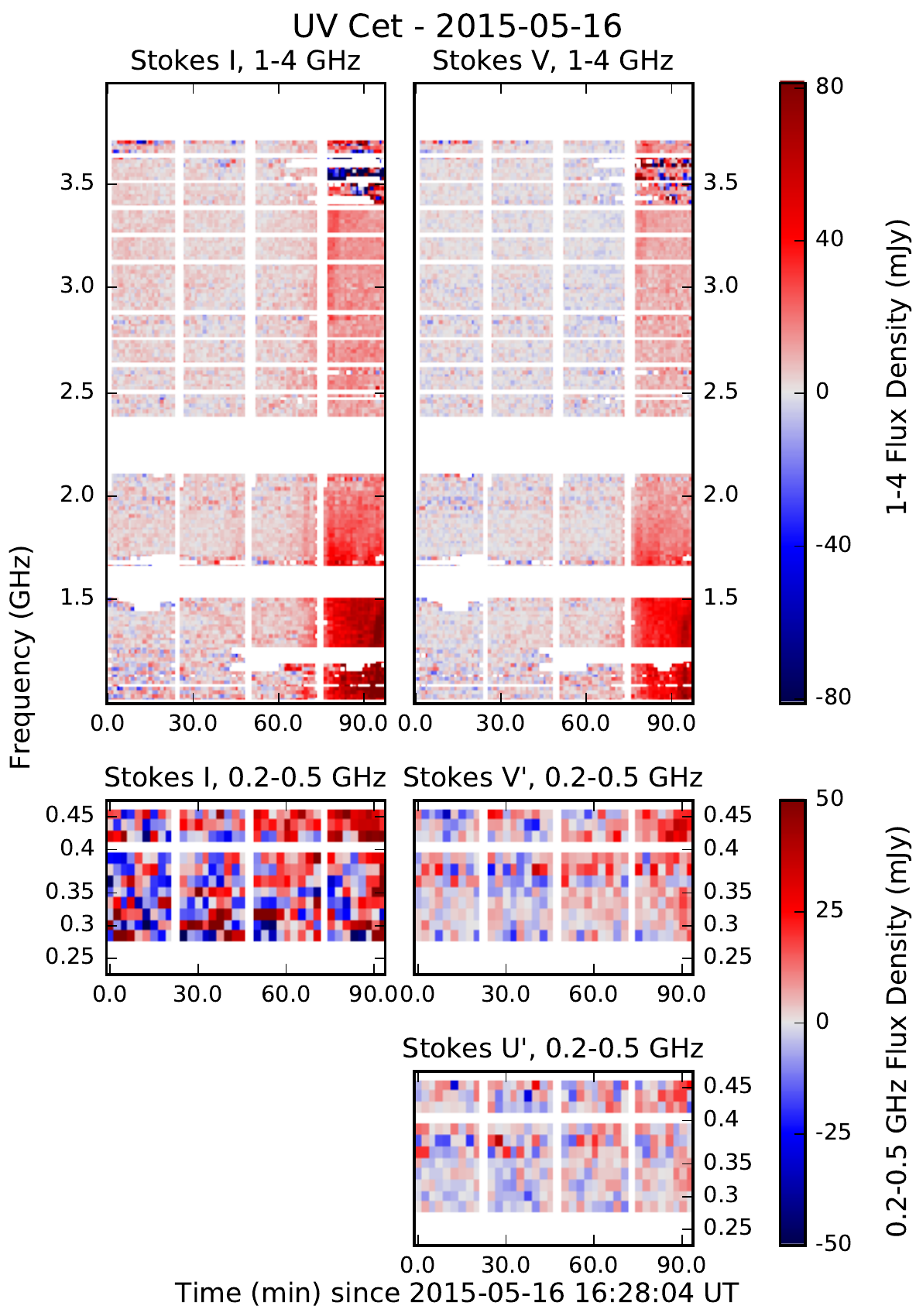}
  \caption{Dynamic spectra of the 2-hour observation of UV~Cet on 2015 May 16, in the style of Figure~\ref{fig:2015_EQPeg_2}. There is a coherent burst in the last 20 minutes apparently spanning the entire observed frequency band (at least 0.3-3.7~GHz). We classify this event as ``long-duration'' due to its morphological similarity to the start of other long-duration events on UV~Cet. \label{fig:2015_UVCet_2}}
\end{figure}

\begin{figure}[h]
  \centering
  \includegraphics[angle=270]{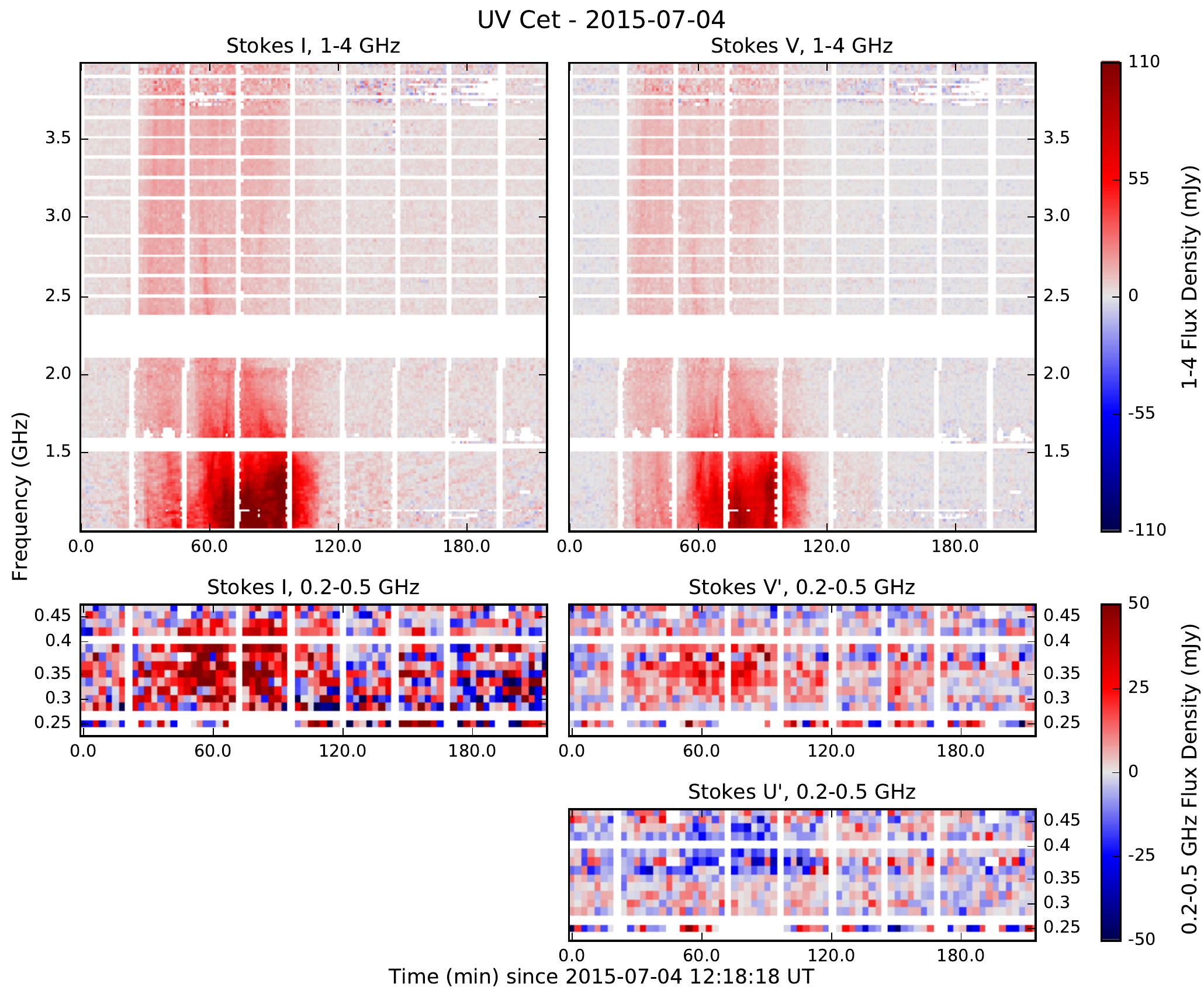}
  \caption{Dynamic spectra of the 4-hour observation of UV~Cet on 2015 July 4, in the style of Figure~\ref{fig:2015_EQPeg_2}. This epoch contains a 90-minute-long burst spanning at least 0.3-4~GHz (also detected at 8.2-8.5~GHz in simultaneous VLBA observations, publication forthcoming). \label{fig:2015_UVCet_3}}
\end{figure}

\begin{figure}[h]
  \centering
  \includegraphics[angle=270]{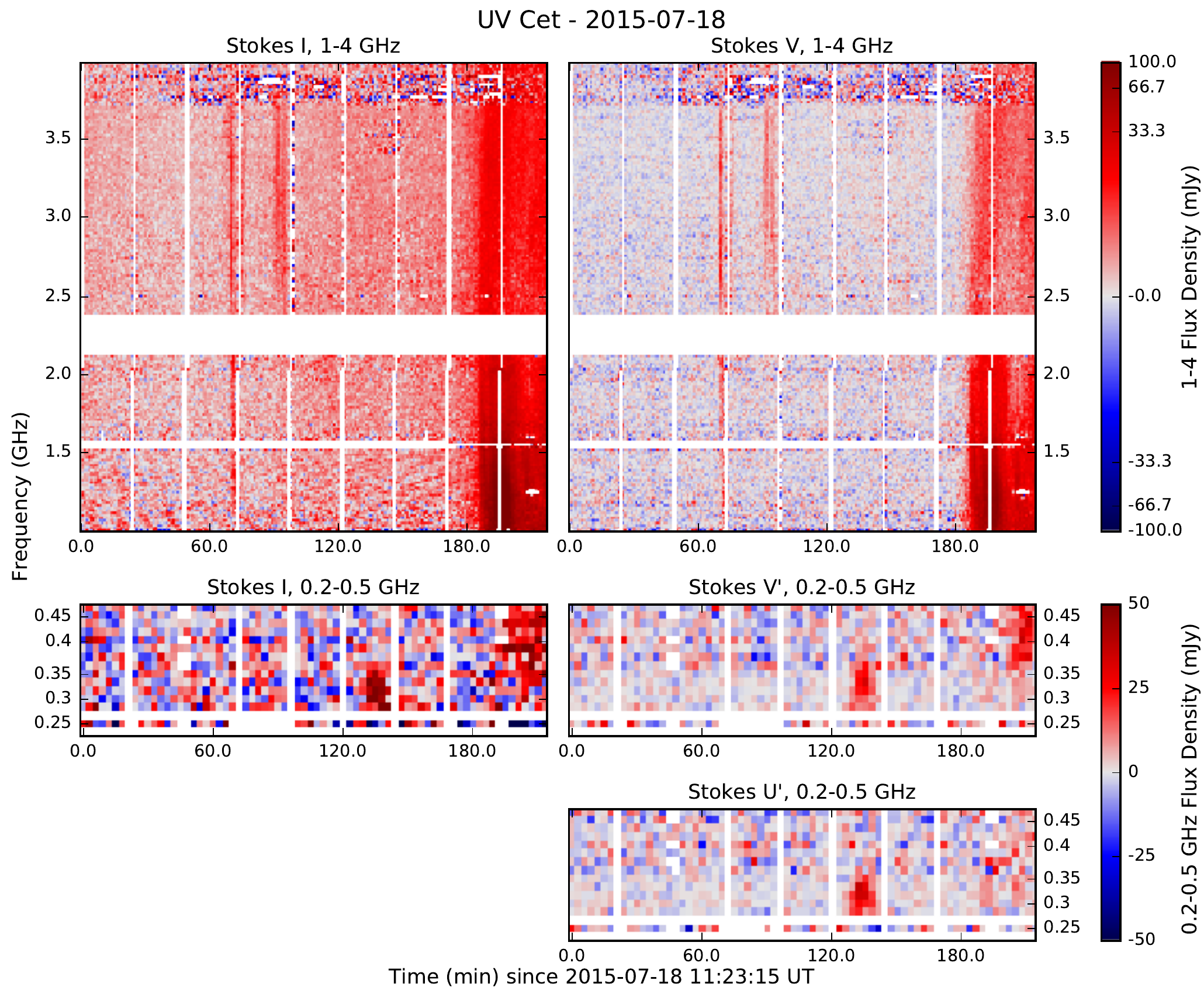}
  \caption{Dynamic spectra of the 4-hour observation of UV~Cet on 2015 July 18, in the style of Figure~\ref{fig:2015_EQPeg_2}. We categorize the bursts in this epoch as 3 events: a long-duration burst spanning 0.35-4~GHz in the last 30 minutes (also detected at 8.2-8.5~GHz in simultaneous VLBA observations), a short-duration event consisting of a series of at least 4 short bursts spanning at least 1-3.7~GHz, at t$\sim$70-95~min, and a short-duration burst at 0.28-0.37~GHz, at t$\sim$135~min.\label{fig:2015_UVCet_4}}
\end{figure}

\begin{figure}[h]
  \centering
  \includegraphics[angle=270]{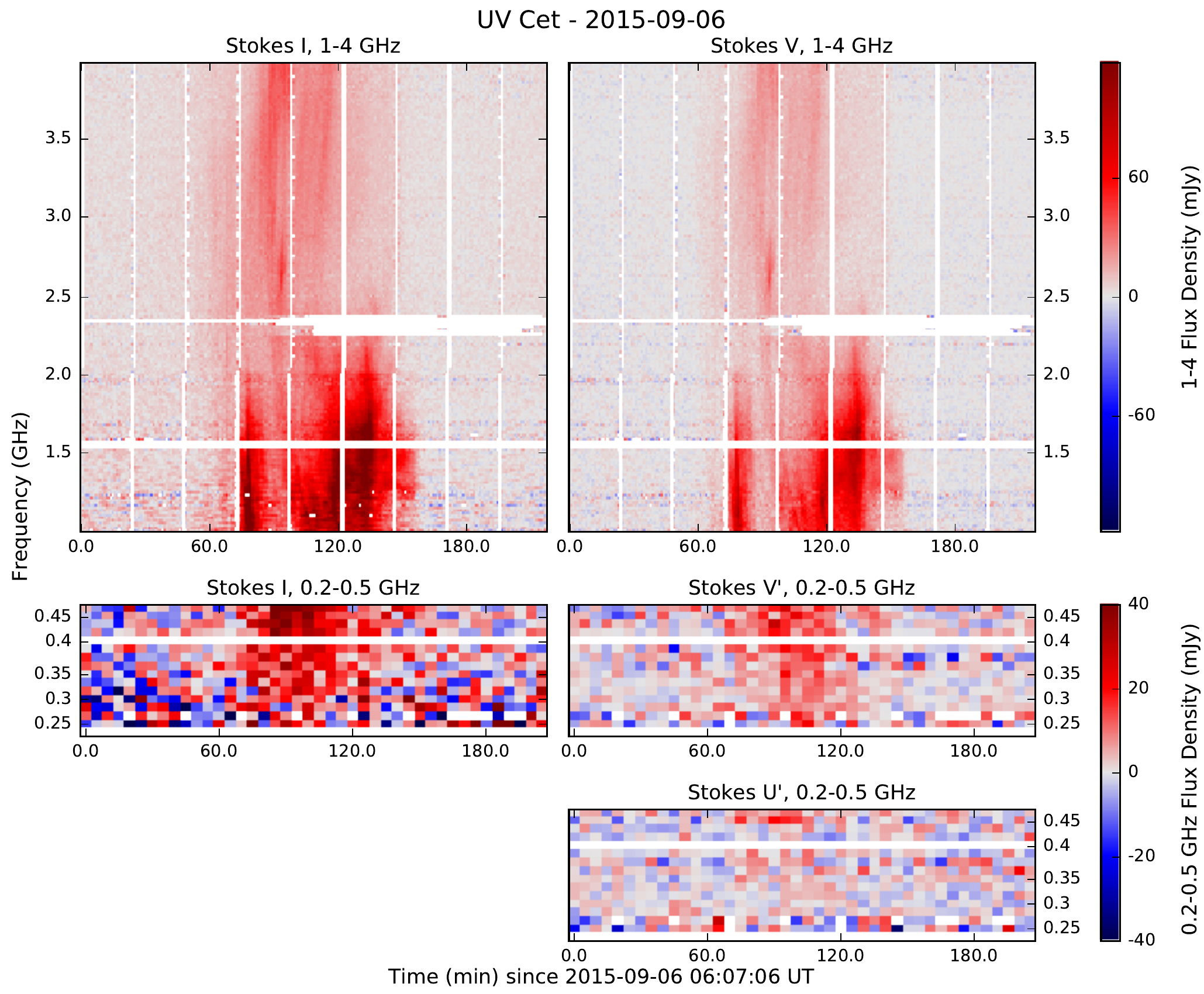}
  \caption{Dynamic spectra of the 4-hour observation of UV~Cet on 2015 September 6, in the style of Figure~\ref{fig:2015_EQPeg_2}. This epoch contains one event, a long-duration burst spanning 0.25-4~GHz (also detected at 8.2-8.5~GHz in simultaneous VLBA observations).\label{fig:2015_UVCet_5}}
\end{figure}

\end{subfigures}

\end{document}